\documentclass[aps,superscriptaddress,twocolumn,showpacs,dvips]{revtex4}
\usepackage{feynmp}
\usepackage{amssymb}
\usepackage{epsfig}
\usepackage{graphicx}
\usepackage{subfigure}
\usepackage{hyperref}
\usepackage{bbm}
\usepackage{gensymb}
\usepackage{float}

\begin{document}

\title{Influence of Coulomb interaction on the anisotropic Dirac cone in graphene}

\author{Jing-Rong Wang}
\affiliation{Department of Modern Physics, University of Science and
Technology of China, Hefei, Anhui 230026, P. R. China}
\affiliation{Max Planck Institut f$\ddot{u}$r Physik komplexer
Systeme, D-01187 Dresden, Germany}
\author{Guo-Zhu Liu}
\affiliation{Department of Modern Physics, University of Science and
Technology of China, Hefei, Anhui 230026, P. R. China}

\begin{abstract}
Anisotropic Dirac cone can appear in a number of correlated electron
systems, such as cuprate superconductor and deformed graphene. We
study the influence of long-range Coulomb interaction on the
physical properties of an anisotropic graphene by using the
renormalization group method and $1/N$ expansion, where $N$ is the
flavor of Dirac fermions. Our explicit calculations reveal that the
anisotropic fermion velocities flow monotonously to an isotropic
fixed point in the lowest energy limit in clean graphene. We then
incorporate three sorts of disorders, including random chemical
potential, random gauge potential, and random mass, and show that
the interplay of Coulomb interaction and disorders can lead to rich
and unusual behaviors. In the presence of strong Coulomb interaction
and random chemical potential, the fermion velocities are driven to
vanish at low energies and the system turns out to be an exotic
anisotropic insulator. In the presence of Coulomb interaction and
other two types of disorders, the system flows to an isotropic
low-energy fixed point more rapidly than the clean case, and
exhibits non-Fermi liquid behaviors. We also investigate the
non-perturbative effects of Coulomb interaction, focusing on how the
dynamical gap is affected by the velocity anisotropy. It is found
that, the dynamical gap is enhanced (suppressed) as the fermion
velocities decrease (increase), but is suppressed as the velocity
anisotropy increases.

\end{abstract}

\pacs{71.10.Hf, 73.43.Nq, 74.62.En}

\maketitle


\section{Introduction}

Massless Dirac fermions with a relativistic dispersion are known to
be the low-energy elementary excitations in a variety of
two-dimensional (2D) condensed matter systems, including $d$-wave
superconductors \cite{LeeRMP, Orenstein}, topological insulators
\cite{topo}, and graphene \cite{Novoselov04, Novoselov05,
CastroNeto, Peres, DasSarma, Kotov}. Different from conventional
Schrodinger electron systems with a finite Fermi surface, 2D Dirac
fermion systems have discrete Fermi points and a vanishing density
of states (DOS) at the lowest energy. Due to this difference, Dirac
fermion systems exhibit nontrivial properties that cannot be
realized in electron systems with finite Fermi surface. These
properties become particularly interesting when massless Dirac
fermions couple to some kind of massless bosonic modes. For
instance, Dirac fermions may interact strongly with gauge field,
long-range Coulomb potential, or critical fluctuation of an order
parameter, depending on the concrete materials.

If a Dirac fermion system has an isotropic Dirac cone, there will be
an uniform fermion velocity $v_F$ that can be defined from the
kinetic energy $\varepsilon(k)$ by the relationship, $v_F \propto
\partial \varepsilon(k)/\partial k |_{k_{F}}$. However, in many
cases the Dirac fermion systems may be spatially anisotropic for
various reasons. A well-known example is the quasi-2D $d_{x^2 -
y^2}$-wave cuprate superconductors \cite{LeeRMP, Orenstein}, where
the massless nodal quasiparticles have a Fermi velocity $v_{F}$ and
a gap velocity $v_{\Delta}$, obtained from the derivatives of Fermi
energy and superconducting gap respectively. These two velocities
are not equal in magnitude \cite{Orenstein}, and their ratio
$v_{F}/v_{\Delta}$ can be as large as $10 \sim 20$. The velocity
ratio is known to strongly affect many observable quantities
\cite{Orenstein}. Moreover, it is recently discovered that the
isotropic Dirac cone of graphene can be made anisotropic once some
external force, which might be uniaxial strain \cite{Pereira09,
Ribeiro09, Goerbig08, Choi10} or external periodic potential
\cite{ParkNP08, ParkPRL08, Rusponi10}, is applied to the originally
ideal honeycomb lattice. When this happens, Dirac fermions have two
different velocities, $v_{1}$ and $v_{2}$, with their ratio $\delta
= v_{2}/v_{1}$ measuring the extent of spatial anisotropy. In
addition, it is also possible to realize anisotropic Dirac cone in
other Dirac fermion systems.

An interesting and widely studied problem is how the velocity
anisotropy in Dirac fermion systems is affected by various
interactions. We would like to know whether it is enhanced,
weakened, or entirely suppressed. These problems deserve serious and
systematic investigations for two reasons. First, the velocity ratio
enters into many observable physical quantities, and hence should
have measurable effects. Second, the interaction-induced nontrivial
renormalization of velocity ratio can lead to a number of unusual
behaviors. In the existing literature, the interactions of Dirac
fermions with two sorts of critical bosonic excitations are broadly
studied: gauge field and order parameter fluctuation.

\emph{Gauge field}: It has been proposed that many unusual physics
of underdoped cuprates can be described by an effective QED$_{3}$
theory \cite{LeeRMP}. Within this effective theory, massless Dirac
fermions couple strongly to an emergent U(1) gauge field, which may
have different physical origins in different models. Detailed
renormalization group (RG) calculations have shown that gauge
interaction drives the anisotropic fermion velocities to flow to an
isotropic fixed point \cite{Vafek02, Lee02, Hermele05}, i.e.,
$v_{F}/v_{\Delta} \rightarrow 1$, in the low-energy regime.
Therefore, the velocity anisotropy is irrelevant, and there might be
a restored relativity \cite{Vafek02}.

\emph{Order parameter fluctuation}: In the close vicinity of certain
quantum phase transitions, massless Dirac fermions may couple to the
fluctuation of some order parameters. For instance, the fermions
interact with the fluctuation of a nematic order parameter at a
nematic quantum critical point \cite{Kim08, Huh, Xu, Fritz, Wang11,
Liu2012, Wang13}, which is supposed to exist in some $d$-wave
superconductors. In contrast to the case of gauge interaction, such
interaction leads to an extreme anisotropy of fermion velocities
\cite{Kim08, Huh}, i.e., $v_{\Delta}/v_{F} \rightarrow 0$. Such
extreme anisotropy can give rise to a series of intriguing
properties, such as non-Fermi liquid behavior \cite{Kim08, Xu},
enhancement of thermal conductivity \cite{Fritz}, and suppression of
superconductivity \cite{Liu2012}. Furthermore, Dirac fermions may
couple to an incommensurate antiferromagnetic order parameter. It
was demonstrated \cite{Pelissetto} that this coupling is very
similar to that between fermions and nematic order parameter, so one
could expect an analogous extreme anisotropy in this case.

In this paper, we further investigate the influence of strong
interactions in anisotropic Dirac fermion systems. Here we consider
the long-range Coulomb interaction in a graphene that exhibits an
anisotropic Dirac cone. RG techniques \cite{Shankar94} will be used
to address this issue. We are mainly interested in how the velocity
ratio $\delta$ flows in the low-energy regime and how such flow
affects the physical properties of the system.

Recently, the influence of Coulomb interaction on Dirac fermions
with anisotropic dispersion is studied by Sharma \emph{et. al.}
\cite{Sharma12}, who have performed RG calculations by making
perturbative expansion in powers of coupling constant $\alpha_{1} =
e^{2}/v_{1}\epsilon$, where $v_{1}$ is supposed to the larger
component of the two velocities. It was argued \cite{Sharma12} that
Coulomb interaction can give rise to unusual behaviors. The RG
scheme adopted in Ref.~\cite{Sharma12} could be improved in two
important aspects. First, in the present problem, the Coulomb
interaction strength is actually determined by both of the two
components of fermion velocities, i.e., $v_{1}$ and $v_{2}$, hence
the RG calculations performed by making expansion in powers of
$\alpha_{1}$ may not be able to capture all the essential features,
especially when the anisotropy becomes strong. Second, the
perturbative expansion in powers of coupling constant $\alpha_{1}$
is valid only in the weak coupling regime, i.e., $\alpha_{1} \ll 1$.
However, the Coulomb interaction is known to play a much more
significant role in the strong coupling regime, which cannot be
accessed by the expansion scheme based on small $\alpha_{1}$.

In order to perform a more general analysis that applies to both
weak and strong couplings, here we will make use of the
$1/N$-expansion method, with $N$ being the flavor of Dirac fermions.
This method proves to be powerful in dealing with field-theoretic
models of strongly interacting fermionic systems. Although the
physical fermion flavor is taken to be $N=2$, to be explained in
Sec.~\ref{sec:Hamiltonian}, we will consider a general large $N$. An
important advantage of this $1/N$-expansion method is that it can be
straightforwardly generalized to include the non-perturbative
effects of strong Coulomb interaction. After performing explicit RG
calculations, we will show that both $v_1$ and $v_2$ increase
monotonously with the decreasing energy and that the velocity ratio
flows to unity, i.e., $v_2/v_1 \rightarrow 1$, in the lowest energy
limit. Apparently, the anisotropic Dirac fermion system is driven to
approach a stable isotropic fixed point at low energies by the
Coulomb interaction.

It is also interesting to study the effects of quenched disorders,
which exist in almost all realistic graphene samples and are known
to govern many low-temperature transport properties
\cite{CastroNeto, DasSarma, Kotov}. The interactions between Dirac
fermions and various disorders have recently stimulated extensive
research works \cite{Ludwig, Stauber05, Herbut08, Vafekdis}.
According to their coupling to Dirac fermions, disorders are usually
divided into three classes: random chemical potential, random gauge
potential, and random mass. We investigate the interplay of Coulomb
interaction and fermion-disorder interaction, and demonstrate that
it leads to a series of unusual behaviors, including breakdown of
Fermi liquid and emergence of non-Fermi-liquid states. Further, it
is shown that random chemical potential exerts very different
influence on the system compared with random gauge potential and
random mass. To understand these behaviors in more details, we
calculate Landau damping rate, DOS and specific heat after taking
into account the effects of singular velocity renormalization and
then discuss the physical properties of these quantities.

When the Coulomb interaction is sufficiently strong, a finite
fermion gap may be dynamically generated through the formation of
excitonic particle-hole pairs \cite{Khveshchenko01, Gorbar02,
Khveshchenko04, Liu09, Khveshchenko09, LiuWang, Gamayun10, Zhang11,
Wang12, Drut09, Hands08, CastroNetoPhys}. The dynamical gap
generation drives an instability of the original semimetal ground
state of graphene and leads to semimetal-insulator quantum phase
transition. Since the conventional perturbative expansion is unable
to study this problem, we will combine $1/N$-expansion with
Dyson-Schwinger (DS) gap equation method, and then analyze the
non-perturbative effects of strong Coulomb interaction. Our main
interest here is the dependence of dynamical gap generation on the
fermion velocities and velocity ratio. In the presence of velocity
anisotropy, the DS gap equation is formally very complicated. To
simplify numerical computations, we introduce a number of
approximations and try to extract some common feature from the
numerical results. Our results show that, the dynamical gap gets
enhanced (suppressed) as the fermion velocities decrease (increase),
whereas the dynamical gap is suppressed as the anisotropy increases.

The rest of paper is organized as follows. In Sec.~\ref{sec:Hamiltonian},
we write down the Hamiltonian and provide the
Feynman rules which are used in the following calculations. Three
sorts of disorders are introduced explicitly in this section. In
Sec.~\ref{sec:Derivation}, we calculate the corrections to the
self-energy function of fermions and the fermion-disorder vertex due
to the interplay of Coulomb interaction and fermion-disorder
interaction. We then derive the RG flow equations for fermion
velocities and disorder strength parameters. In Sec.~\ref{Sec:NumResults},
we present numerical solutions of RG equations
at four different limits and give a detailed interpretation of the
results. In Sec.~\ref{Sec:InfluenceObQu}, we compute a number of
physical quantities after taking into account the velocity
renormalization. In Sec.~\ref{Sec:Gap}, we consider the effects
of anisotropy on dynamical gap generation after including the
non-perturbative effects of Coulomb interaction. In Sec.~\ref{Sec:Summ},
we summarize our results and discuss their physical implications.

\section{Model Hamiltonian\label{sec:Hamiltonian}}

After monolayer graphene was successfully separated in laboratories
\cite{Novoselov04, Novoselov05}, a great deal of experimental and
theoretical efforts have been devoted to explore its novel and
fascinating properties \cite{CastroNeto, DasSarma, Kotov}. Compared
to the conventional metals, the most remarkable new feature of
graphene is that its low-energy excitations are massless Dirac
fermions having a linear dispersion. Since the fermion DOS vanishes
at the neutral Dirac points, the Coulomb interaction between Dirac
fermions remains long-ranged after including the dynamical screening
due to particle-hole excitations. It is thus widely expected that
such long-range Coulomb interaction is responsible for many unusual
behaviors of graphene \cite{CastroNeto, DasSarma, Kotov}.

The physical effects of Coulomb interaction have been extensively
investigated, with those on fermion velocity renormalization
\cite{Gonzalez9394, Gonzalez99, Son07}, thermodynamics
\cite{Vafek07, Sheehy07, Hwang07}, and electric conductivity
\cite{Herbut08, Fritz08, Mish08, Sheehy09, Juricic10, Kotov08,
Sodemann12, Rosenstein13, Gazzola13} being particularly intriguing.
Here we are mainly interested in the singular fermion velocity
renormalization caused by Coulomb interaction. If the Dirac cone is
isotropic, the uniform velocity $v_{F}$ will be strongly
renormalized by Coulomb interaction, and driven to diverge in the
lowest energy limit \cite{Gonzalez9394, Gonzalez99, Son07}. It is
remarkable that the predicted singular renormalization of fermion
velocity has already been observed in ultra clean suspended graphene
\cite{Elias11}, in graphene placed on boron nitride (BN) substrate
\cite{Yu13}, and in ARPES measurements of quasi-freestanding
graphene on silicon carbide (SiC) \cite{Siegel12}. However, when
graphene exhibits an anisotropic Dirac cone, the two components of
fermions velocities should be renormalized separately. In this case,
the velocity ratio may flow to some nontrivial fixed point.

We now write down the total Hamiltonian of the system. The free
Hamiltonian of massless Dirac fermions with anisotropic dispersion
is
\begin{eqnarray}
H_{0} = i\sum_{\sigma=1}^{N}\int d^{2}\mathbf{x}
\bar{\Psi}_{\sigma}(\mathbf{x}) \left[v_{1}\gamma_{1}\nabla_{1} +
v_{2}\gamma_{2}\nabla_{2}\right] \Psi_{\sigma}(\mathbf{x}),
\end{eqnarray}
where $\bar{\Psi} = \Psi^{\dag}\gamma_{0}$. Here we have defined $4
\times 4$ matrices $\gamma_{0,1,2} =
(\tau_{3},-i\tau_{2},i\tau_{1})\otimes\tau_{3}$ in terms of Pauli
matrices $\tau_{i}$ with $i = 1,2,3$, which satisfy the
anticommutation relation $\left\{\gamma_{\mu},\gamma_{\nu}\right\} =
2\mathrm{diag}(1,-1,-1)$. The spin index is $\sigma$, which takes
all integers from $1$ to $N$. The physical value of spin degeneracy
is $N = 2$. However, to perform $1/N$-expansion, it is convenient to
generalize the flavor to a large, general $N$. The two spatial
components of the anisotropic fermion velocities are $v_{1}$ and
$v_{2}$, respectively. The massless Dirac fermions couple to each
other through the long-range Coulomb interaction
\begin{eqnarray}
H_{\mathrm{ee}} = \frac{1}{4\pi}\sum_{\sigma,\sigma'=1}^{N} \int
d^{2}\mathbf{x} d^{2}\mathbf{x}'\rho_{\sigma}(\mathbf{x})
\frac{e^2}{\epsilon\left|\mathbf{x}-\mathbf{x}'\right|}
\rho_{\sigma'}(\mathbf{x}'),
\end{eqnarray}
where $\rho_{\sigma}(\mathbf{x}) = \bar{\Psi}_{\sigma}(\mathbf{x})
\gamma_{0}\Psi_{\sigma}(\mathbf{x})$ and $\epsilon$ is dielectric
constat whose magnitude is determined by the substrate.

The disorder scattering process can be described by coupling the
Dirac fermions to a random field $A(\mathbf{x})$ in the following
manner \cite{Stauber05}
\begin{eqnarray}
H_{\mathrm{dis}} = v_{\Gamma}\sum_{\sigma=1}^{N}\int d^2\mathbf{x}
\bar{\Psi}_{\sigma}(\mathbf{x})\Gamma\Psi_{\sigma}(\mathbf{x})
A(\mathbf{x}).
\end{eqnarray}
The random field $A(\mathbf{x})$ is a quenched, Gaussian variable
which satisfies
\begin{eqnarray}
\left< A(\mathbf{x})\right> = 0,\qquad \left<
A(\mathbf{x})A(\mathbf{x}') \right> =
\Delta\delta^2(\mathbf{x}-\mathbf{x}'),
\end{eqnarray}
where $\Delta$ is a dimensionless variance. Here we consider three
types of disorders distinguished by the definitions of the $\Gamma$
matrix \cite{Stauber05}. More concretely, $\Gamma = \gamma_{0}$ for
random chemical potential, $\Gamma=(\gamma_{1},\gamma_{2})$ and
$v_{\Gamma}=(v_{\Gamma1},v_{\Gamma2})$ for random gauge potential,
and $\Gamma = \mathbbm{1}_{4 \times 4}$ for random mass. Physically,
the random chemical potential may be induced by local defects,
neutral impurity atoms or neutral absorbed atoms in the plane of
graphene \cite{Peres, Mucciolo10}; random gauge field can be
generated by ripples of graphene \cite{CastroNeto, Herbut08,
Vafekdis, Meyer07}; random mass may be produced by the random
configurations of the substrates \cite{Champel10, Kusminskiy11}.

Starting from $H_{0}$, it is easy to obtain the free Dirac fermion
propagator
\begin{eqnarray}
G_{0}(i\omega,\mathbf{k}) = \frac{1}{-i\omega\gamma_{0} +
v_{1}k_{1}\gamma_{1} + v_{2}k_{2}\gamma_{2}}.\label{eqn:FreeFermion}
\end{eqnarray}
The bare Coulomb interaction is
\begin{eqnarray}
D_{0}(q) = \frac{2\pi e^{2}}{\epsilon|\mathbf{q}|}.
\end{eqnarray}
At the one-loop level, the polarization is given by
\begin{eqnarray}
\Pi(i\Omega,\mathbf{q}) &=& -N\int\frac{d\omega}{2\pi}
\frac{d^2\mathbf{k}}{(2\pi)^2} \mathrm{Tr}\left[\gamma_{0}
G_{0}\left(i\omega,\mathbf{k}\right)\gamma_{0}\right. \nonumber
\\
&&\times \left.G_{0}(i\omega+i\Omega,\mathbf{k}+\mathbf{q})\right]
\nonumber \\
&=& \frac{N}{8v_1v_2}\frac{v_{1}^2q_{1}^{2} +
v_{2}^{2}q_{2}^{2}}{\sqrt{\Omega^2 +
v_{1}^2q_{1}^{2}+v_{2}^{2}q_{2}^{2}}},\label{eqn:Polarization}
\end{eqnarray}
It is  consistent with the polarization obtained by Sharma
\emph{et al.} \cite{Sharma12}. It is shown perviously
\cite{Pellegrino11, LeBlanc13} that the polarization in strained
graphene is related to the polarization in unstrained graphene by an
additional prefactor and a linear transformation for the momenta.
Here, the polarization is calculated by starting directly from a
fermion propagator with an anisotropic dispersion, namely
Eq.~(\ref{eqn:FreeFermion}). The polarization obtained in Ref.
\cite{Pellegrino11, LeBlanc13} is basically equivalent to
Eq.~(\ref{eqn:Polarization}) at the zero chemical potential limit.

\begin{figure}[htbp]
\center
\includegraphics[width=3.3in]{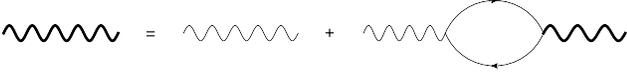}
\caption{One-loop Feynman diagram for dynamical screening of Coulomb
interaction, where solid line represents free propagator of
Dirac fermions, thin wavy line represents bare Coulomb
interaction function, and thick wavy line represents dynamically
screened Coulomb interaction function.}\label{Fig:DressedCoulomb}
\end{figure}

\begin{figure}[htbp]
\center
\includegraphics[width=3in]{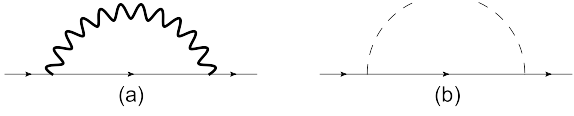}
\caption{Fermion self-energy correction due to (a) Coulomb
interaction and (b) disorder. Dashed line represents
disorder scattering. }\label{Fig:FermionSelfEnergy}
\end{figure}

According to the diagram shown in
Fig.~\ref{Fig:DressedCoulomb}, the dressed Coulomb interaction
should be written as
\begin{eqnarray}
D^{-1}(i\Omega,\mathbf{q}) = \frac{\epsilon|\mathbf{q}|}{2\pi
e^{2}}+\frac{N}{8v_1v_2}
\frac{v_{1}^{2}q_{1}^{2}+v_{2}^{2}q_{2}^{2}}
{\sqrt{\Omega^2+v_{1}^{2}q_{1}^{2}+v_{2}^{2}q_{2}^{2}}}.
\label{eqn:DressedBoson}
\end{eqnarray}
In the isotropic case, $v_{1} = v_{2} = v$, hence the strength of
Coulomb interaction can be well described by a single parameter
$\alpha=e^2/v\epsilon$. In the anisotropic case, however, the
Coulomb interaction are actually characterized by two parameters,
i.e., the coupling $\alpha_{1} = e^2/v_{1}\epsilon$ and the velocity
ratio $\delta=v_{2}/v_{1}$. Therefore, at any given coupling
$\alpha_{1}$, the effective interaction strength is changing as one
tunes the ratio $\delta$.

\section{Renormalization group analysis to the leading order of $1/N$ expansion
\label{sec:Derivation}}

\begin{figure}[htbp]
\center
\includegraphics[width=2.6in]{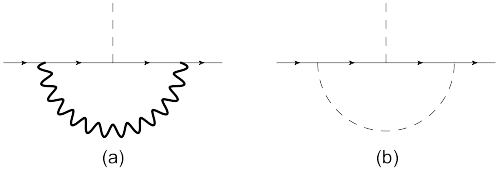}
\caption{Fermion-disorder vertex correction due to (a) Coulomb
interaction and (b) disorder.}\label{Fig:VertexCorrection}
\end{figure}

In this section, we first calculate the self-energy corrections of
Dirac fermions caused by the interplay of Coulomb interaction and
disorder scattering, and then calculate the corrections to the
fermion-disorder vertex. On the basis of these results, we will be
able to derive the analytical expressions of RG flow equations for
fermion velocities and disorder strength parameters. Our
calculations are done to the leading order of $1/N$ expansion.

\subsection{Fermion self-energy and vertex correction \label{subsec:FSCorrectVXCorrect}}

The Dirac fermions receive self-energy corrections from both the
Coulomb interaction and the fermion-disorder interaction, described
by diagrams shown in Fig.~\ref{Fig:FermionSelfEnergy}.

According to Fig.~\ref{Fig:FermionSelfEnergy}(a), the fermion
self-energy due to Coulomb interaction is given by
\begin{eqnarray}
\Sigma_{C}(i\omega,\mathbf{k})
&=&-\int\frac{d^2{\mathbf{q}}}{(2\pi)^{2}}
\int\frac{d\Omega}{2\pi}\gamma_{0}
G_{0}(i\Omega+i\omega,\mathbf{q}+\mathbf{k})\gamma_{0}\nonumber
\\
&&\times D(i\Omega,\mathbf{q}).
\end{eqnarray}
After substituting Eq.~(\ref{eqn:FreeFermion}) and
Eq.~(\ref{eqn:DressedBoson}) into this expression and performing
tedious analytical calculations, which are detailed in
Appendix~\ref{Sec:AppendixA}, we obtain
\begin{eqnarray}
\frac{d \Sigma_{C}(i\omega,\mathbf{k})}{d\ln\Lambda} =
-i\omega\gamma_{0}C_{0} + v_{1}k_{1}\gamma_{1}C_{1} +
v_{2}k_{2}\gamma_{2}C_{2}, \nonumber \\ \label{eqn:SelfEnergyCloumb}
\end{eqnarray}
where
\begin{eqnarray}
C_{0}&=&\frac{1}{8\pi^3}\int_{-\infty}^{+\infty}dx\int_{0}^{2\pi}d\theta
\nonumber \\
&&\times\frac{-x^2+\cos^2\theta+(v_{2}/v_{1})^{2}\sin^2\theta}
{\left(x^2+\cos^2\theta+(v_{2}/v_{1})^{2}\sin^2\theta\right)^{2}}
\mathcal{G}(x,\theta), \label{eq:C0}
\\
C_{1}&=&\frac{1}{8\pi^3}\int_{-\infty}^{+\infty}dx\int_{0}^{2\pi}d\theta
\nonumber \\
&&\times\frac{-x^2+\cos^2\theta-(v_{2}/v_{1})^{2}\sin^2\theta}
{\left(x^2+\cos^2\theta+(v_{2}/v_{1})^{2}\sin^2\theta\right)^{2}}
\mathcal{G}(x,\theta), \label{eq:C1}
\\
C_{2}&=&\frac{1}{8\pi^3}\int_{-\infty}^{+\infty}dx\int_{0}^{2\pi}d\theta
\nonumber \\
&&\times\frac{-x^2-\cos^2\theta+(v_{2}/v_{1})^{2}\sin^2\theta}
{\left(x^2+\cos^2\theta+(v_{2}/v_{1})^{2}\sin^2\theta\right)^{2}}
\mathcal{G}(x,\theta),\label{eq:C2}
\end{eqnarray}
and
\begin{equation}
\mathcal{G}^{-1}(x,\theta)=\frac{1}{2\pi\alpha_{1}} +
\frac{N}{8v_2/v_{1}}\frac{\cos^2\theta+(v_{2}/v_{1})^{2}\sin^2\theta}
{\sqrt{x^2+\cos^2\theta+(v_{2}/v_{1})^{2}\sin^2\theta}} \label{eq:GDef}
\end{equation}
with $\alpha_{1}=\frac{e^{2}}{v_{1}\epsilon}$. Here we point out
that Eqs.(\ref{eq:C0})-(\ref{eq:GDef}) can also be written in
symmetric forms presented in Appendix~\ref{Sec:AppendixC}. In order
to directly compare our results to those obtained based on
perturbative expansion in powers of $\alpha_{1}$ \cite{Sharma12,
Sharma13}, we will use the non-symmetric expressions of $C_{0,1,2}$
and $\mathcal{G}$.

According to Fig.~\ref{Fig:FermionSelfEnergy}(b), the fermion
self-energy induced by disorder takes the form
\begin{eqnarray}
\Sigma_{\mathrm{dis}}(i\omega) &=& \Delta v_{\Gamma}^{2}\int\frac{d^2
\mathbf{q}}{(2\pi)^2}\Gamma G_{0}(i\omega,\mathbf{q})\Gamma \nonumber \\
&=&i\omega v_{\Gamma}^{2}\Delta\int\frac{d^2 \mathbf{q}}{(2\pi)^2}
\frac{\Gamma\gamma_{0}\Gamma} {\left(\omega^{2}+v_{1}^{2}q_{1}^{2} +
v_{2}^{2}q_{2}^{2}\right)}.
\end{eqnarray}
Different from the case of Coulomb interaction,
$\Sigma_{\mathrm{dis}}(i\omega)$ is independent of momentum, which
reflects the fact that disorders are static. We now have
\begin{eqnarray}
\frac{d\Sigma_{\mathrm{dis}}(i\omega)}{d\ln\Lambda} = C_g
i\omega\gamma_{0},
\end{eqnarray}
where
\begin{eqnarray}
C_g = \frac{v_{\Gamma}^{2}\Delta}{2\pi v_{1}v_{2}}
\end{eqnarray}
for random chemical potential and random mass, and
\begin{eqnarray}
C_g = \frac{\left(v_{\Gamma1}^{2} +
v_{\Gamma2}^{2}\right)\Delta}{2\pi v_{1}v_{2}}
\end{eqnarray}
for random gauge potential.

We next consider the corrections to the fermion-disorder vertex,
which receives contributions from both Coulomb interaction and
fermion-disorder interaction, as described by the diagrams shown in
Fig.~\ref{Fig:VertexCorrection}. According to
Fig.~\ref{Fig:VertexCorrection}(a), at zero external momentum and
frequency, the vertex correction due to Coulomb interaction is
calculated as follows
\begin{eqnarray}
V_{C} &=&-\int\frac{d\Omega}{2\pi}\int\frac{d^2\mathbf{q}}{(2\pi)^2}
\gamma_{0}G_{0}(i\Omega,\mathbf{q})v_\Gamma \Gamma
G_{0}(i\Omega,\mathbf{q})\gamma_{0} \nonumber \\
&& \times D(i\Omega,\mathbf{q}).
\end{eqnarray}
After analytical calculations detailed in
Appendix~\ref{Sec:AppendixB}, we have
\begin{eqnarray}
\frac{d V_{C}}{d\ln\Lambda} &=& v_\Gamma\gamma_{0}\left(-C_0\right)
\end{eqnarray}
for random chemical potential;
\begin{eqnarray}
\frac{d V_{C}}{d\ln\Lambda} &=&
v_{\Gamma1}\gamma_{1}\left(-C_1\right) \\
\frac{d V_{C}}{d\ln\Lambda} &=&
v_{\Gamma2}\gamma_{2}\left(-C_2\right)
\end{eqnarray}
for the $\gamma_{1}$ and $\gamma_{2}$ components of random gauge
potential, respectively; and
\begin{eqnarray}
\frac{d V_{C}}{d\ln\Lambda} &=&
v_\Gamma\mathbbm{1}\left(C_{0}-C_{1}-C_{2}\right)
\end{eqnarray}
for random mass;

According to Fig.~\ref{Fig:VertexCorrection}(b), at zero momentum
the vertex correction due to averaging over disorders are found to be
\begin{eqnarray}
V_{\mathrm{dis}}&=& \Delta
v_{\Gamma}^{2}\int\frac{d^2\mathbf{q}}{(2\pi)^2}\Gamma
G_0(i\omega,\mathbf{q})v_\Gamma\Gamma G_0(i\omega,\mathbf{q})\Gamma.
\end{eqnarray}
It is shown that
\begin{eqnarray}
\frac{dV_{\mathrm{dis}}}{d\ln\Lambda} &=&
v_{\Gamma}\gamma_{0}C_{g}
\end{eqnarray}
for random chemical potential;
\begin{eqnarray}
\frac{dV_{\mathrm{dis}}}{d\ln\Lambda}=0
\end{eqnarray}
for random gauge potential; and
\begin{eqnarray}
\frac{dV_{\mathrm{dis}}}{d\ln\Lambda} &=&
-v_{\Gamma}\mathbbm{1}C_{g}
\end{eqnarray}
for random mass.

\subsection{Derivation of the RG equations\label{Subsec:DerivationRG}}

The fermion self-energy corrections and fermion-disorder corrections
obtained in the last subsection will be used to derive the relevant
RG equations. According to the renormalization group theory \cite{Huh, Wang11, Shankar94}, after integrating out
the fields in the momentum shell $\Lambda/b<k<\Lambda$ with $b>1$, where $b$ can be written as $b=e^{l}$
with a running scale $l>0$ , we can get following
action for the fermion
\begin{eqnarray}
S_{\Psi}
&=&\sum_{\sigma=1}^{N}\int\frac{d\omega}{2\pi}\frac{d^{2}\mathbf{k}}{(2\pi)^{2}}
\bar{\Psi}_{\sigma}(i\omega,\mathbf{k})\left[G_{0}^{-1}(i\omega,\mathbf{k})
- \Sigma_{C}(i\omega,\mathbf{k})\right. \nonumber \\
&&\left.-\Sigma_{dis}(i\omega,\mathbf{k})\right]
\Psi_{\sigma}(i\omega,\mathbf{k})\nonumber
\\
&\approx&\sum_{\sigma=1}^{N}\int\frac{d\omega}{2\pi}\frac{d^{2}\mathbf{k}}{(2\pi)^{2}}
\bar{\Psi}_{\sigma}(i\omega,\mathbf{k})
\left[-i\omega\gamma_{0}e^{\int_{0}^{l}dl\left(-C_{0}+C_{g}\right)}\right.
\nonumber \\
&&\left.+v_{1}k_{1}e^{\int_{0}^{l}dl\left(-C_{1}\right)}
+v_{2}k_{2}e^{\int_{0}^{l}dl\left(-C_{2}\right)}\right]
\Psi_{\sigma}(i\omega,\mathbf{k}).
\end{eqnarray}
In the spirit of RG theory \cite{Huh, Wang11, Shankar94}, one can
perform the following re-scaling transformation
\begin{eqnarray}
k_{i}&=&k'_{i}e^{-l},\label{eqn:ScalingM}
\\
\omega&=&\omega'e^{-l},\label{eqn:ScalingE}
\\
\Psi_{\sigma}(i\omega,\mathbf{k}) &=&
\Psi_{\sigma}'(i\omega',\mathbf{k}')
e^{\frac{1}{2}\int_{0}^{l}dl\left(4 + C_{0} - C_{g}\right)},
\label{eqn:ScalingPsi}
\\
v_{1} &=& v_{1}'e^{\int_{0}^{l}dl\left(-C_{0}+C_{1}+C_{g}\right)},
\label{eqn:ScalingV1}
\\
v_{2} &=& v_{2}'e^{\int_{0}^{l}dl\left(-C_{0}+C_{2}+C_{g}\right)},
\label{eqn:ScalingV2}
\end{eqnarray}
which should keep the kinetic term of fermions invariant, namely
\begin{eqnarray}
S_{\Psi'} &=& \sum_{\sigma=1}^{N}\int
\frac{d\omega'}{2\pi}\frac{d^{2}\mathbf{k}'}{(2\pi)^{2}}
\bar{\Psi}_{\sigma}'(i\omega',\mathbf{k}') \left[-i\omega'\gamma_{0}
+ v_{1}'k_{1}'\right.
\nonumber \\
&&\left.+v_{2}'k_{2}'\right] \Psi_{\sigma}'(i\omega',\mathbf{k}').
\end{eqnarray}

After including the influence of interaction, the action for the
disorder scattering to the fermion becomes
\begin{eqnarray}
S_{dis}
&=&\sum_{\sigma=1}^{N}\int\frac{d\omega}{2\pi}\frac{d^{2}\mathbf{k}}{(2\pi)^{2}}
\int\frac{d^{2}\mathbf{k}_{1}}{(2\pi)^{2}}
\bar{\Psi}_{\sigma}(i\omega,\mathbf{k})\left(\Gamma+V_{C}\right.\nonumber
\\
&&\left.+V_{dis}\right) \Psi_{\sigma}(i\omega,\mathbf{k}_{1})
A(\mathbf{k}-\mathbf{k}_{1}).
\end{eqnarray}
Specifically,
\begin{eqnarray}
S_{dis}
&\approx&\sum_{\sigma=1}^{N}\int\frac{d\omega}{2\pi}\frac{d^{2}\mathbf{k}}{(2\pi)^{2}}
\int\frac{d^{2}\mathbf{k}_{1}}{(2\pi)^{2}}
\bar{\Psi}_{\sigma}(i\omega,\mathbf{k})v_{\Gamma}\gamma_{0}\nonumber
\\
&&\times e^{\int_{0}^{l}dl\left(-C_{0} +C_{g} \right)}
\Psi_{\sigma}(i\omega,\mathbf{k}_{1}) A(\mathbf{k}-\mathbf{k}_{1});
\end{eqnarray}
for random chemical potential;
\begin{eqnarray}
S_{dis}
&\approx&\sum_{\sigma=1}^{N}\int\frac{d\omega}{2\pi}\frac{d^{2}\mathbf{k}}{(2\pi)^{2}}
\int\frac{d^{2}\mathbf{k}_{1}}{(2\pi)^{2}}
\bar{\Psi}_{\sigma}(i\omega,\mathbf{k})v_{\Gamma}\gamma_{1}\nonumber
\\
&&\times e^{\int_{0}^{l}dl\left(-C_{1}\right)}
\Psi_{\sigma}(i\omega,\mathbf{k}_{1}) A(\mathbf{k}-\mathbf{k}_{1}),
\\
S_{dis}
&\approx&\sum_{\sigma=1}^{N}\int\frac{d\omega}{2\pi}\frac{d^{2}\mathbf{k}}{(2\pi)^{2}}
\int\frac{d^{2}\mathbf{k}_{1}}{(2\pi)^{2}}
\bar{\Psi}_{\sigma}(i\omega,\mathbf{k})v_{\Gamma}\gamma_{2}\nonumber
\\
&&\times e^{\int_{0}^{l}dl\left(-C_{2}\right)}
\Psi_{\sigma}(i\omega,\mathbf{k}_{1}) A(\mathbf{k}-\mathbf{k}_{1})
\end{eqnarray}
for the $\gamma_{1}$ component and $\gamma_{2}$ component of the
random gauge potential respectively; and
\begin{eqnarray}
S_{dis}
&=&\sum_{\sigma=1}^{N}\int\frac{d\omega}{2\pi}\frac{d^{2}\mathbf{k}}{(2\pi)^{2}}
\int\frac{d^{2}\mathbf{k}_{1}}{(2\pi)^{2}}
\bar{\Psi}_{\sigma}(i\omega,\mathbf{k})v_{\Gamma}\mathbbm{1}\nonumber
\\
&&\times e^{\int_{0}^{l}dl\left(C_{0}-C_{1}-C_{2} -C_{g}\right)}
\Psi_{\sigma}(i\omega,\mathbf{k}_{1})\nonumber
\\
&&\times A(\mathbf{k}-\mathbf{k}_{1})
\end{eqnarray}
for the random mass. Carry out the scaling (\ref{eqn:ScalingM}),
(\ref{eqn:ScalingE}), (\ref{eqn:ScalingPsi}) along with
\begin{eqnarray}
A(\mathbf{k})&=&A'(\mathbf{k'})e^{l},
\end{eqnarray}
and
\begin{eqnarray}
v_{\Gamma}=v'_{\Gamma}\label{eqn:ScalingCP}
\end{eqnarray}
for random chemical potential;
\begin{eqnarray}
v_{\Gamma1}&=&v_{\Gamma1}'e^{\int_{0}^{l}dl\left(-C_{0}+C_{1}+C_{g}\right)},
\label{eqn:ScalingGP1}
\\
v_{\Gamma2}&=&v_{\Gamma2}'e^{\int_{0}^{l}dl\left(-C_{0}+C_{2}+C_{g}\right)}
\label{eqn:ScalingGP2}
\end{eqnarray}
for $\gamma_{1}$ and $\gamma_{2}$ components of random gauge
potential respectively; and
\begin{eqnarray}
v_{\Gamma}=v_{\Gamma}'e^{\int_{0}^{l}dl\left(-2C_{0}+C_{1}+C_{2}
+2C_{g}\right)}\label{eqn:ScalingMass}
\end{eqnarray}
for random mass. Then the corresponding action can keep the
invariant form as
\begin{eqnarray}
S_{dis} &=&\sum_{\sigma=1}^{N}\int\frac{d\omega'}{2\pi}
\frac{d^{2}\mathbf{k}'}{(2\pi)^{2}}
\int\frac{d^{2}\mathbf{k}_{1}'}{(2\pi)^{2}}
\bar{\Psi}_{\sigma}'(i\omega',\mathbf{k}')v_{\Gamma}'\Gamma
\nonumber \\
&&\times\Psi_{\sigma}'(i\omega',\mathbf{k}_{1}')
A'(\mathbf{k}'-\mathbf{k}_{1}').
\end{eqnarray}
From Eqs.~(\ref{eqn:ScalingV1}), (\ref{eqn:ScalingV2}) and
Eqs.~(\ref{eqn:ScalingCP}), (\ref{eqn:ScalingGP1}),
(\ref{eqn:ScalingGP2}), (\ref{eqn:ScalingMass}), we can get the
renormalization group equations
\begin{eqnarray}
\frac{dv_{1}}{dl}&=&\left(C_{0}-C_{1}-C_{g}\right)v_{1}\label{eqn:RGV1},
\\
\frac{dv_{2}}{dl}&=&\left(C_{0}-C_{2}-C_{g}\right)v_{2}\label{eqn:RGV2},
\\
\frac{d\left(v_{2}/v_{1}\right)}{dl} &=&\left(C_{1} -
C_{2}\right)\frac{v_{2}}{v_{1}}\label{eqn:RGVRatio},
\end{eqnarray}
and
\begin{eqnarray}
\frac{d v_{\Gamma}}{dl}=0\label{eqn:RGVGammaCP}
\end{eqnarray}
for random chemical potential;
\begin{eqnarray}
\frac{dv_{\Gamma1}}{dl}&=&\left(C_{0}-C_{1}-C_{g}\right)v_{\Gamma1}
\label{eqn:RGVGammaGP1}
\\
\frac{dv_{\Gamma2}}{dl}&=&\left(C_{0}-C_{2}-C_{g}\right)v_{\Gamma2}
\label{eqn:RGVGammaGP2}
\end{eqnarray}
for $\gamma_{1}$ and $\gamma_{2}$ components of random gauge
potential respectively;
\begin{eqnarray}
\frac{dv_{\Gamma}}{dl}=\left(2C_{0}-C_{1}-C_{2}
-2C_{g}\right)v_{\Gamma}\label{eqn:RGVGammaMass}
\end{eqnarray}
for random mass.

\section{Numerical Results\label{Sec:NumResults}}

In this section, we present numerical solutions of RG equations
obtained in Sec.~\ref{sec:Derivation} and discuss their physical
implications. In order to examine the effects of various physical
mechanisms and parameters, it is helpful to analyze the results at
different limits. First, we consider the case of isotropic Dirac
cone in the absence of disorders. Second, we consider the case of
anisotropic Dirac cone in the absence of disorders. Third, we
consider isotropic Dirac cone in the presence of disorders. Finally,
we turn to anisotropic Dirac cone in the presence of disorders.

\subsection{Clean isotropic case\label{subsec:ClIso}}

\begin{figure}[htbp]
\center
\includegraphics[width=2.8in]{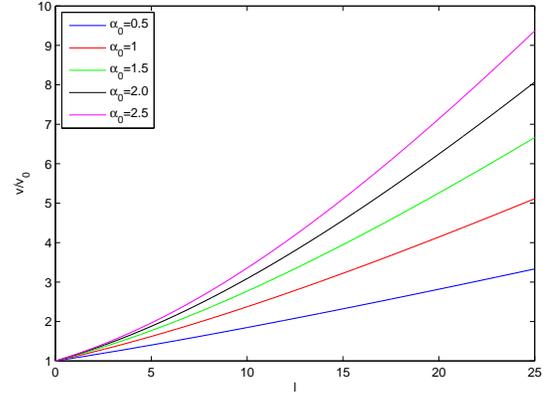}
\caption{Renormalized fermion velocity for isotropic Dirac cone and
without disorder.}\label{eqn:VRGIsCl}
\end{figure}

We first consider graphene with isotropic Dirac cone and uniform
velocity, $v_{1} = v_{2} = v$, and assume the sample is clean. In
this case, the velocity flows as
\begin{eqnarray}
\frac{dv}{dl} &=& Cv, \label{eqn:RGVIsCl}
\end{eqnarray}
where
\begin{eqnarray}
C&=&\frac{4}{N\pi^2}\left[1-\frac{1}{\lambda}\frac{\pi}{2}
+\frac{1}{\lambda}\left\{
\begin{array}{ll}
\frac{1}{\sqrt{1-\lambda^2}}\arccos\left(\lambda\right) & \lambda<1
\\
\\
\frac{1}{\sqrt{\lambda^2-1}}\mathrm{arccosh}\left(\lambda\right) &
\lambda>1
\\
\\
1 & \lambda=1
\end{array}
\right.\right]\nonumber
\end{eqnarray}
with $\lambda = \frac{N\pi \alpha}{4}$. This result has previously
been obtained by Son \cite{Son07}. The renormalized fermion
velocity, shown in Fig.~\ref{eqn:VRGIsCl}, increases monotonically
in the low-energy regime. It is interesting that this behavior is
recently observed in experiments \cite{Elias11, Yu13, Siegel12}.

\subsection{Clean anisotropic case}

\begin{figure}[htbp]
\centering \subfigure{\includegraphics[width=2.8in]{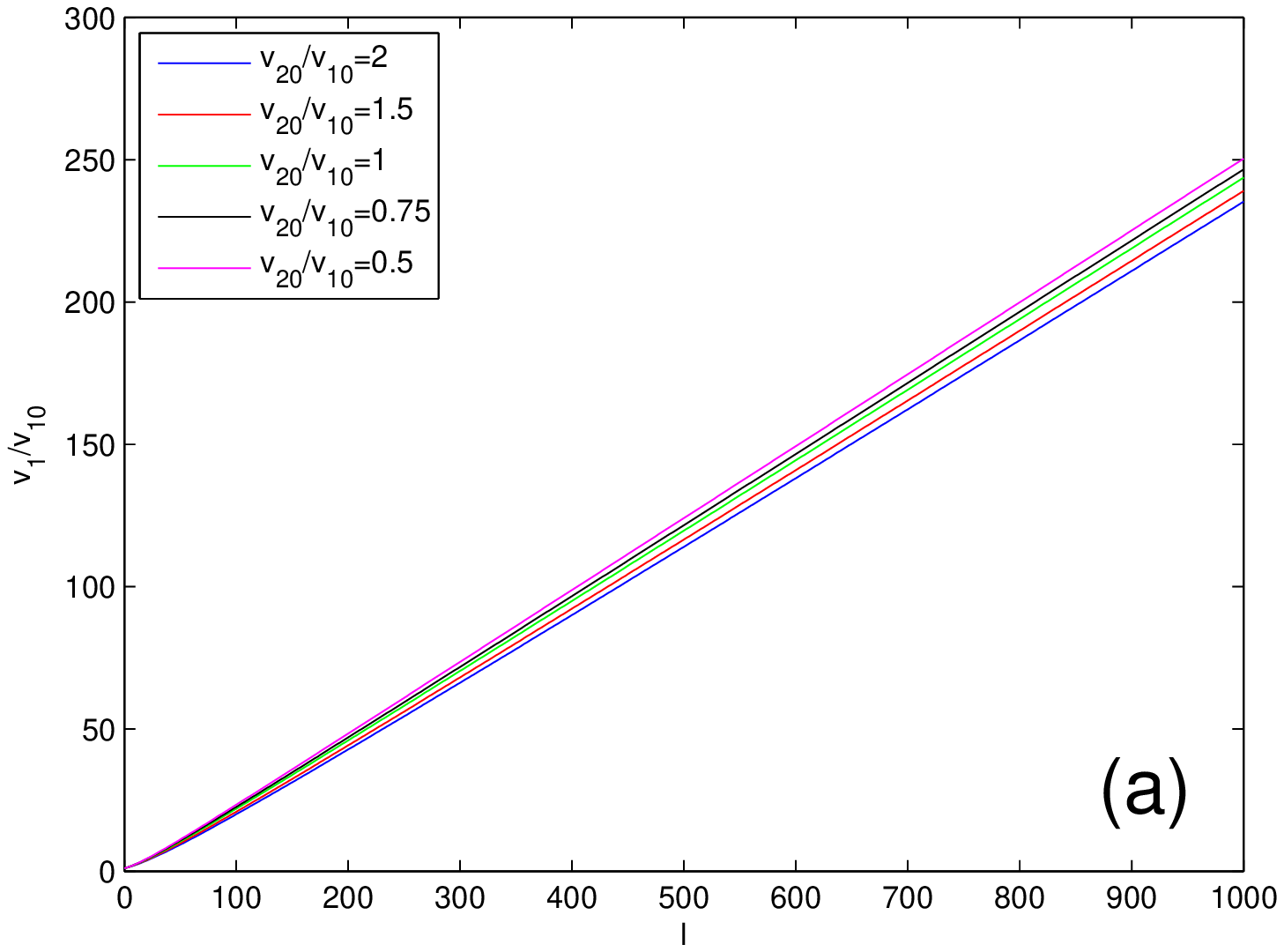}}
\subfigure{\includegraphics[width=2.8in]{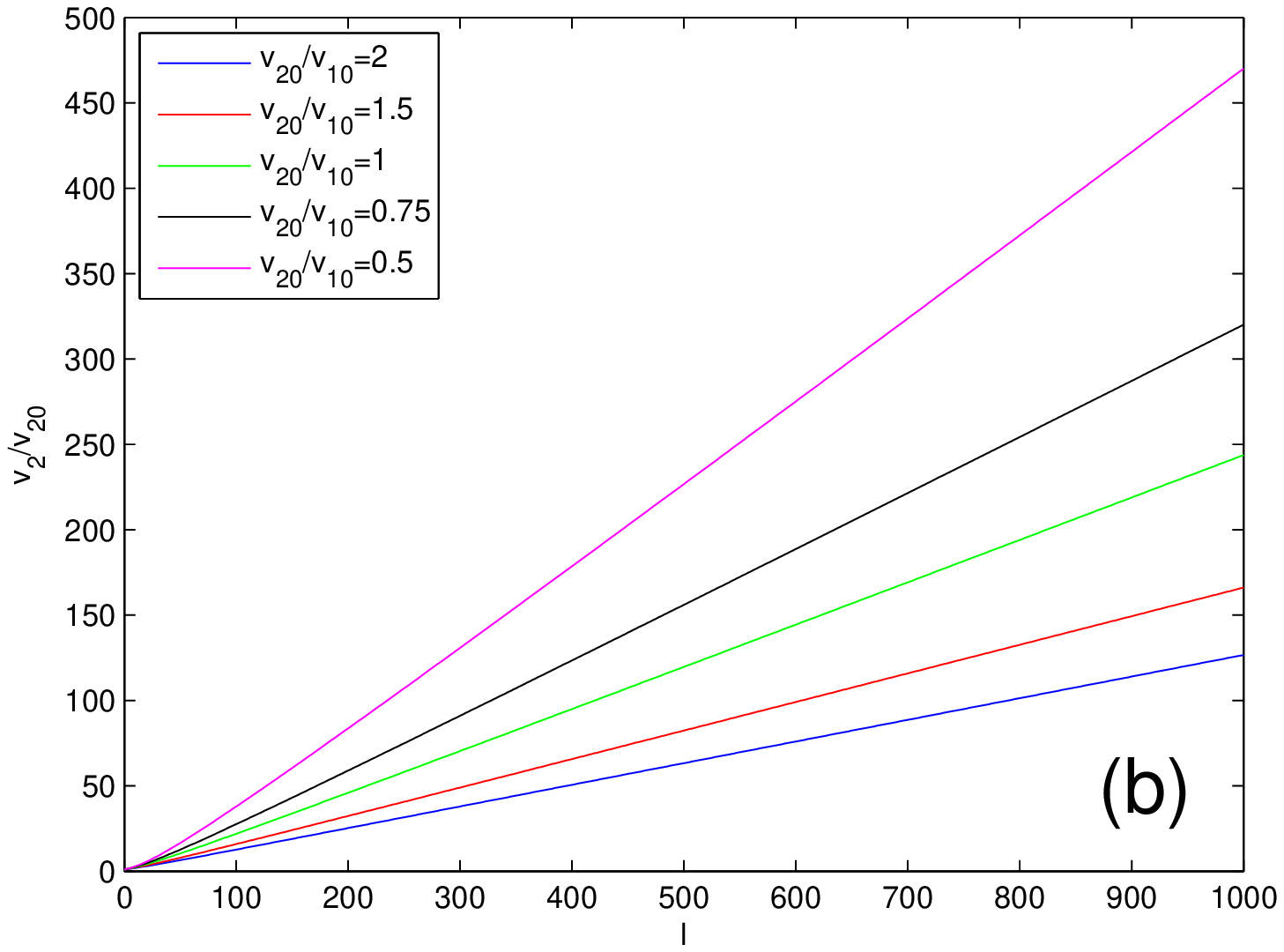}}
\subfigure{\includegraphics[width=2.8in]{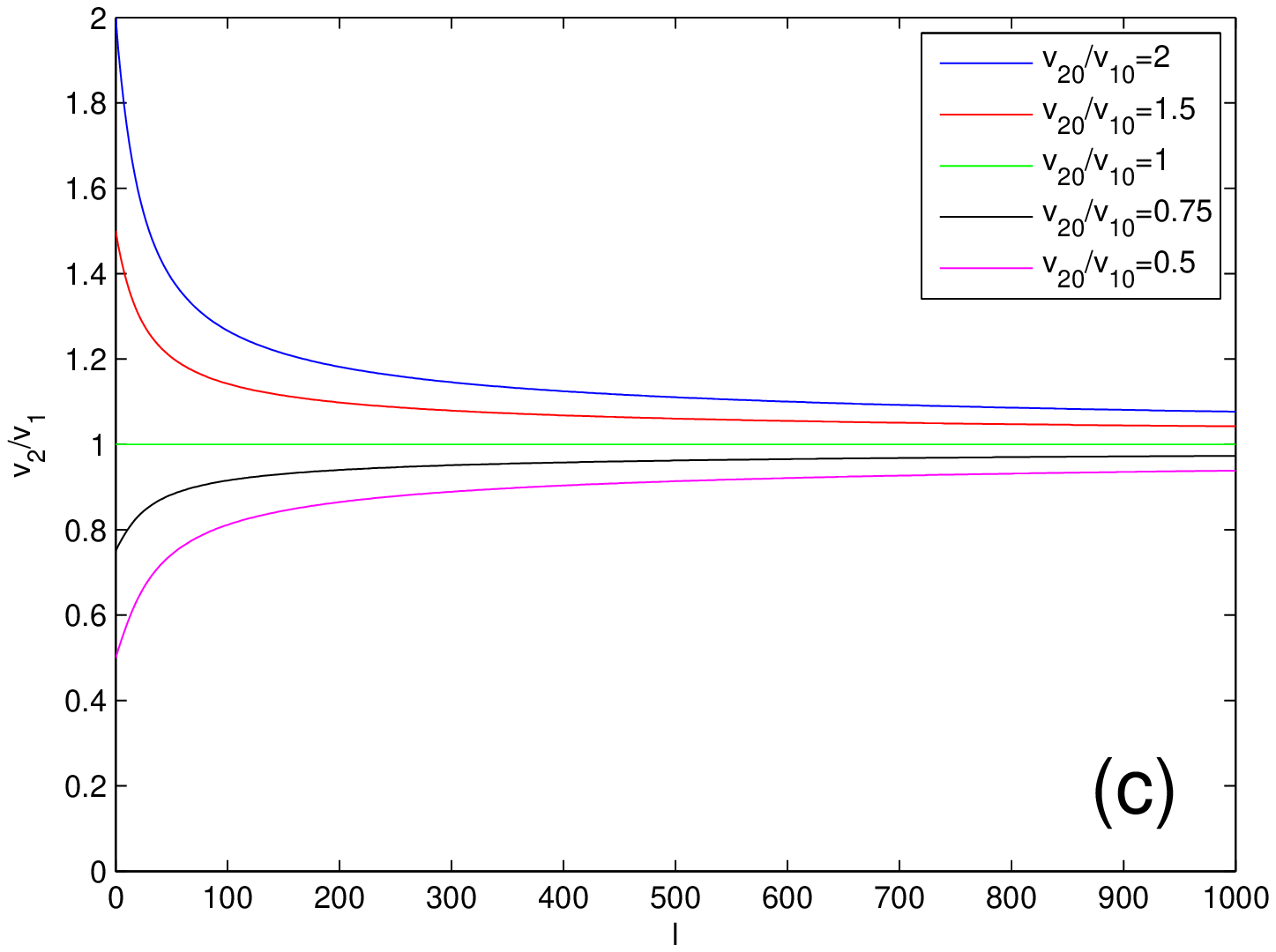}}
\caption{Renormalized $v_{1}$, $v_{2}$ and $v_{2}/v_{1}$ at fixed
coupling $\alpha_{10} = e^2/\epsilon v_{10} = 1$ in the absence of
disorders.} \label{fig:VRGAsCl}
\end{figure}

We then consider clean graphene with an anisotropic Dirac cone. We
obtain the following flow equations of fermion velocities $v_{1,2}$
and their ratio,
\begin{eqnarray}
\frac{dv_{1}}{dl}&=&\left(C_{0}-C_{1}\right)v_{1},\label{eqn:RGV1AsCl}
\\
\frac{dv_{2}}{dl}&=&\left(C_{0}-C_{2}\right)v_{2},\label{eqn:RGV2AsCl}
\\
\frac{d\left(v_{2}/v_{1}\right)}{dl}
&=&\left(C_{1}-C_{2}\right)\frac{v_{2}}{v_{1}},\label{eqn:RGVRatioAsCl}
\end{eqnarray}
where $C_{0,1,2}$ are given in Sec.~\ref{sec:Derivation}. From Fig.~\ref{fig:VRGAsCl},
it is easy to see that both $v_{1}$ and $v_{2}$ increase
monotonically with the decreasing energy scale, and that the velocities
flow to the isotropic limit at the lowest energy, i.e., $v_{2}/v_{1}
\rightarrow 1$ as $l \rightarrow \infty$. Apparently, the velocity
anisotropy is irrelevant, analogous to the case of QED$_3$
\cite{Vafek02, Lee02, Hermele05}. Notice this conclusion is
different from the non-monotonic flow of velocity ratio claimed in
Ref.~\cite{Sharma12}.

It is now necessary to make a comparison between our results with
those of Ref.~\cite{Sharma12}. Sharma \emph{et. al.} investigated
the influence of Coulomb interaction on Dirac fermion systems with
an anisotropic dispersion by performing perturbative expansion in
powers of $\alpha_{1} = e^2/v_{1}\epsilon$ \cite{Sharma12}. They
found that the system will flow to three different fixed points,
depending on the initial values of $\alpha_{1}$ and $\delta =
v_{2}/v_{1}-1$(notice that the meaning of $\delta$ in our paper is
different from Ref.~\cite{Sharma12}), where it is assumed that
$v_{2}<v_{1}$ with $\delta = -1$ representing an infinite
anisotropic limit. When $\alpha_{1}$ is small and $|\delta|$ is
smaller than certain critical value, the flow of $\delta$ is not
monotonic. In particular, the anisotropy of fermion dispersion
initially increases with decreasing energy in a range of energy
scale, but finally goes to an isotopic limit at the lowest energy.
If $\alpha_{1}$ is strong enough, the system can become an
anisotropic insulator. If $\alpha_{1}$ is small and $|\delta|$ is
larger than certain critical value, the system will become a
quasi-one-dimensional non-Fermi liquid. It is obvious that these
results differ significantly from ours.

We would point out that the perturbative expansion presented in
Ref.~\cite{Sharma12} is valid if the Coulomb interaction is weak.
This expansion scheme is no longer valid when the Coulomb
interaction becomes strong. In addition, the Coulomb interaction
strength actually depend on both $\alpha_{1}$ and $\delta$. Since
$\delta$ itself also flows strongly with the varying energy, it
seems questionable to make perturbative expansion in powers of
$\alpha_{1}$ alone. In order to avoid these problems and make RG
analysis reliable for both weak and strong couplings, we have chosen
to study the influence of Coulomb interaction on velocity anisotropy
by means of the $1/N$ expansion. As demonstrated in our results, the
anisotropic system flows to an isotropic fixed point. An earlier
calculation of Aleiner \emph{et al.} has showed that the long-range
Coulomb interaction tends to suppress the strength of a trigonal
wrapping term, which otherwise can make the system anisotropic
\cite{Aleiner}. Our conclusion, though based on different approach,
agrees with that of Ref.~\cite{Aleiner}.

When the Coulomb interaction becomes sufficiently strong, the
perturbative $1/N$ expansion is also invalid since the interaction
may lead to an excitonic instability. We do agree with
Ref.~\cite{Sharma12} on the opinion that the excitonic insulating
transition should be investigated by means of a non-perturbative
method. This issue will be addressed in Sec.~\ref{Sec:Gap} by
constructing and solving the self-consistent DS gap equation.

\subsection{Disordered and isotropic case }

Now we come to the case of isotropic Dirac cone in the presence of
disorders. After assuming $v_{1} = v_{2} = v$ and introducing
disordered potentials, we find that
\begin{eqnarray}
\frac{dv}{dl} &=& \left(C-C_{g}\right)v, \label{eqn:RGVIsDis}
\end{eqnarray}
and that
\begin{eqnarray}
\frac{d v_{\Gamma}}{dl} = 0 \label{eqn:RGVGammaCPIsDis}
\end{eqnarray}
for random chemical potential;
\begin{eqnarray}
\frac{dv_{\Gamma}}{dl}=\left(C-C_{g}\right)v_{\Gamma}
\label{eqn:RGVGammaGPIsDis}
\end{eqnarray}
for random gauge potential;
\begin{eqnarray}
\frac{dv_{\Gamma}}{dl}=2\left(C -C_{g}\right)v_{\Gamma}
\label{eqn:RGVGammaMassIsDis}
\end{eqnarray}
for random mass. Apparently, the flows of $v_{\Gamma}$ in random
gauge potential and random mass are very similar, but are quite
different from random chemical potential.

\begin{figure}[htbp]
\center \subfigure{
\includegraphics[width=2.8in]{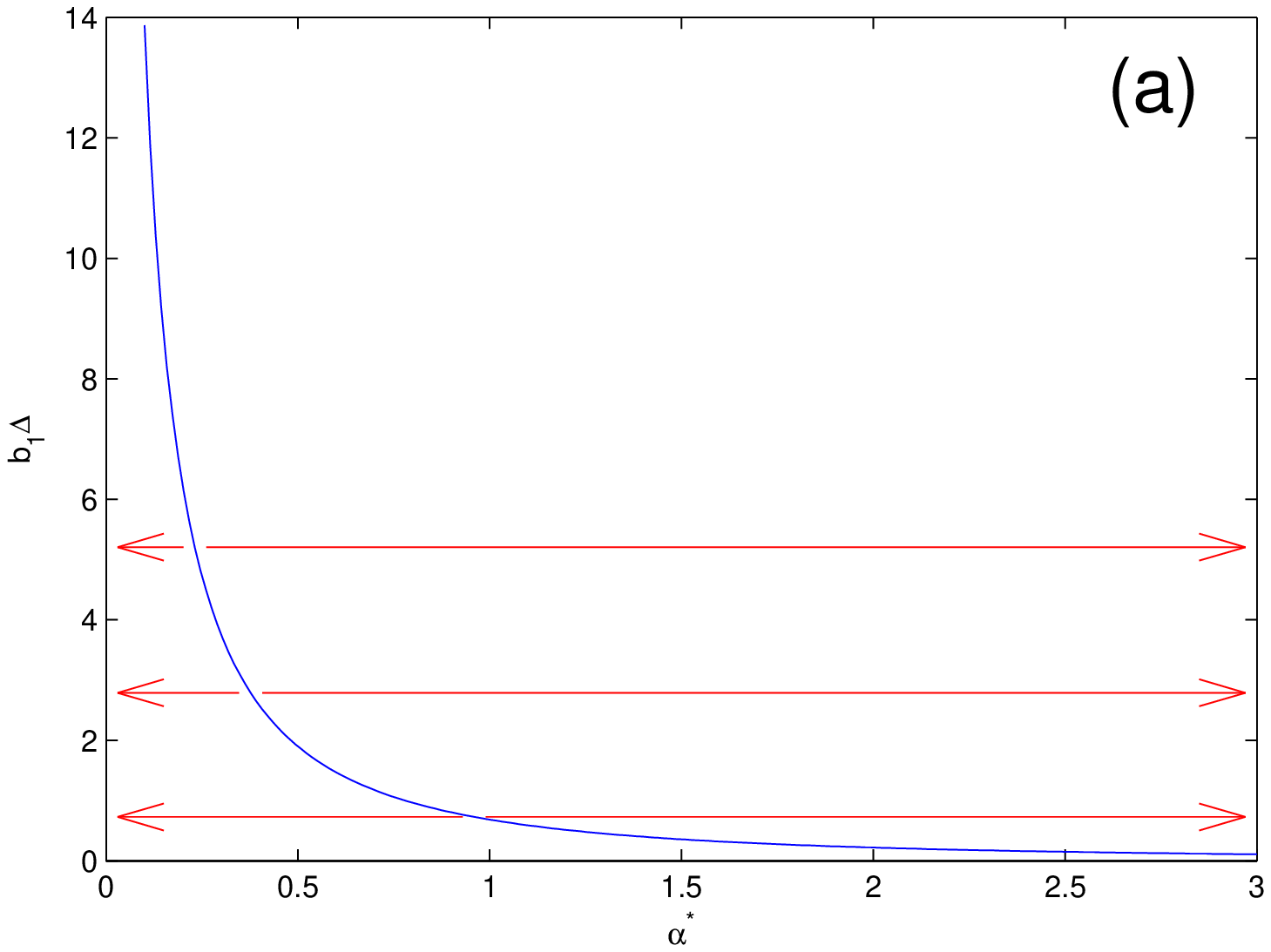}}
\subfigure{
\includegraphics[width=2.8in]{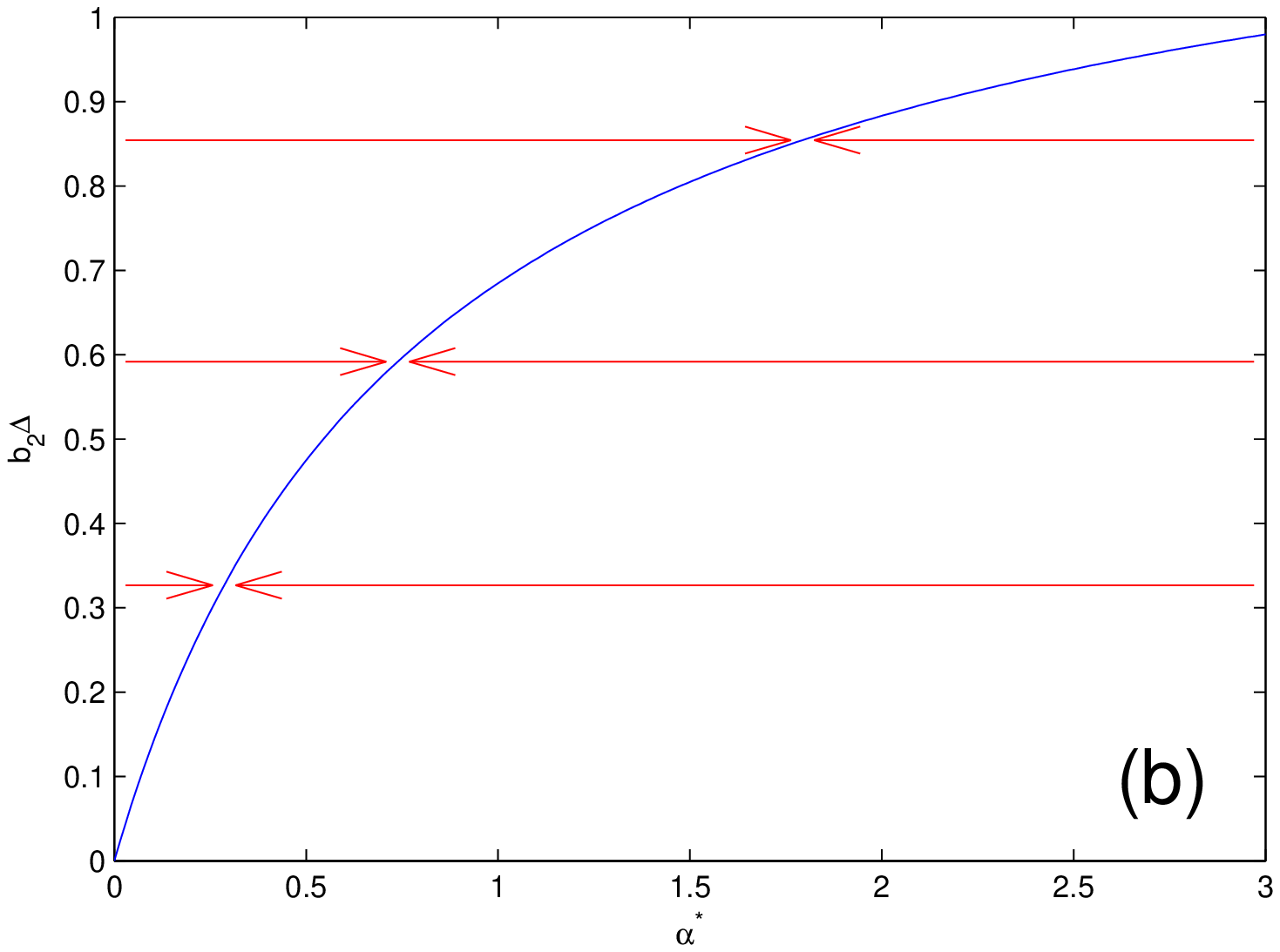}}
\subfigure{
\includegraphics[width=2.8in]{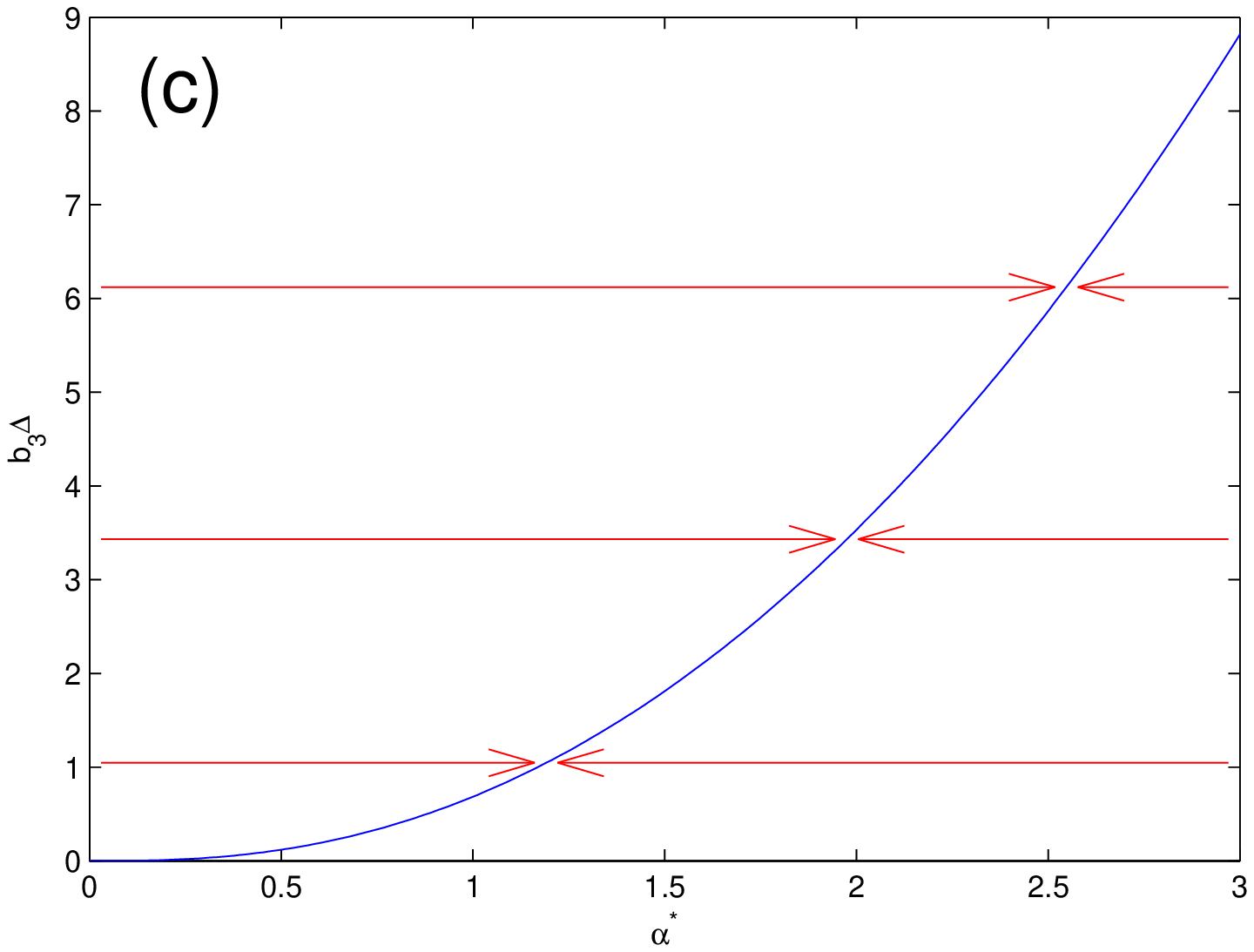}}
\caption{(a) Fixed points for random chemical potential with
$b_{1}=\frac{\epsilon^2v_{\Gamma0}^{2}}{e^4}$; (b) Fixed point for
random gauge potential with $b_{2}=\frac{v_{\Gamma_{0}}^{2}}{v_{0}^{2}}$;
 (c) Fixed point for
random mass with $b_{3}=\frac{v_{\Gamma0}^{2}e^4}{\epsilon^2v_{0}^4}$.}
\label{Fig:FixedLine}
\end{figure}

The fixed point $\alpha^{*}$ for random chemical potential, random
gauge potential, and random mass can be obtained by setting
\begin{eqnarray}
C(\alpha^{*}) - C_{g}(\alpha^{*}) = 0.
\end{eqnarray}
The expression of $C$ as a function of $\alpha$ is presented in
Sec.~\ref{subsec:ClIso}. As will be shown in Eqs. (\ref{eqn:CgCP}),
(\ref{eqn:CgGP}), and (\ref{eqn:CgMass}), $C_{g}$ can  also be
written as a function of $\alpha$, in different forms for different
kinds of disorders. This issue has been studied earlier by Stauber
\emph{et al.} \cite{Stauber05}, who have made perturbative
expansions in powers of interaction strength. It was discovered in
Ref.~\cite{Stauber05} that $\alpha^{*} \propto \Delta^{-1}$ for
random chemical potential, $\alpha^{*} \propto \Delta$ for random
gauge potential, and $\alpha^{*} \propto \Delta^{3}$ for random
mass. Our calculations are performed by means of $1/N$ expansion
approach and have reached quantitatively different results, depicted
in Fig.~\ref{Fig:FixedLine}. However, in agreement with the
qualitative results of Stauber \emph{et al.} \cite{Stauber05}, we
find that, the fixed points for both random gauge potential and
random mass are stable, whereas there is no stable fixed point for
random chemical potential. We now present our results for random
chemical potential, random gauge potential, and random mass
respectively in order.

\begin{figure}[htbp]
\center \subfigure{
\includegraphics[width=2.8in]{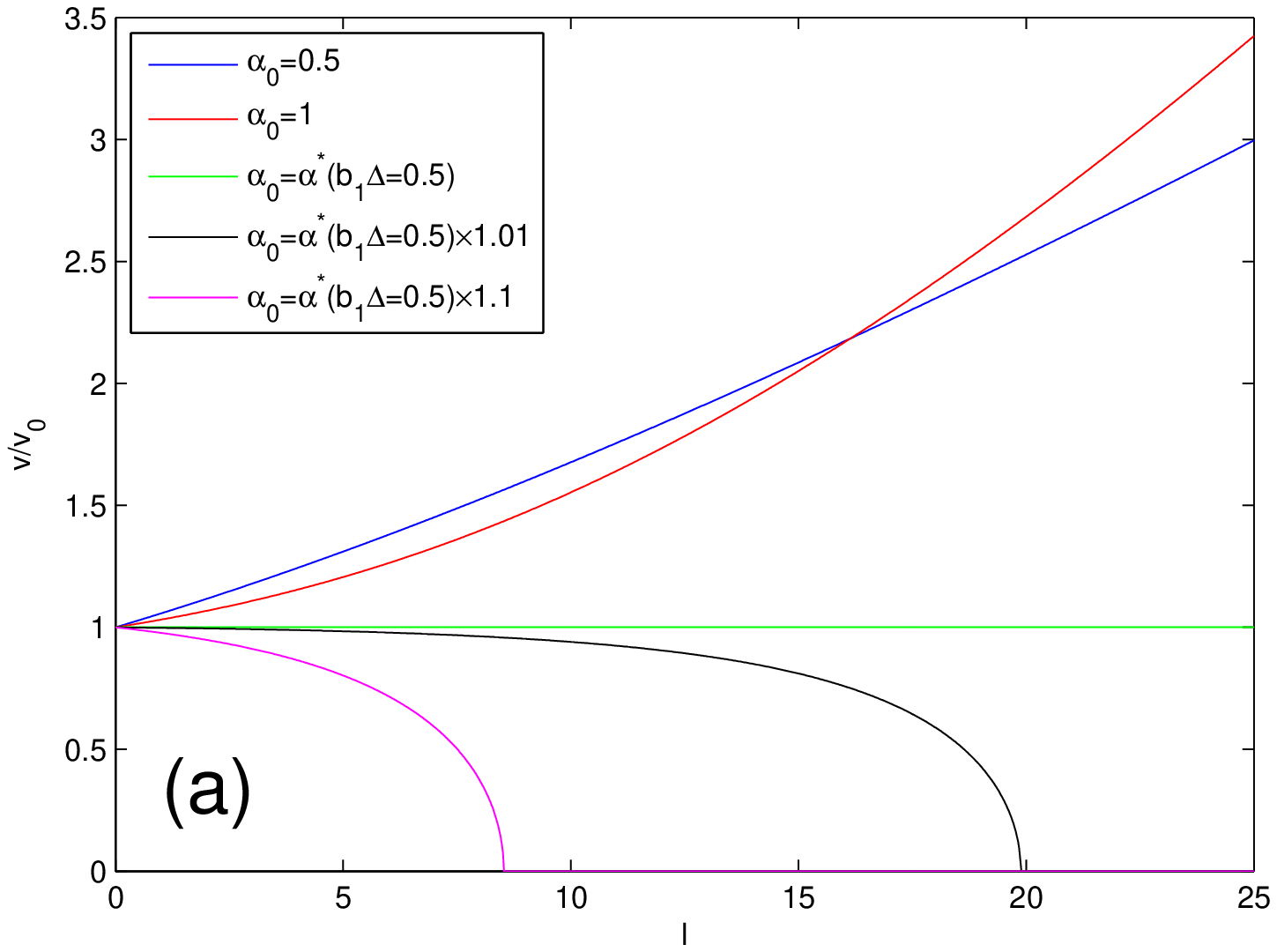}}
\subfigure{
\includegraphics[width=2.8in]{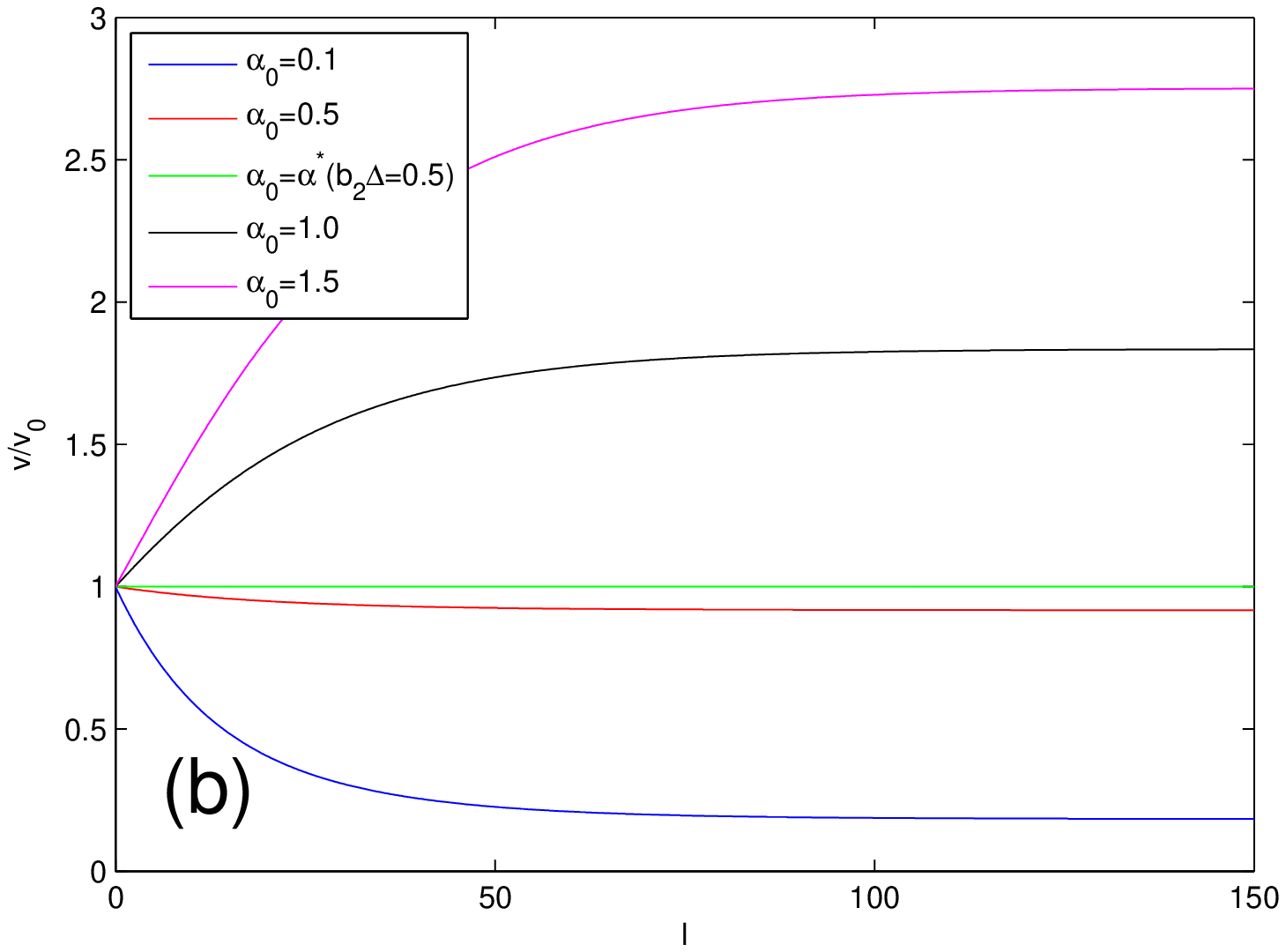}}
\subfigure{
\includegraphics[width=2.8in]{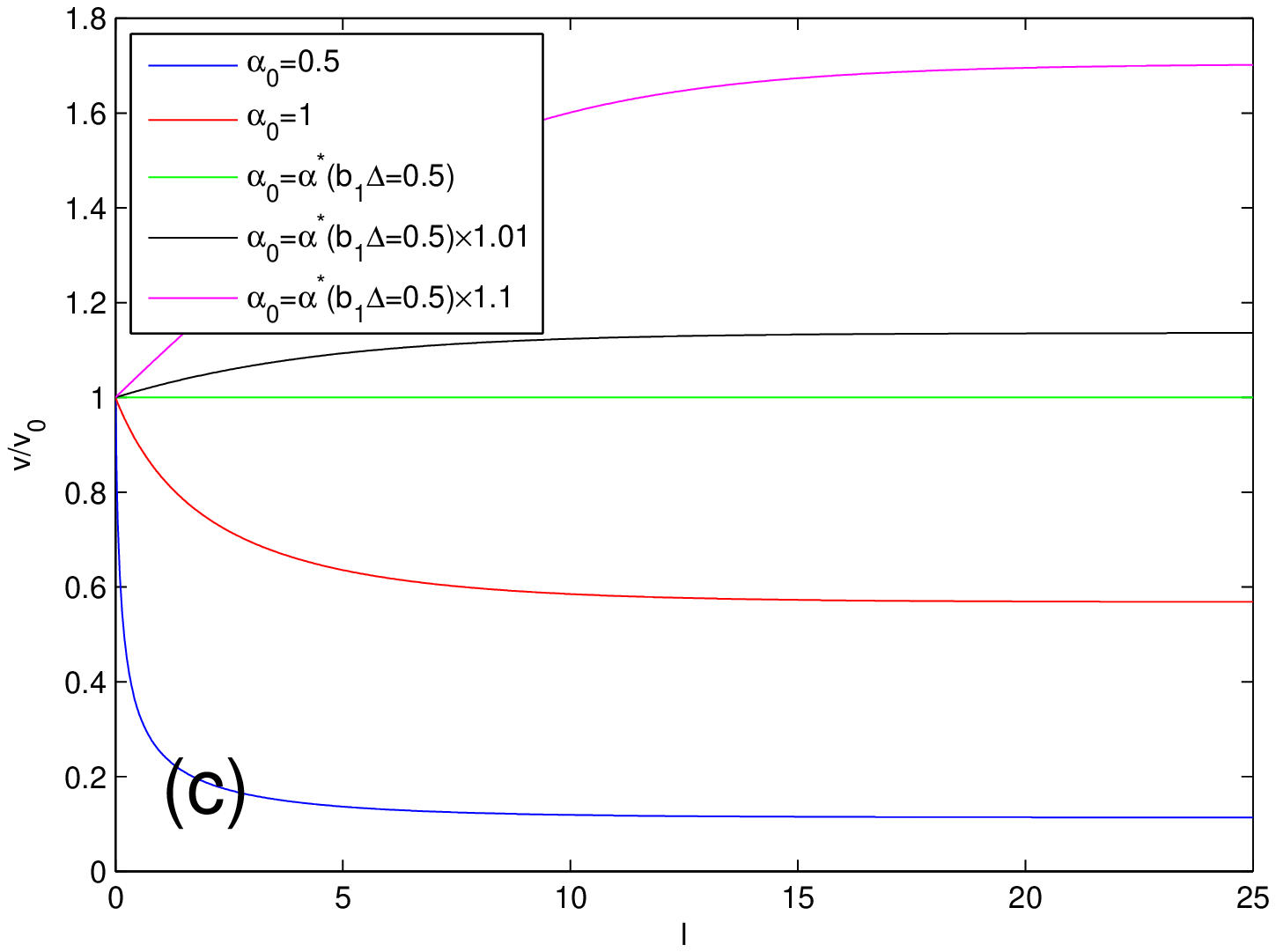}}
\caption{(a) Renormalized fermion velocity for random chemical
potential with $b_{1}\Delta=0.5$; (b) Renormalized fermion velocity
for random gauge potential with $b_{2}\Delta=0.5$ ; (c) Renormalized
fermion velocity for random mass with $b_{3}\Delta =
0.5$.}\label{Fig:RGVIsDis}
\end{figure}

For random chemical potential, Eq.~(\ref{eqn:RGVGammaCPIsDis})
implies that
\begin{eqnarray}
v_{\Gamma} = v_{\Gamma0},\label{eqn:RGVRelationCP}
\end{eqnarray}
then $C_{g}$ can be written as the function of $\alpha$
\begin{eqnarray}
C_{g}=\frac{v_{\Gamma}^{2}\Delta}{2\pi v^2} =
\frac{v_{\Gamma0}^{2}\Delta}{2\pi v^2} =
\left(\frac{v_{\Gamma0}^{2}\Delta\epsilon^2}{2\pi
e^{4}}\right)\alpha^2.\label{eqn:CgCP}
\end{eqnarray}
The lines of fixed points are shown in Fig.~\ref{Fig:FixedLine}(a).
Apparently, the fixed points are unstable in this case.
Fig.~\ref{Fig:RGVIsDis}(a) shows the velocity flow at different
values of $\alpha_0$. If $\alpha_{0}$ is smaller than some critical
value $\alpha^{*}(\Delta)$, the velocity increases continuously as
the energy scale is decreasing, and the effective strength of
Coulomb interaction flows to the infinitely weak coupling limit. In
this case, the weak Coulomb interaction is obviously irrelevant.
However, when $\alpha_{0}
> \alpha^{*}(\Delta)$, the fermion velocity decreases with decreasing
energy scale, and finally vanishes at certain finite energy scale,
which means the effective strength of Coulomb interaction is greatly
enhanced and flows to an infinitely strong coupling limit before $l$
approaches infinity. Such unusual behaviors may be interpreted as a
signature for the emergence of an interaction-driven insulating
phase \cite{Stauber05}.

For random gauge potential, Eq.~(\ref{eqn:RGVIsDis}) and
Eq.~(\ref{eqn:RGVGammaGPIsDis}) combine to yield
\begin{eqnarray}
\frac{v_{\Gamma}}{v} =
\frac{v_{\Gamma0}}{v_{0}},\label{eqn:RGVRelationGP}
\end{eqnarray}
now $C_{g}$ can be written as
\begin{eqnarray}
C_{g}=\frac{v_{\Gamma}^{2}\Delta}{\pi v^2} =
\frac{v_{\Gamma0}^{2}\Delta}{\pi v_{0}^{2}},\label{eqn:CgGP}
\end{eqnarray}
which is a constant. The lines of fixed points are shown in
Fig.~\ref{Fig:FixedLine}(b). The fixed points are stable in this
case. We see from Fig.~\ref{Fig:RGVIsDis}(b) that the fermion
velocity either increases or decreases with the decreasing energy
scale, depending on the concrete value of $\alpha_0$, but finally
are saturated to certain finite values. This behavior is consistent
with previous results obtained in Ref.~\cite{Stauber05} and
Ref.~\cite{Herbut08}. It is argued in Ref.~\cite{Herbut08} that such
disorder dependent fixed point can give rise to a number of
interesting properties, such as nonuniversal minimum dc
conductivity.

For random mass, Eq.~(\ref{eqn:RGVIsDis}) and
Eq.~(\ref{eqn:RGVGammaMassIsDis}) lead to
\begin{eqnarray}
\frac{v_{\Gamma}}{v^2}=\frac{v_{\Gamma0}}{v_{0}^2},
\label{eqn:RGVRelationMass}
\end{eqnarray}
so $C_{g}$ becomes a function of $\alpha$,
\begin{eqnarray}
C_{g}=\frac{v_{\Gamma}^{2}\Delta}{2\pi v^2}
=\frac{v_{\Gamma0}^{2}\Delta v^2}{2\pi v_{0}^{4}}
=\left(\frac{v_{\Gamma0}^{2}\Delta e^{4}}{2\pi
\epsilon^2}\right)\alpha^{-2}. \label{eqn:CgMass}
\end{eqnarray}

As shown in Fig.~\ref{Fig:FixedLine}(c), the fixed points are
stable. According to Fig.~\ref{Fig:RGVIsDis}(c), the fermion
velocity is also saturated to finite values at the low-energy limit,
similar to the case of random gauge potential. An apparent
conclusion is that random chemical potential leads to very different
behaviors compared with random gauge potential and random mass.

It is necessary to remark on the insulating behavior happening in
the presence of random chemical potential. This presumed insulating
state is formed by the interplay of strong Coulomb interaction and
random chemical potential, and is signalled by the absence of a
stable fixed point and the divergence of interaction strength. At
this stage, it is premature to judge whether this insulator is
associated with an excitonic pairing instability
\cite{Khveshchenko01, Gorbar02, Khveshchenko04, Liu09,
Khveshchenko09, LiuWang, Gamayun10, Zhang11, Wang12, Drut09,
Hands08, CastroNetoPhys} or a disorder-driven localization-like
state. We feel that the present RG scheme alone is unable to uncover
the fundamental nature and detailed properties of such an insulating
state. Further research effort is called for to investigate this
problem.

\subsection{Disordered and anisotropic case}

We finally come to the general and most interesting case in which
both anisotropy and disorder are present. We will show that Coulomb
interaction and fermion-disorder coupling can result in rich
behaviors. The physical properties are very complicated and
determined by several parameters, including Coulomb coupling
$\alpha_{10}$ and bare velocity ratio $\delta_{10} = v_{20}/v_{10}$.
To simplify our analysis, we fix the coupling at $\alpha_{10} = 1$
and examine how the two velocities and their ratio flow as
$\delta_{0}$ is varying.

\begin{figure}[htbp]
\centering
\subfigure{\includegraphics[width=2.8in]{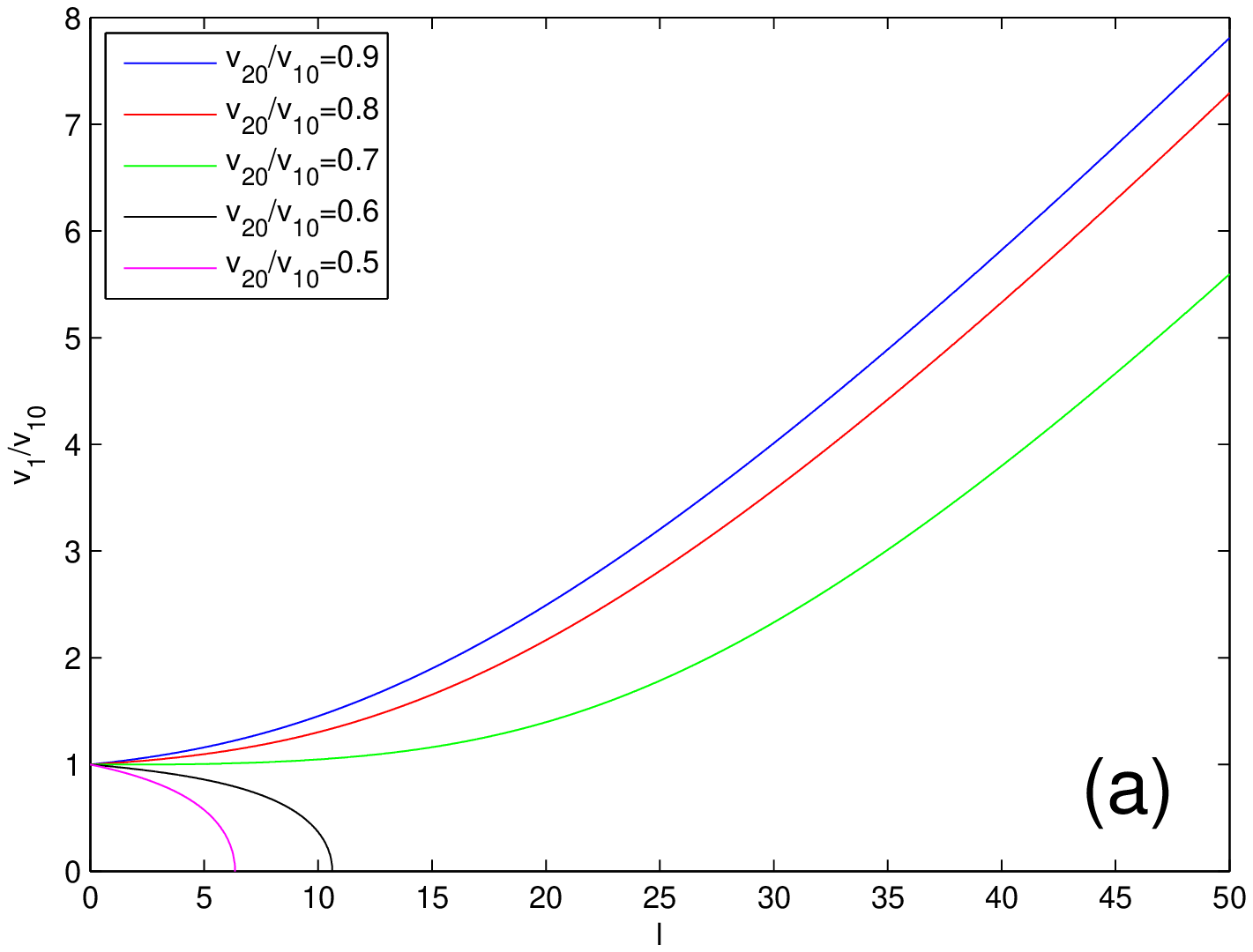}}
\subfigure{\includegraphics[width=2.8in]{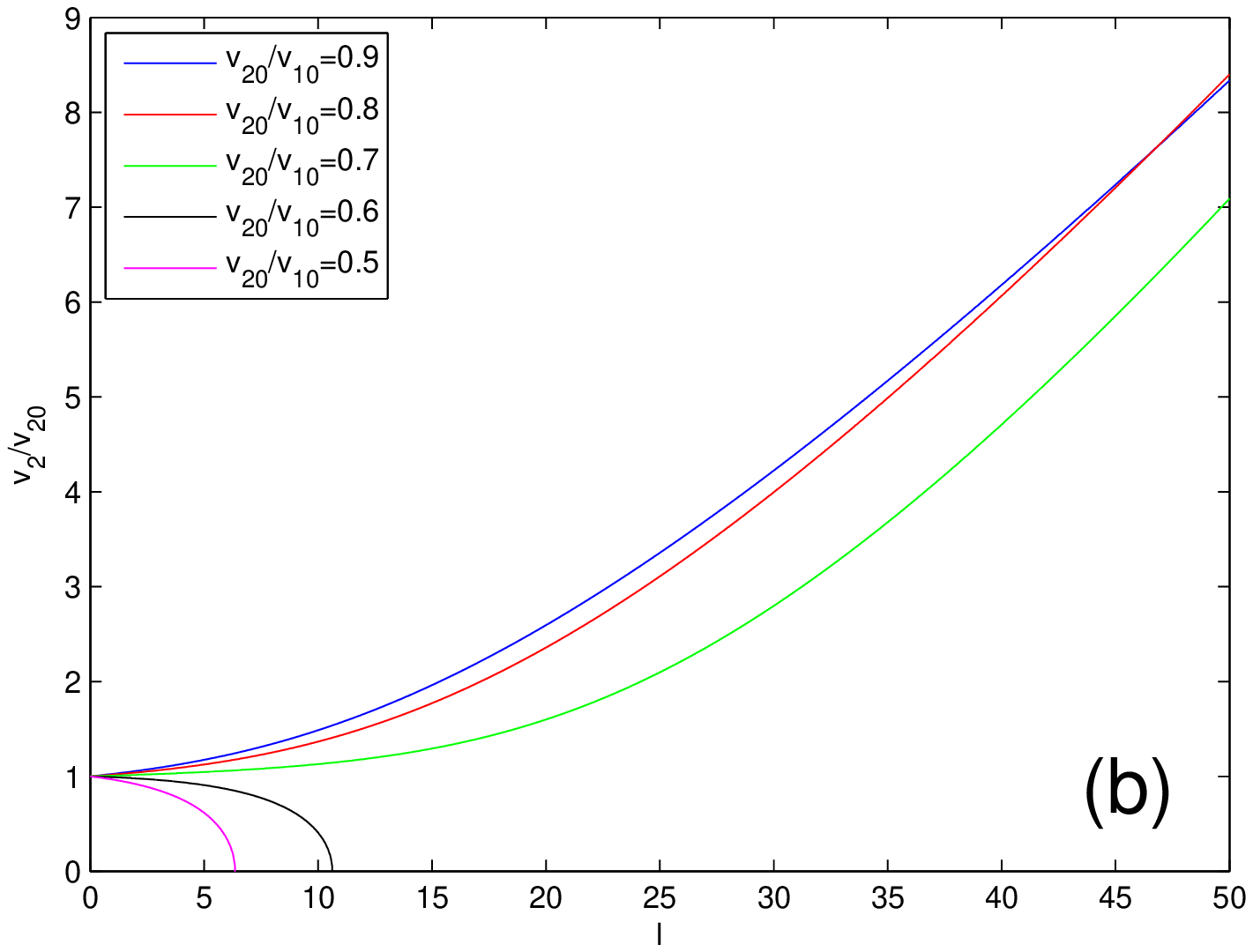}}
\subfigure{\includegraphics[width=2.8in]{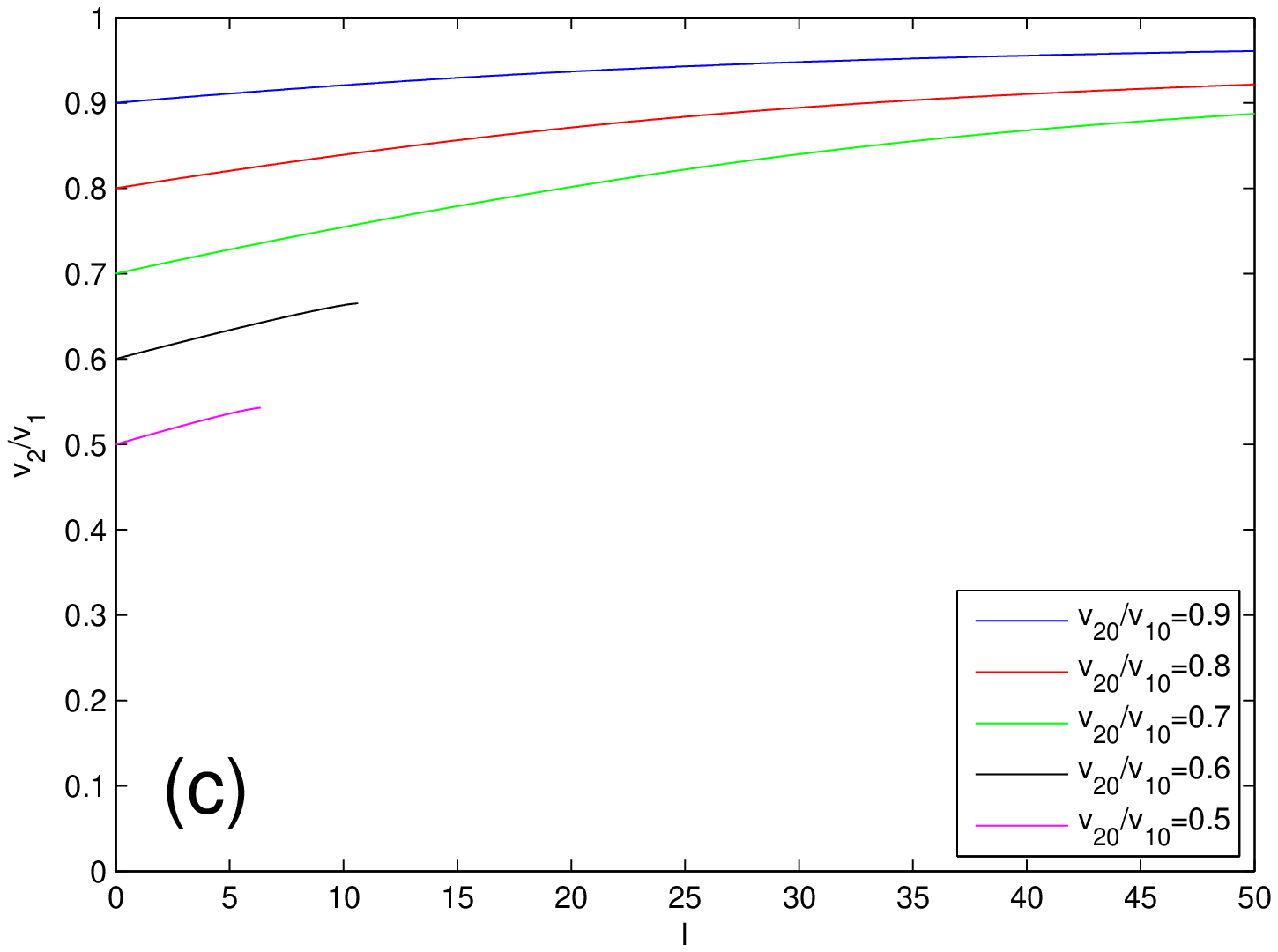}}
\caption{Renormalized $v_{1}$, $v_{2}$ and $v_{2}/v_{1}$ at fixed
coupling $\alpha_{10} = 1$ in the presence of random chemical
potential with $v_{\Gamma0}^{2}\Delta/v_{10}^{2} =
0.05$.}\label{fig:VRGAsDisCP}
\end{figure}

For random chemical potential, we remember that $v_{\Gamma}$ remains
a constant, i.e., $v_{\Gamma} = v_{\Gamma0}$. At a fixed Coulomb
coupling $\alpha_{10} = 1$, the renormalized velocities $v_{1,2}$
and the ratio $v_{2}/v_{1}$ for different bare values of ratio
$\delta_0$ are presented in Fig.~\ref{fig:VRGAsDisCP}. From
Fig.~\ref{Fig:RGVIsDis}, we already know that the isotropic velocity
$v$ increases monotonically with the growing $l$ at fixed coupling
$\alpha_{10} = 1$. In the anisotropic case, there is a critical
value $\delta_{0c}$ lying in the range $0.6 < \delta_{0c} < 0.7$.
When $\delta_{0c} < \delta_{0} \leq 1$, the corresponding Coulomb
interaction is relatively weak. Both $v_{1}$ and $v_{2}$ increase
monotonically as the energy scale is decreasing, whereas the
velocity ratio $v_{2}/v_{1} \rightarrow 1$ at the lowest energy,
which corresponds to an isotropic fixed point. On the other hand, if
$0.5 \le \delta_0 < \delta_{0c}$, the Coulomb interaction becomes
sufficiently strong. In this case, both $v_{1}$ and $v_{2}$ decrease
rapidly as the energy is lowering, and are driven to vanish at
certain finite energy scale. The latter behavior suggests the
disappearance of well-defined quasiparticles, and probably indicate
the appearance of an anisotropic insulating phase.

\begin{figure}[htbp]
\centering
\subfigure{\includegraphics[width=2.8in]{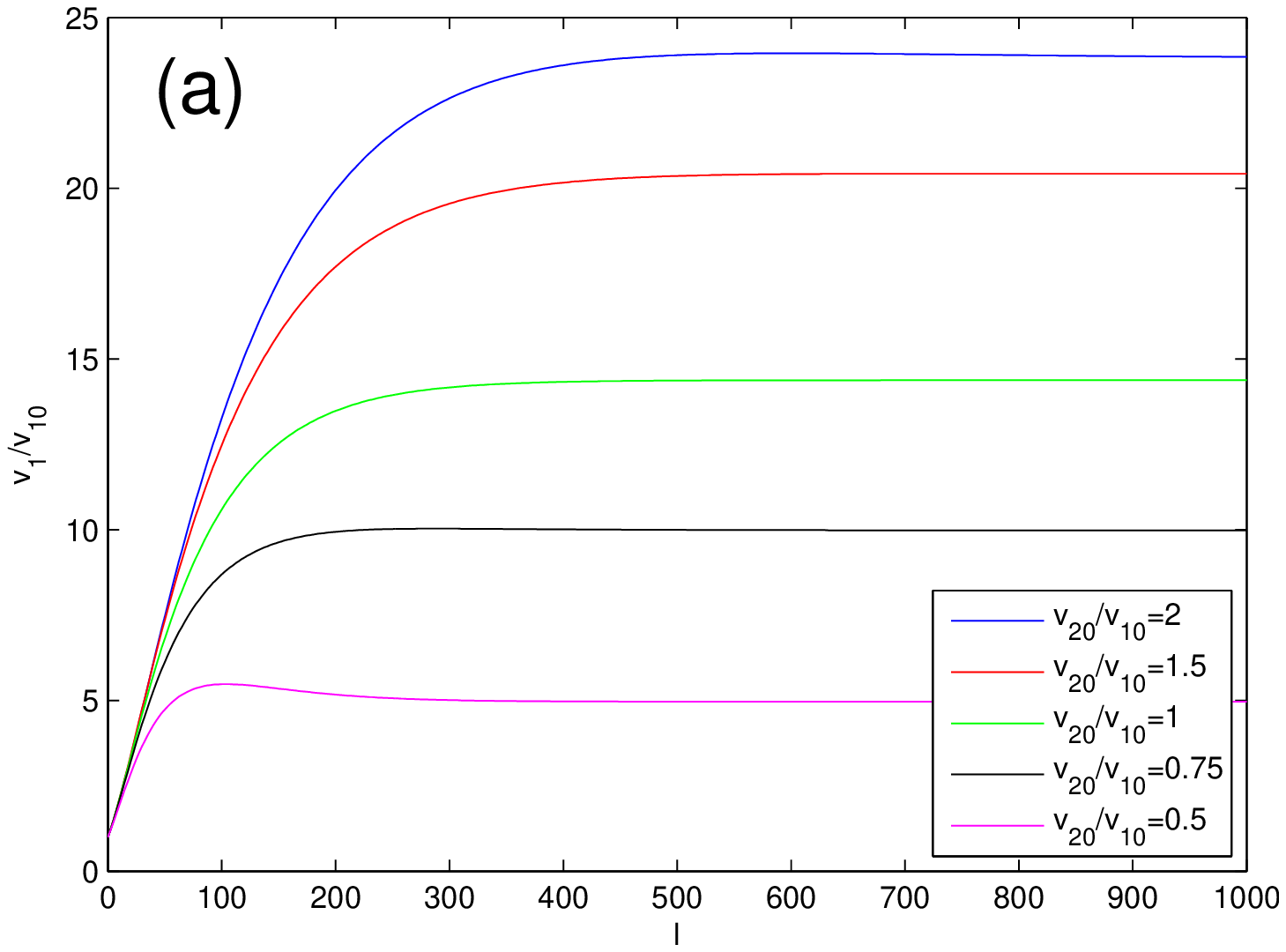}}
\subfigure{\includegraphics[width=2.8in]{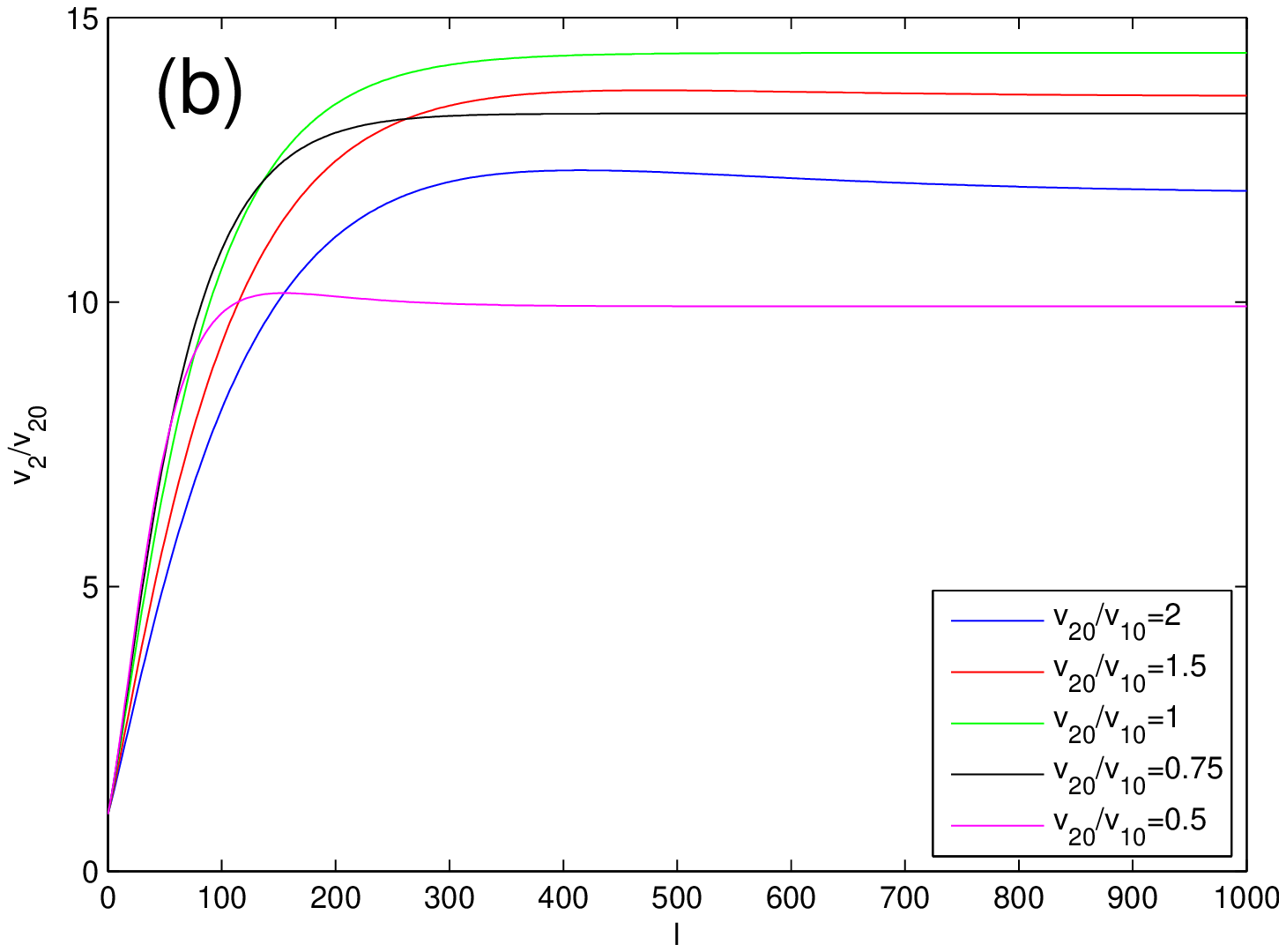}}
\subfigure{\includegraphics[width=2.8in]{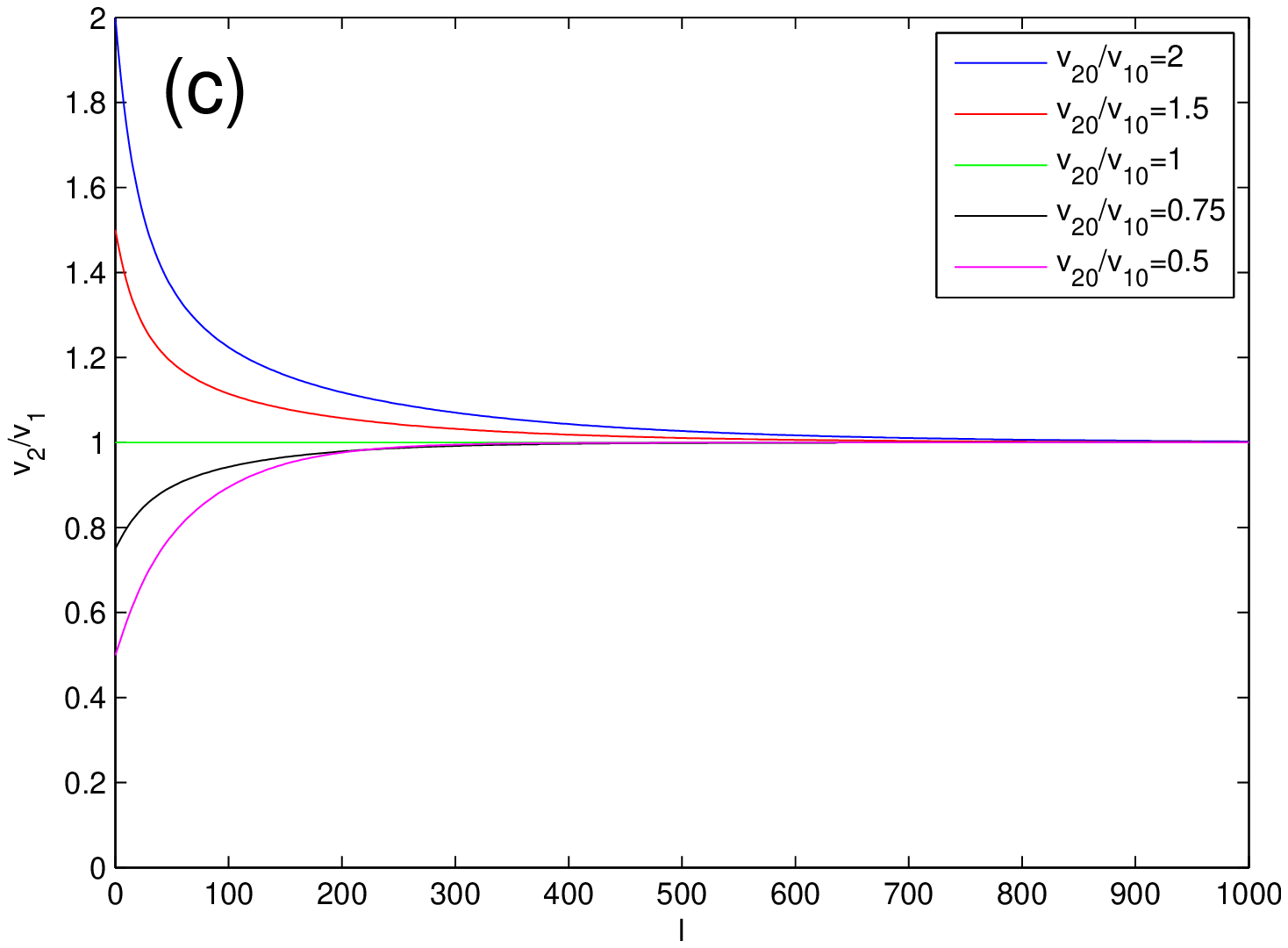}}
\caption{Renormalized $v_{1}$, $v_{2}$ and $v_{2}/v_{1}$ at fixed
coupling $\alpha_{10} = 1$ in the presence of random gauge potential
with $v_{\Gamma0}^{2}\Delta/v_{10}^{2}=0.05$. Here we take
$v_{\Gamma10} = v_{\Gamma20} = v_{\Gamma0}$}\label{fig:VRGAsDisGP}
\end{figure}

In the presence of random gauge potential, we know from
Eqs.~(\ref{eqn:RGV1}), (\ref{eqn:RGV2}), (\ref{eqn:RGVGammaGP1}),
and (\ref{eqn:RGVGammaGP2}) that
\begin{eqnarray}
\frac{v_{\Gamma1}}{v_{1}} = \frac{v_{\Gamma10}}{v_{10}}
\quad\mbox{and} \quad \frac{v_{\Gamma2}}{v_{2}} =
\frac{v_{\Gamma20}}{v_{20}}.
\end{eqnarray}
The RG flows of velocities $v_{1,2}$ and ratio $v_{2}/v_{1}$ are
presented in Fig.~\ref{fig:VRGAsDisGP}. It is easy to observe that,
both $v_{1}$ and $v_{2}$ increase initially but finally approach
certain finite values. In addition, the ratio
$v_{2}/v_{1}$ eventually flows to an isotropic limit, i.e.,
$v_{2}/v_{1} \rightarrow 1$, at the lowest energy. Comparing to the
clean and anisotropic case, $v_{2}/v_{1}$ flows to unity more
rapidly.

\begin{figure*}[hbtp]
\centering
\subfigure{\includegraphics[width=2.8in]{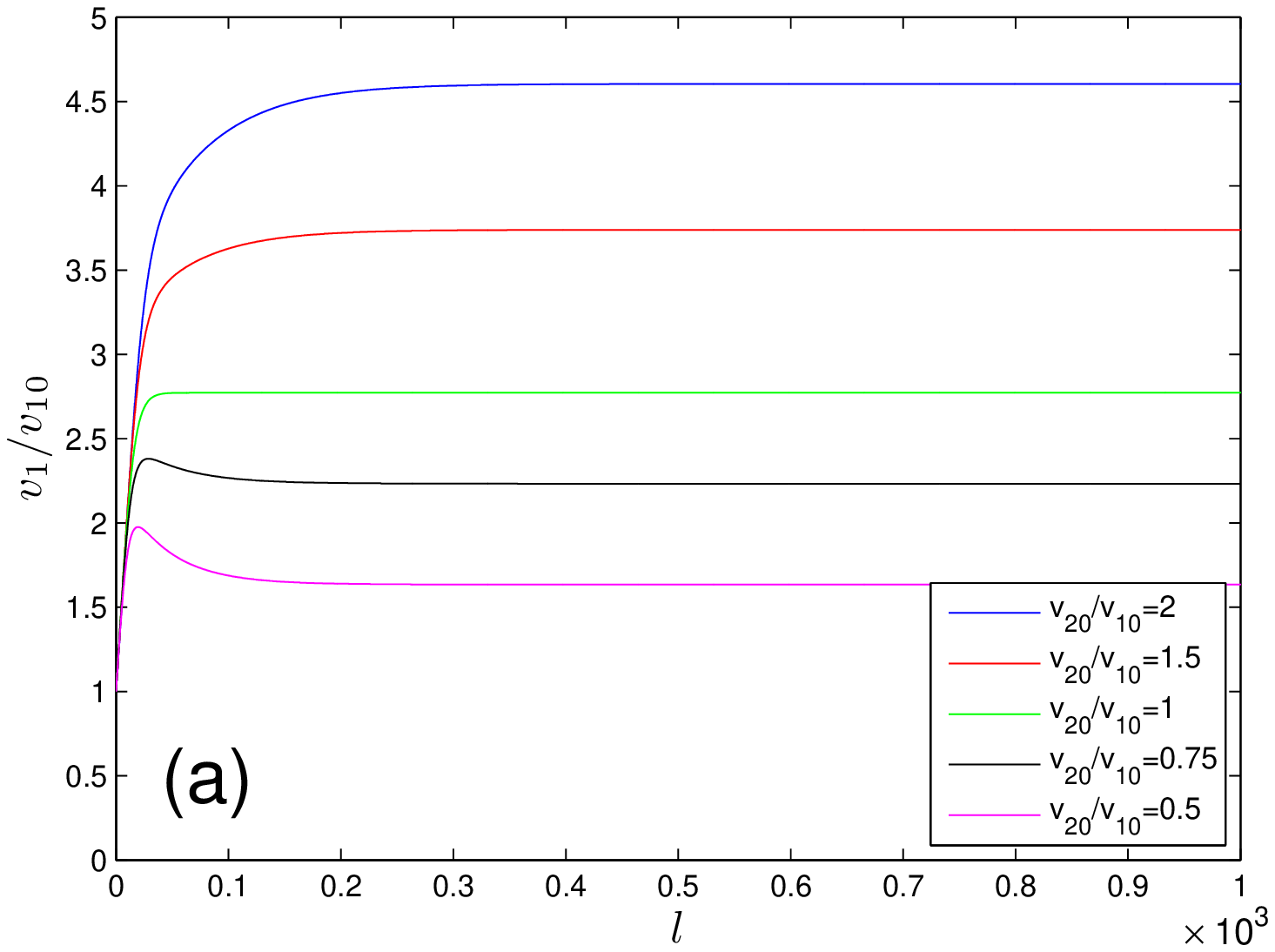}}
\subfigure{\includegraphics[width=2.8in]{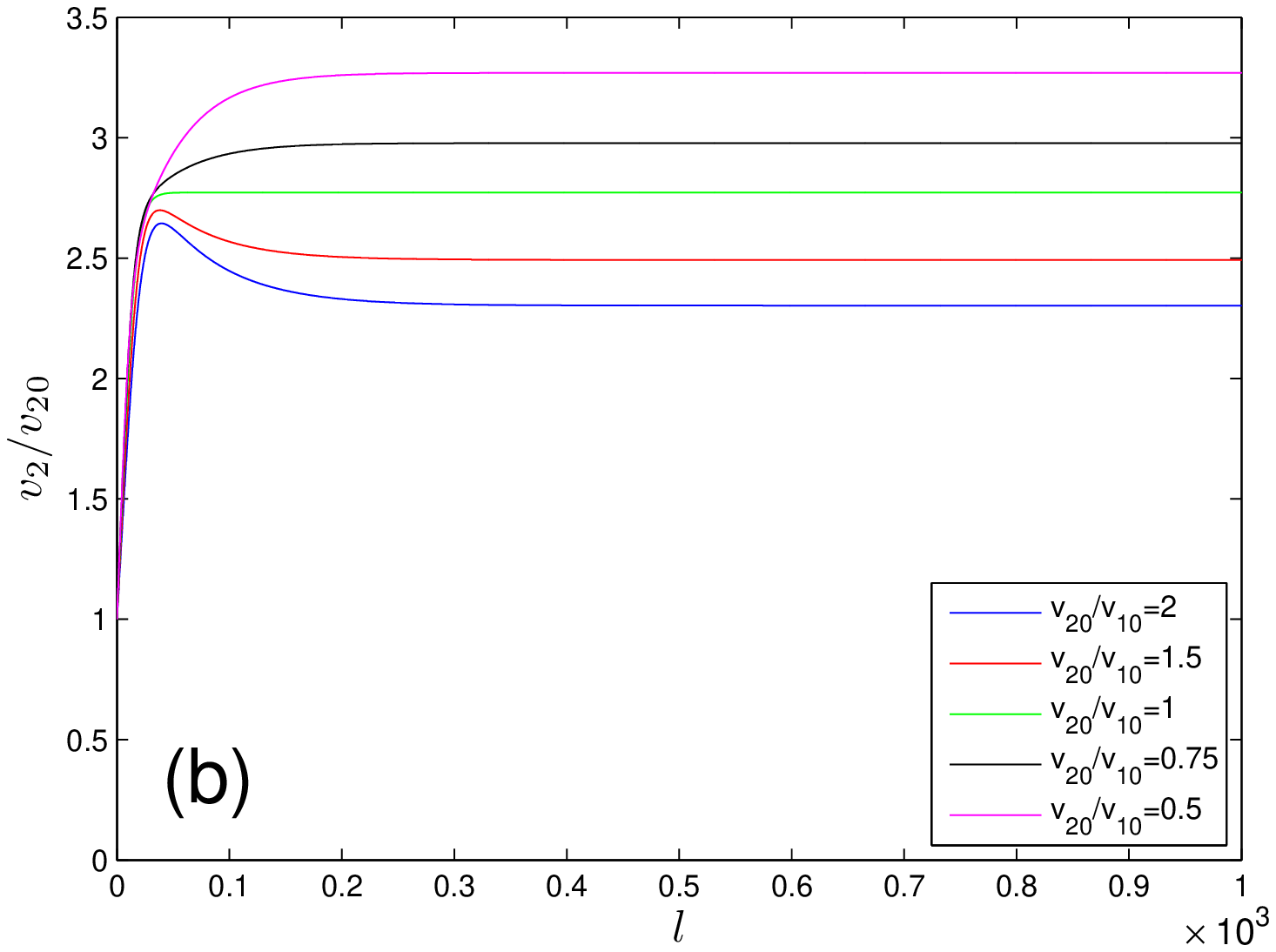}}
\subfigure{\includegraphics[width=2.8in]{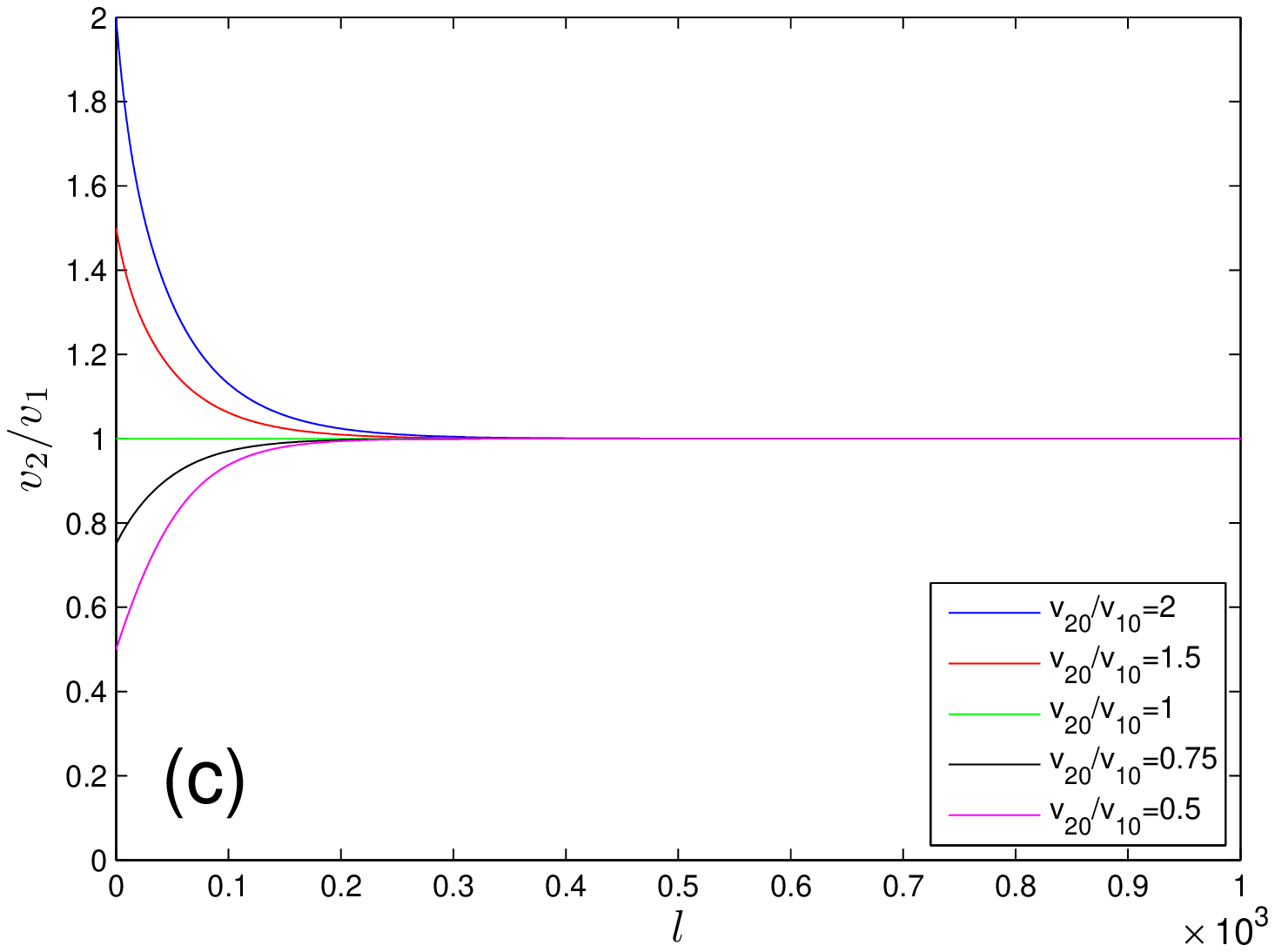}}
\subfigure{\includegraphics[width=2.8in]{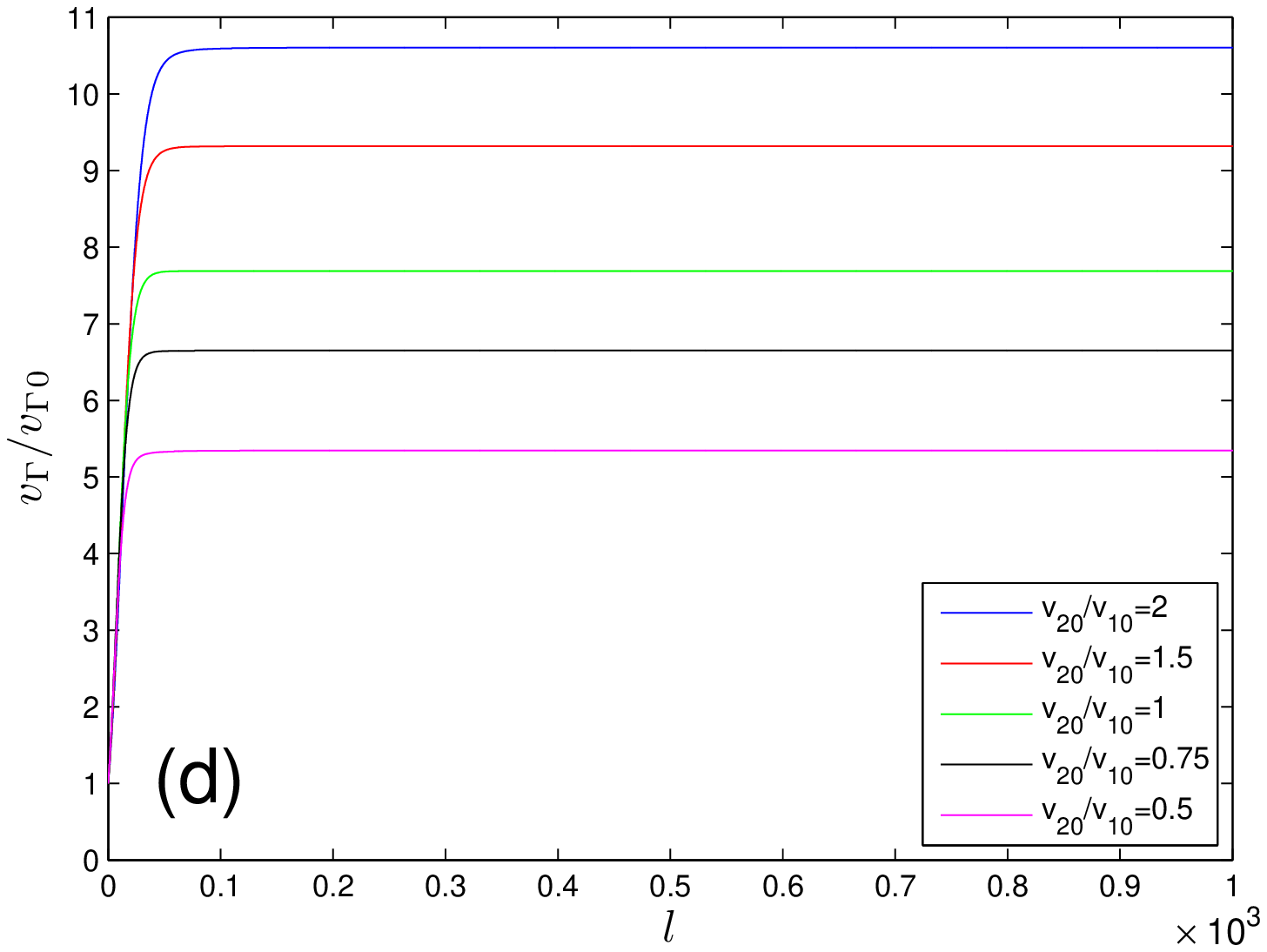}}
\caption{Renormalized $v_{1}$, $v_{2}$ and $v_{2}/v_{1}$ with
$\alpha_{10}=1$ where $\alpha_{10}=e^2/\epsilon v_{10}$ if there is
random mass with
$v_{\Gamma0}^{2}\Delta/v_{10}^{2}=0.05$.}\label{fig:VRGAsDisMass}
\end{figure*}

For random mass, the flows of $v_{1}$, $v_{2}$, $v_{2}/v_{1}$ and
$v_{\Gamma}$ are depicted in Fig.~\ref{fig:VRGAsDisMass}. $v_{1,2}$
and $v_{\Gamma}$ are all saturated to finite values and $v_{2}/v_{1}
\rightarrow 1$ in the lowest energy limit. These properties are
qualitatively very similar to those in the case of random gauge
potential, but is apparently distinct from those in the case of
random chemical potential.

\section{Influence of fermion velocity renormalization \label{Sec:InfluenceObQu}}

In the last section, we have already shown that the long-range
Coulomb interaction, sometimes in collaboration with disorders, can
have remarkable effects on the low-energy properties of fermion
velocities and velocity ratio. These effects should manifest
themselves in various physical quantities. In order to make these
effects more transparent, here we calculate several physical
quantities, including quasiparticle damping rate, DOS
and specific heat, and discuss the physical implications of the
results.

\subsection{Landau damping rate}

Landau damping rate is an important quantity that is frequently used
to characterize the effects of inter-particle interactions and to
judge whether an interacting many-body system is a Fermi liquid or
not. This quantity is intimately related to the wave renormalization
function, which can be calculated as follows,
\begin{eqnarray}
Z_{f}(\omega) = \frac{1}{\left|1 - \frac{\partial}{\partial\omega}
\mathrm{Re}\Sigma^{R}(\omega)\right|},
\end{eqnarray}
where $\mathrm{Re}\Sigma^{R}(\omega)$ is the real part of the
retarded fermion self-energy function. However, taking advantage of
the RG scheme used in this paper, it is more convenient to write it
in the following form
\begin{eqnarray}
Z_{f} = e^{\int_{0}^{l}\left(C_{0} - C_{g}\right)dl}.
\end{eqnarray}
Using the results obtained in Sec.~\ref{sec:Derivation}, it is easy to get
\begin{eqnarray}
\frac{dZ_{f}}{dl}=\left(C_{0} - C_{g}\right)Z_{f},
\end{eqnarray}
where $C_{0}$ and $C_{g}$, given in Sec.~\ref{subsec:FSCorrectVXCorrect},
represent effects of Coulomb interaction and disorders,
respectively.

\begin{figure}[htbp] \center
\includegraphics[width=2.8in]{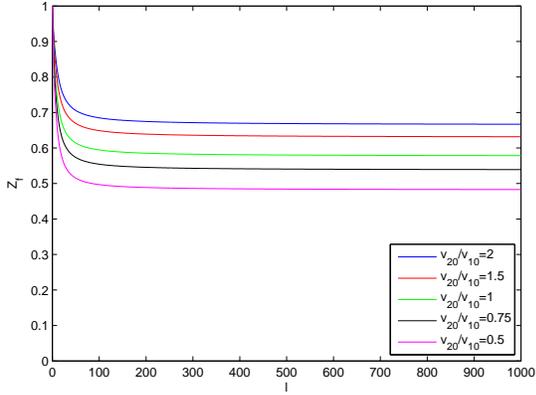}
\caption{Wave renormalization factor for different $v_{20}/v_{10}$
at fixed coupling $\alpha_{10}=1$ in the absence of
disorders.}\label{Fig:ZCl}
\end{figure}

In the clean limit, $C_{g} = 0$, we have
\begin{eqnarray}
\frac{dZ_{f}}{dl} = C_{0}Z_{f}.
\end{eqnarray}
As shown in Fig.~\ref{Fig:ZCl}, $Z_{f}$ initially decreases with
growing $l$, but is saturated to a finite value as $l \rightarrow
\infty$, independent of the values of bare velocity ratio. The
finiteness of $Z_{f}$ indicates that the Dirac quasiparticles are
well-defined. These results are well consistent with previous RG
analysis presented in Ref.~\cite{Gonzalez99}.

\begin{figure}[htbp]
\center\includegraphics[width=2.8in]{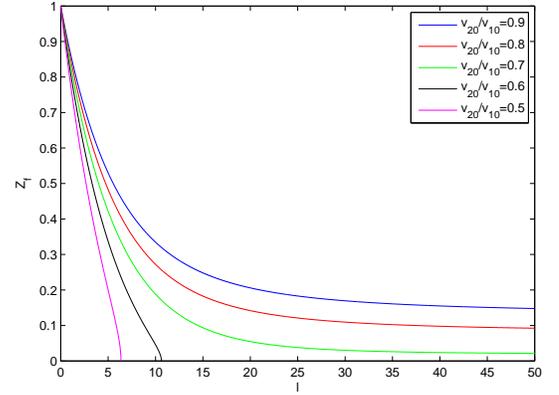} \caption{Wave
renormalization factor for different $v_{20}/v_{10}$ at fixed
coupling $\alpha_{10}=1$ in the presence of random chemical
potential with $v_{\Gamma0}^{2} \Delta/v_{10}^{2} =
0.5$.}\label{Fig:ZDisCP}
\end{figure}

In the presence of disorders, the initial value of $C_{g}$ becomes
finite. We show the flowing behaviors of $Z_f$ with growing $l$ in
the presence of random chemical potential in Fig.~\ref{Fig:ZDisCP}.
At fixed coupling $\alpha_{10}=1$, there is a critical value
$\delta_{0c}$ in the range $0.6 < \delta_{0} < 0.7$. In the range
$\delta_{0c} < \delta_{0} \le 1$ where the Coulomb interaction is
relatively weak, $Z_{f}$ approaches certain finite value as $l
\rightarrow \infty$ and the system is a stable Fermi liquid. In such
a case, the fermion velocities and $Z_f$ all flow in the same way as
that in the clean limit, so the observable quantities (such as DOS
and specific heat) are also very similar to those in clean graphene.
However, in the range $0.5 < \delta_0 \le \delta_{0c}$ where the
Coulomb interaction becomes sufficiently strong, $Z_{f}$ vanishes even
when $l$ is still finite. In addition, the fermion velocities and
$Z_f$ decrease rapidly to zero at certain finite energy scale. These
unusual behaviors indicate the instability of Fermi liquid and may,
as aforementioned, correspond to the formation of an insulating
state, where observable quantities (including DOS and specific heat)
should all vanish in the low-energy regime.

\begin{figure}[htbp]
\center \subfigure{\includegraphics[width=2.8in]{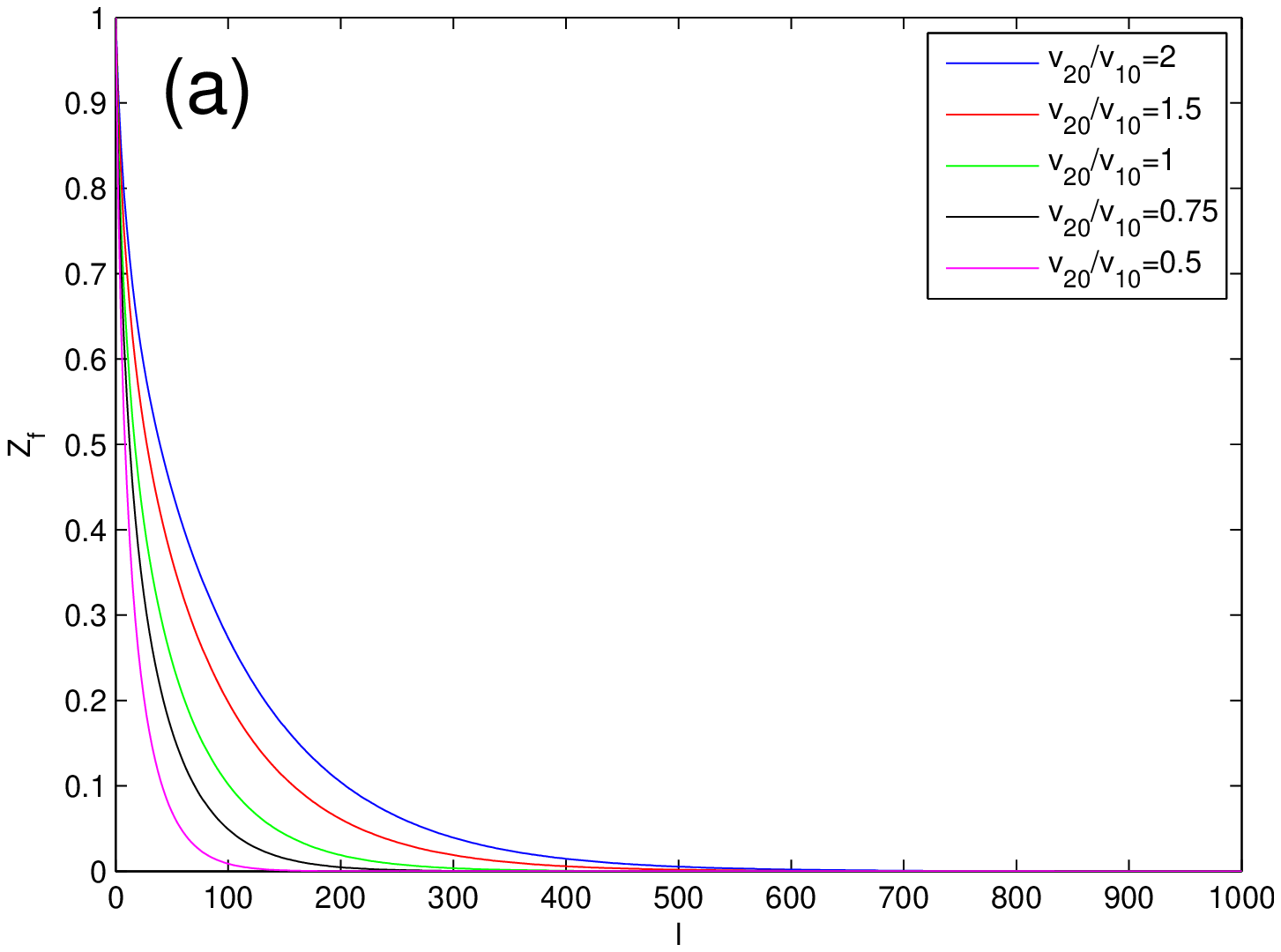}}
\subfigure{\includegraphics[width=2.9in]{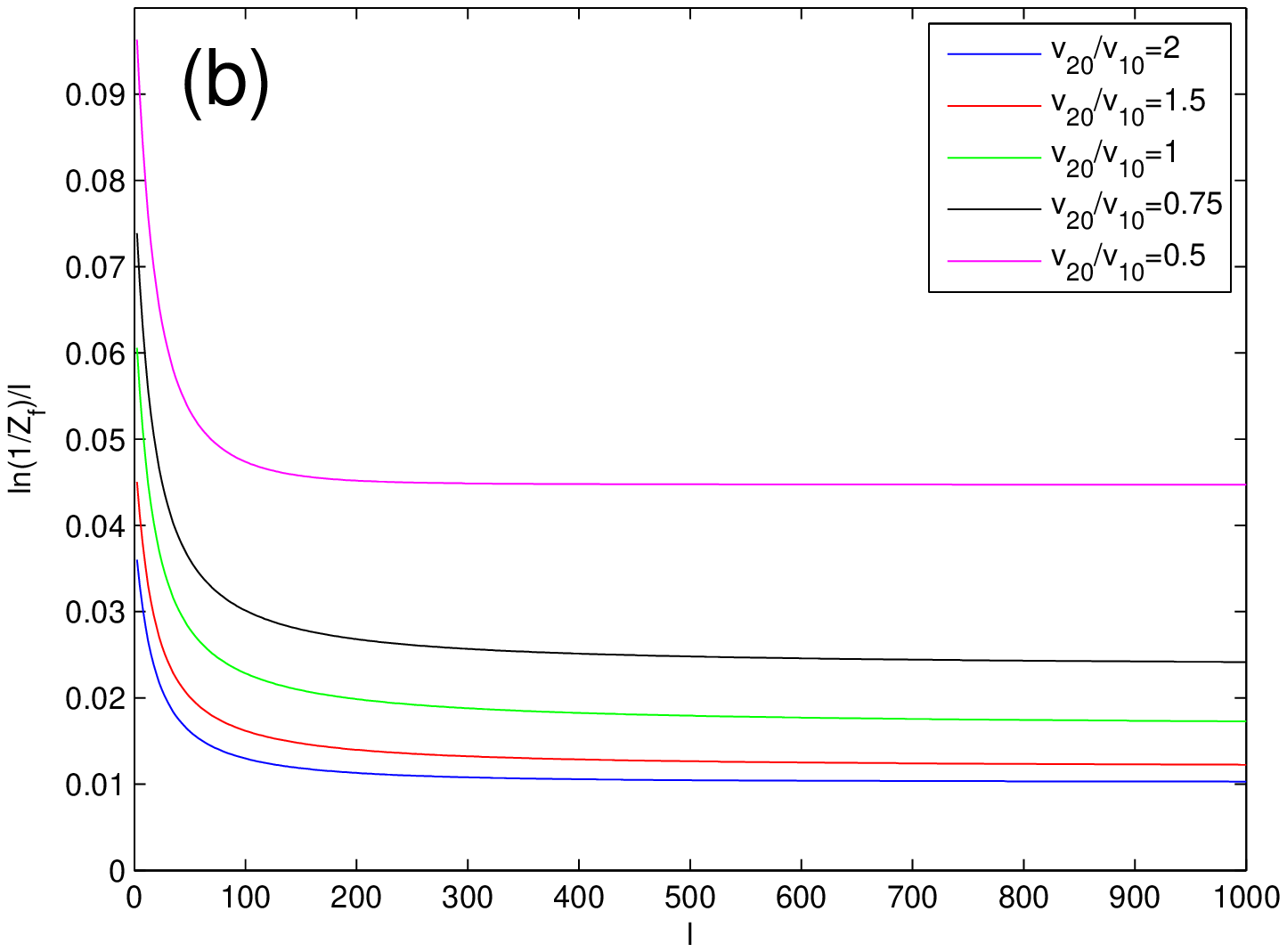}}
\caption{Wave renormalization factor for different $v_{20}/v_{10}$
at fixed coupling $\alpha_{10} = 1$ in the presence of random gauge
potential with $v_{\Gamma0}^{2}\Delta/v_{10}^{2}=0.05$.
}\label{Fig:ZDisGP}
\end{figure}

\begin{figure}[htbp]
\center \subfigure{\includegraphics[width=2.8in]{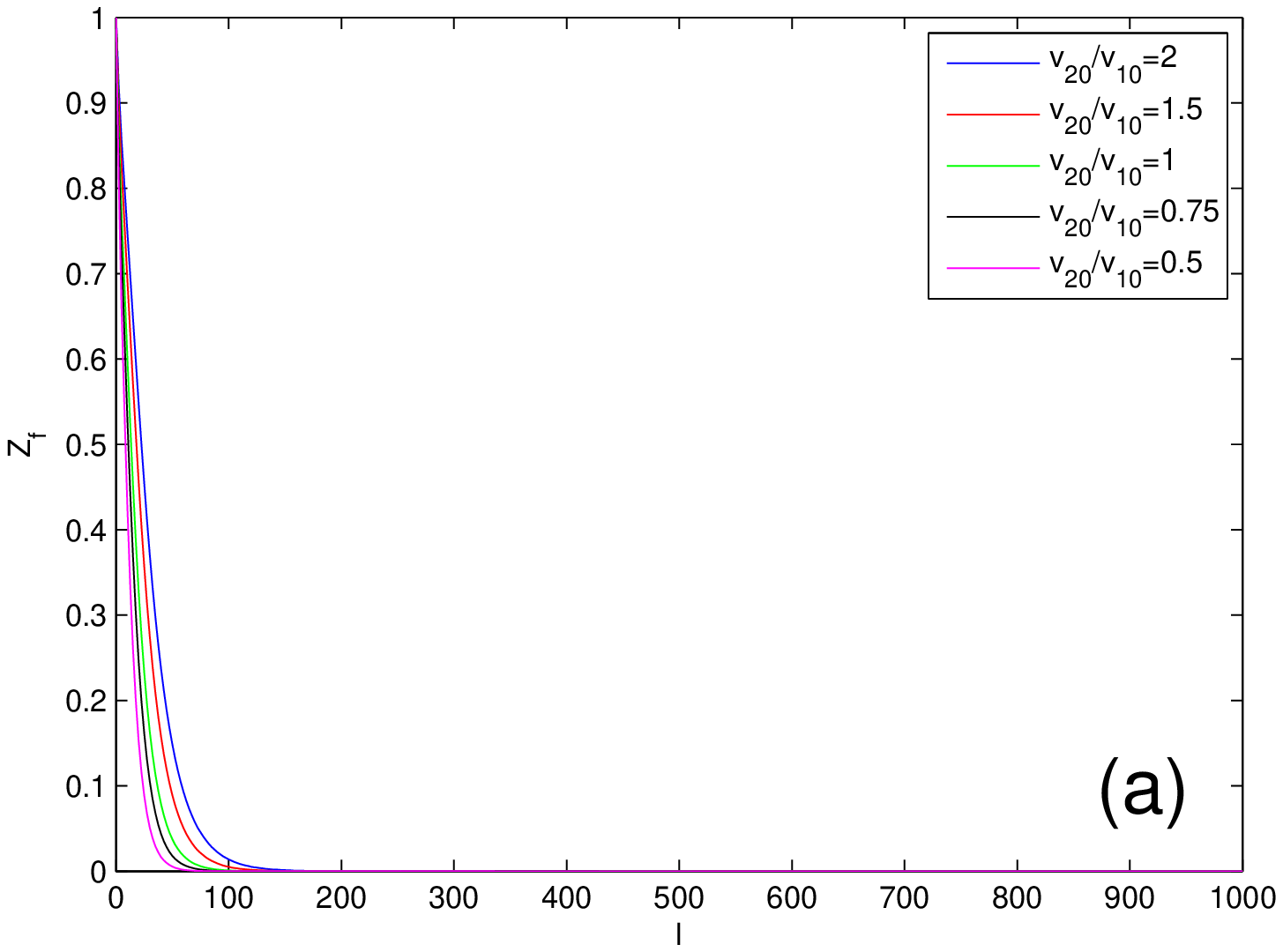}}
\subfigure{\includegraphics[width=2.85in]{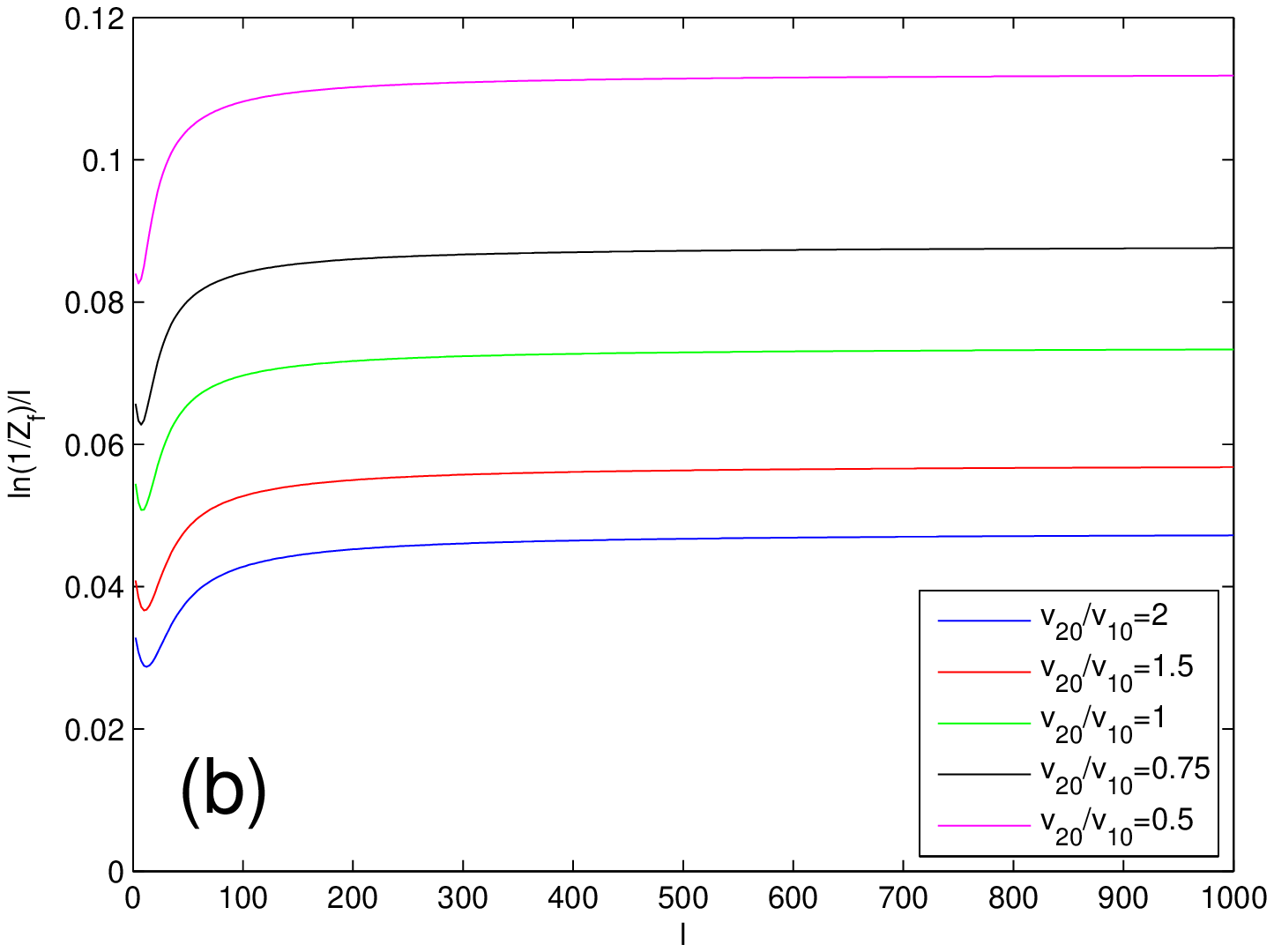}}
\caption{Wave renormalization factor for different $v_{20}/v_{10}$
at fixed coupling $\alpha_{10} = 1$ in the presence of random mass
with $v_{\Gamma0}^{2}\Delta/v_{10}^{2}=0.05$. }\label{Fig:ZDisMass}
\end{figure}

The $l$-dependence of $Z_f$ in the presence of random gauge
potential and random mass are shown in Fig.~\ref{Fig:ZDisGP} and
Fig.~\ref{Fig:ZDisMass}, respectively.
The most noticeable common
feature of these figures is that $Z_f$ vanishes as $l \rightarrow
\infty$, which is independent of the concrete values of bare
velocity ratio $\delta_0$.
This property is a signature of the
emergence of non-Fermi liquid behaviors. More concretely, $Z_{f}$
behaves as
\begin{eqnarray}
\lim_{l \rightarrow \infty}\ln(1/Z_{f})/l = \eta \quad (0<\eta<1)
\end{eqnarray}
in the low energy limit. Here, the magnitude of
constant $\eta$ is
determined by the parameters $\alpha_{10}$, $v_{20}/v_{10}$ and
$v_{\Gamma0}^{2}\Delta/v_{10}^{2}$. We can further write
\begin{eqnarray}
Z_{f}\propto \left(e^{-l}\right)^{\eta}.\label{eqn:ZDisGPMass}
\end{eqnarray}
Rewriting the energy as $\omega = \omega_{0}e^{-l}$, it is then easy
to obtain the real part of retarded self-energy
\begin{eqnarray}
\mathrm{Re}\Sigma^{R}(\omega)\propto \omega^{1-\eta}.
\end{eqnarray}
Using the Kramers-Kronig relation, we obtain the imaginary part of
retarded self-energy
\begin{eqnarray}
\mathrm{Im}\Sigma^{R}(\omega)\propto \omega^{1-\eta},
\end{eqnarray}
which is typical non-Fermi liquid behavior since $0<\eta<1$.
Therefore, both random gauge potential and random mass can lead to
breakdown of Fermi liquid and emergence of non-Fermi liquid ground
state.

We have seen in this subsection that the interplay of Coulomb
interaction and fermion-disorder interaction can lead to a series of
interesting behaviors in graphene samples with an anisotropic Dirac
cone. The system may be a normal Fermi liquid, a non-Fermi liquid,
or an exotic insulator, depending on the concrete values of bare
velocity ratio and the sorts of disorders. In particular, random
chemical potential behaves quite differently from random gauge
potential and random mass.

\subsection{Density of States\label{Subec:DOSBehavior}}

We now study the influence of Coulomb interaction on DOS using the
method employed by Xu \emph{et. al.} \cite{Xu}. The DOS
$\rho(\omega)$ is defined as
\begin{eqnarray}
\rho(\omega)&=&N\int\frac{dk_{1}dk_{2}}{(2\pi)^{2}}
\mathrm{Tr}\left\{\mathrm{Im}
\left[G^{R}(\omega,v_{1}k_{1},v_{2}k_{2})\right] \right\}
\nonumber \\
&=& \frac{N}{v_{1}v_{2}}\int\frac{dk_{1}'dk_{2}'}{(2\pi)^{2}}
\mathrm{Tr}\left\{\mathrm{Im}
\left[G^{R}(\omega,k_{1}',k_{2}')\right] \right\},
\end{eqnarray}
where $G^{R}(\omega,k_1,k_2)$ is the retarded two-point Green's
function (propagator) of Dirac fermions. In the absence of Coulomb
interaction, $G^{R}(\omega,k_1,k_2)$ is simply the retarded free
fermion propagator, and it is well-known that DOS exhibits a linear
energy dependence, i.e., $\rho(\omega) \propto \omega$. This
behavior can be significantly affected by Coulomb interaction and
random gauge potential (random mass). In the present problem, the
interaction effects are manifested in the RG flows of $v_{1,2}$ and
the anomalous dimension of fermion propagator. After straightforward
calculations shown in Appendix~\ref{sec:AppendixD}, we find that
\begin{eqnarray}
\frac{d\ln\rho}{d\ln\omega} &=&
\frac{1+3C_{0}-C_{1}-C_{2}-3C_{g}}{1-C_{0}+C_{1}+C_{g}}
\end{eqnarray}
for $v_{1}>v_{2}$, and that
\begin{eqnarray}
\frac{d\ln\rho}{d\ln\omega} &=&
\frac{1+3C_{0}-C_{1}-C_{2}-3C_{g}}{1-C_{0}+C_{2}+C_{g}}
\end{eqnarray}
for $v_{2}>v_{1}$.

\begin{figure}[htbp]
\center
\subfigure{\includegraphics[width=2.8in]{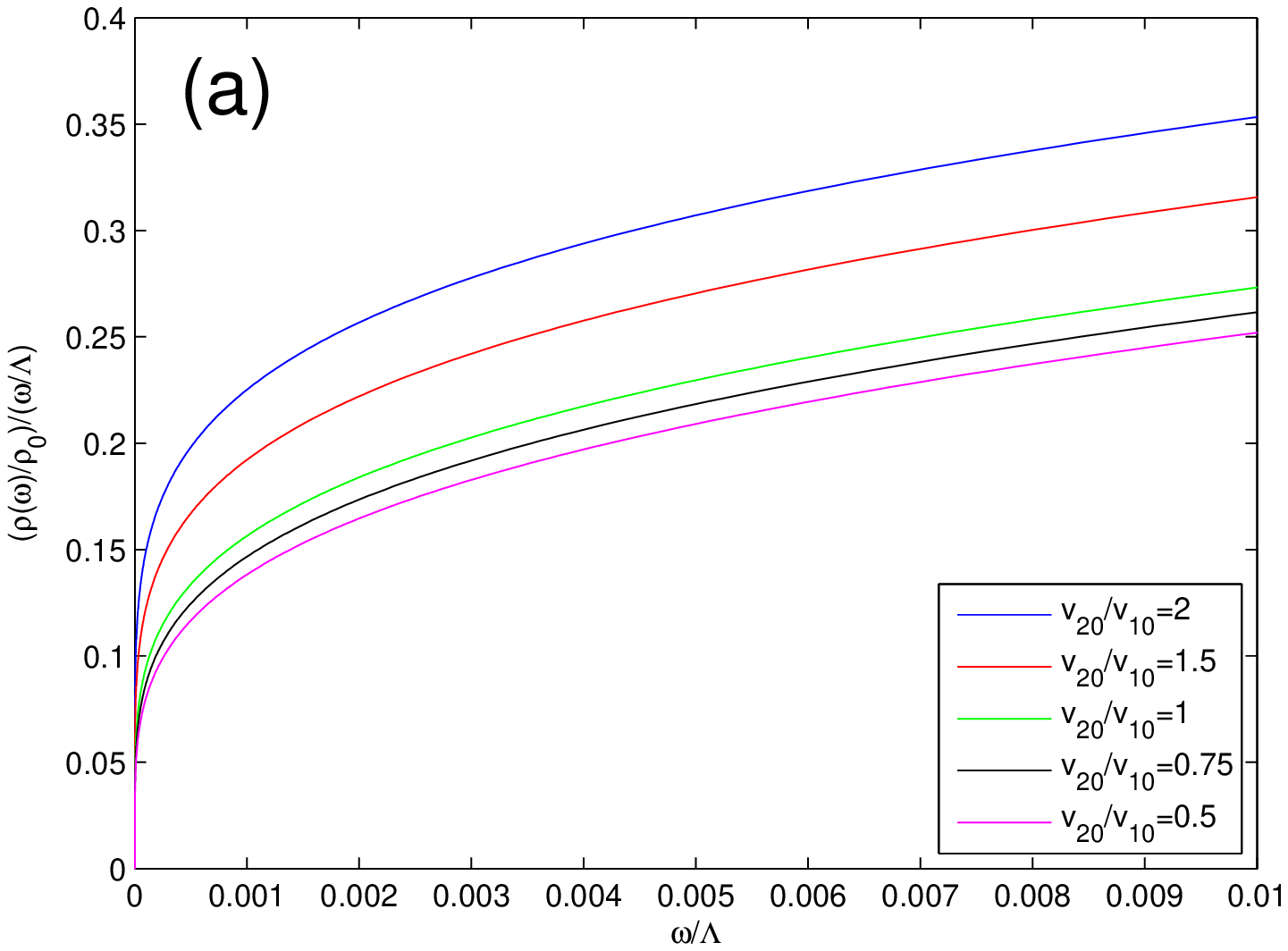}}
\subfigure{\includegraphics[width=2.8in]{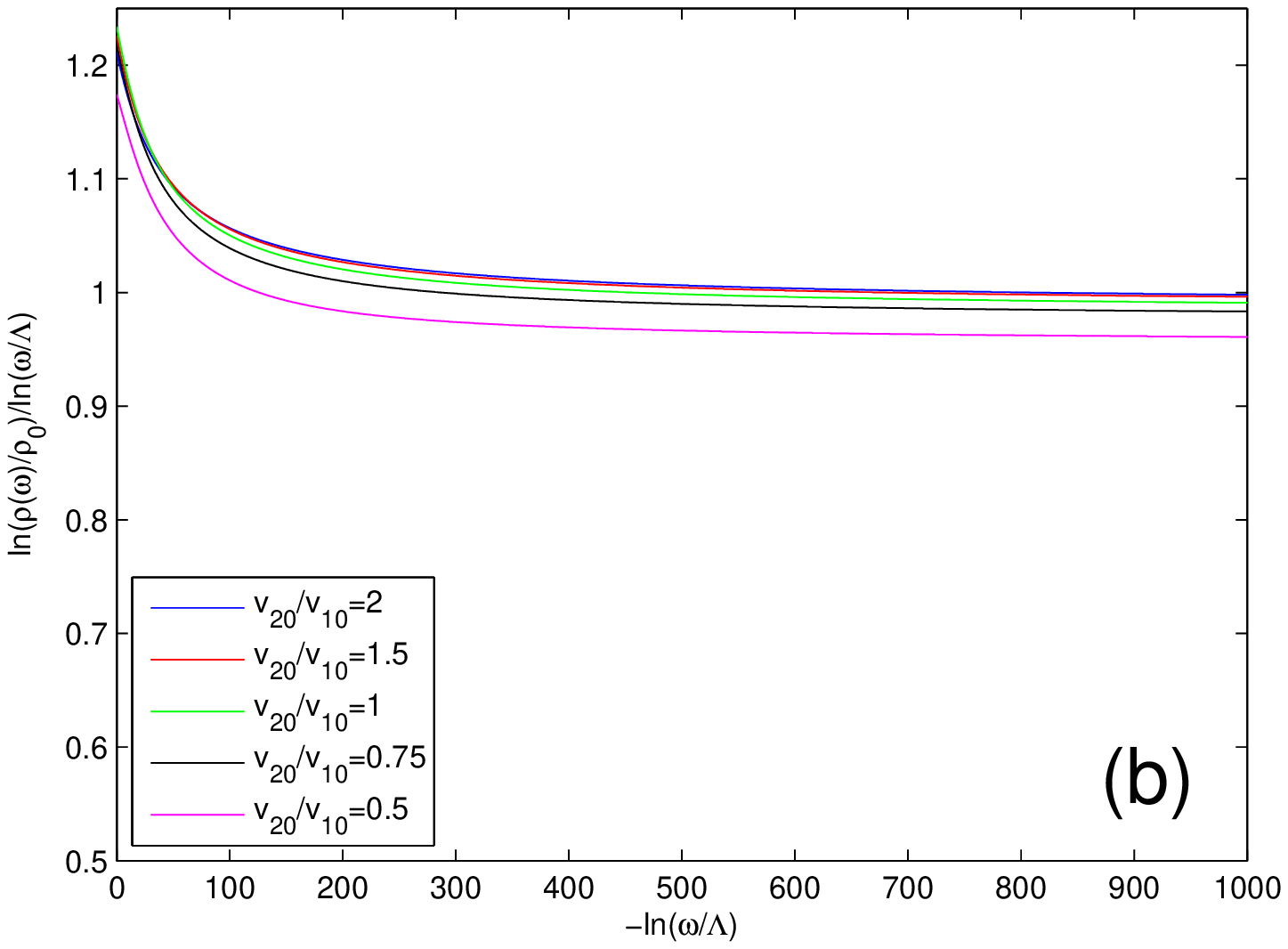}}
\subfigure{\includegraphics[width=2.8in]{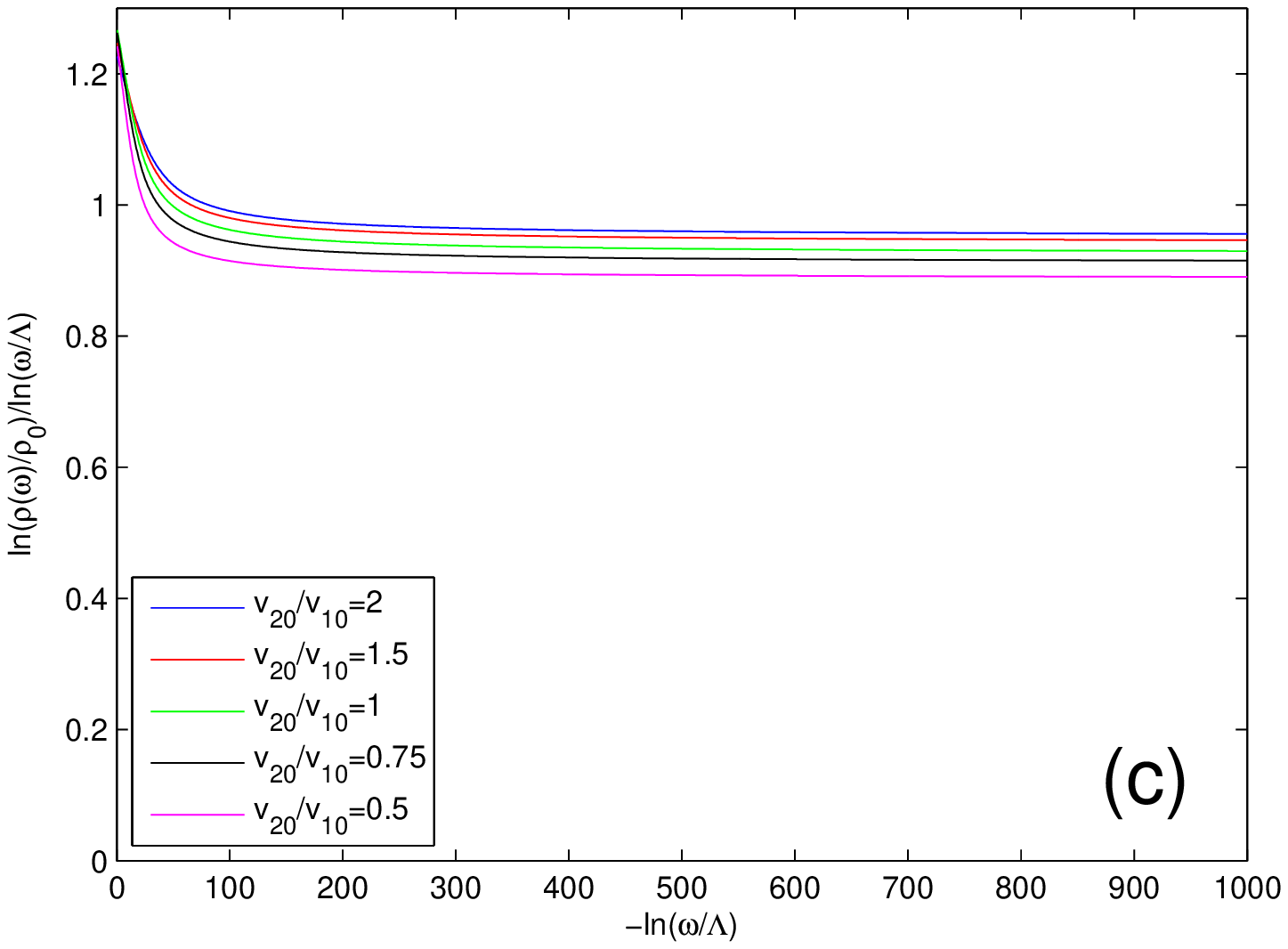}}
\caption{(a) DOS for different $v_{20}/v_{10}$ at fixed coupling
$\alpha_{10} = 1$ in the absence of disorders; (b) DOS for different
$v_{20}/v_{10}$ at fixed coupling $\alpha_{10}=1$ in the presence of
random gauge potential with $v_{\Gamma0}^{2}\Delta/v_{10}^{2} =
0.05$; (c) Wave renormalization factor for different $v_{20}/v_{10}$
at fixed coupling $\alpha_{10}=1$ in the presence of random mass
with $v_{\Gamma0}^{2} \Delta/v_{10}^{2} = 0.05$.} \label{Fig:DOS}
\end{figure}

In the clean limit with $C_{g} = 0$, the $\omega$-dependence of
$\rho(\omega)$ for different bare ratios are presented in
Fig.~\ref{Fig:DOS}(a). In the low-energy regime, $\omega \rightarrow
0$, we have
\begin{eqnarray}
\frac{\rho(\omega)}{\omega}\sim\frac{1}{\ln(\omega)}.
\end{eqnarray}

When there is random gauge potential or random mass, the
corresponding $\rho(\omega)$ for different bare ratios are shown in
Fig.~\ref{Fig:DOS}(b) and (c). It can be found that $\rho(\omega)$
behaves as
\begin{eqnarray}
\rho(\omega) \sim \omega^{1-\eta} \qquad(0<\eta<1)
\label{eqn:DOSDisGPMass}
\end{eqnarray}
in the limit $\omega \rightarrow 0$. Comparing this expression to
the linear $\omega$-dependence of DOS obtained in the
non-interacting case, we know that $\eta$ reflects the corrections
arising from Coulomb interaction and disorder scattering.

The dynamical exponent describes how the energy should be rescaled
relative to the momenta \cite{Hertz76, Lohneysen07, Herbutbook}. In
our notations, $z$ is encoded in velocities of the Dirac fermions.
For a free anisotropic Dirac fermion system, the fermion velocities
are constants, so the dynamical exponent $z=1$. Including the
interaction effects, we have two dynamic exponents defined as
\begin{eqnarray}
z_{1}&=&1-\frac{d \ln v_{1}(l)}{d l},
\\
z_{2}&=&1-\frac{d \ln v_{2}(l)}{d l}.
\end{eqnarray}
Due to interplay of Coulomb interaction and random gauge
field (random mass), $v_{1}(l)/v_{2}(l)\rightarrow1$ in the lowest
energy limit $l \rightarrow \infty$, which corresponds to an
isotropic fixed point. At the same time, $v_{1}(l)$ and $v_{2}(l)$
approach a finite constant. Therefore, $z$ satisfies
\begin{eqnarray}
z = z_{1,2}(l\rightarrow \infty) = 1.
\end{eqnarray}

Our result of $\rho(\omega)$ is different from that obtained in
Ref.~\cite{Herbut08}, where it is shown that the fermion velocity is
saturated to a finite value in the presence of random gauge
potential and that $\rho(\omega) \sim \omega$ since $z=1$ in the
low-energy regime. However, we notice that the non-trivial
corrections to DOS actually come from both the change of dynamic
exponent and the non-trivial wave function renormalization. Although
the dynamical exponent $z=1$, the wave renormalization function
receives a non-trivial correction given in
Eq.~(\ref{eqn:ZDisGPMass}) and lead to non-Fermi liquid like damping
rate of Dirac fermions. Consequently, the behaviors of low-energy
DOS are disorder dependent, as can be seen from
Eq.~(\ref{eqn:DOSDisGPMass}).

\begin{figure}[htbp]
\center \subfigure{
\includegraphics[width=2.8in]{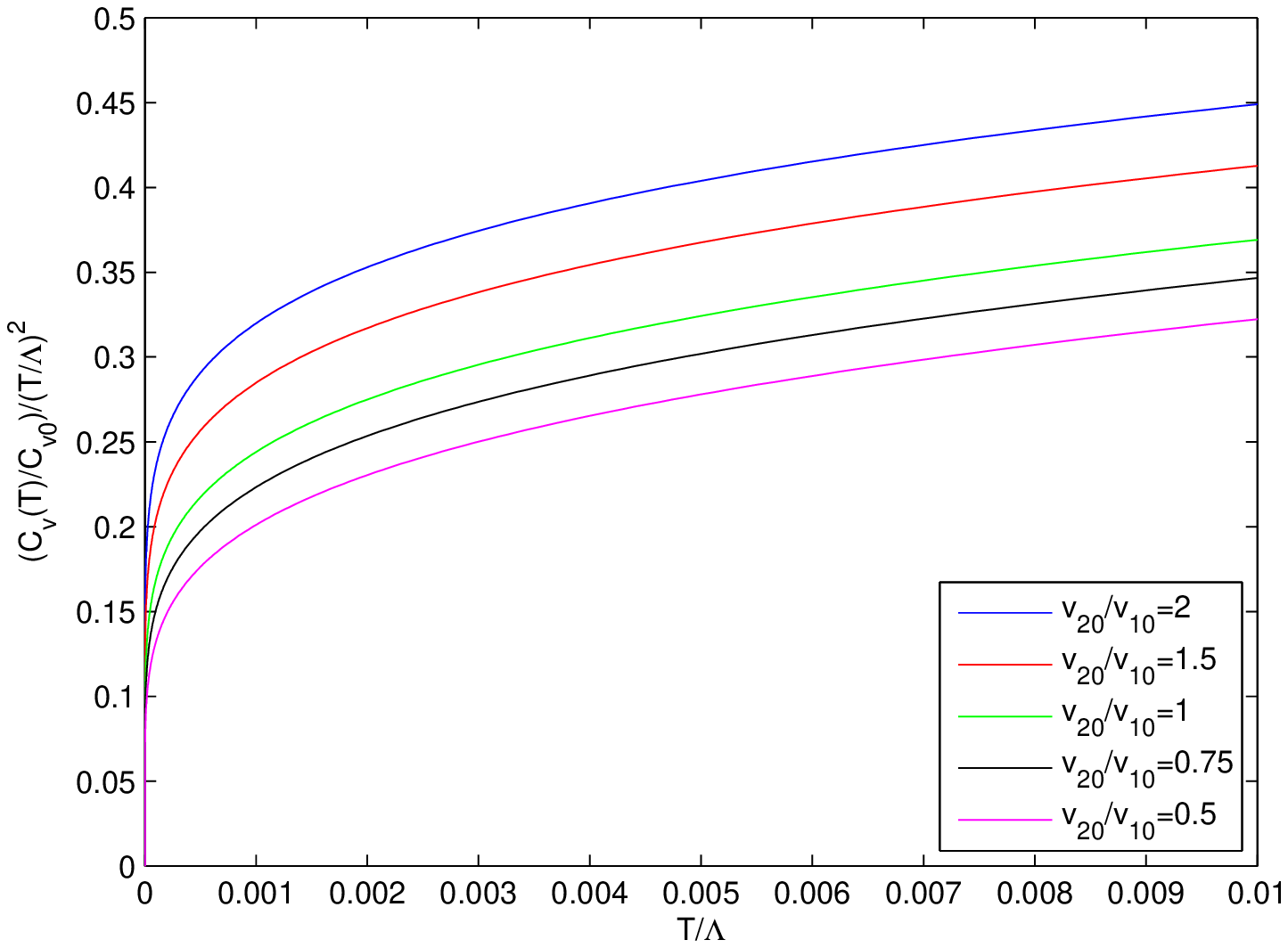}}
\subfigure{\includegraphics[width=2.8in]{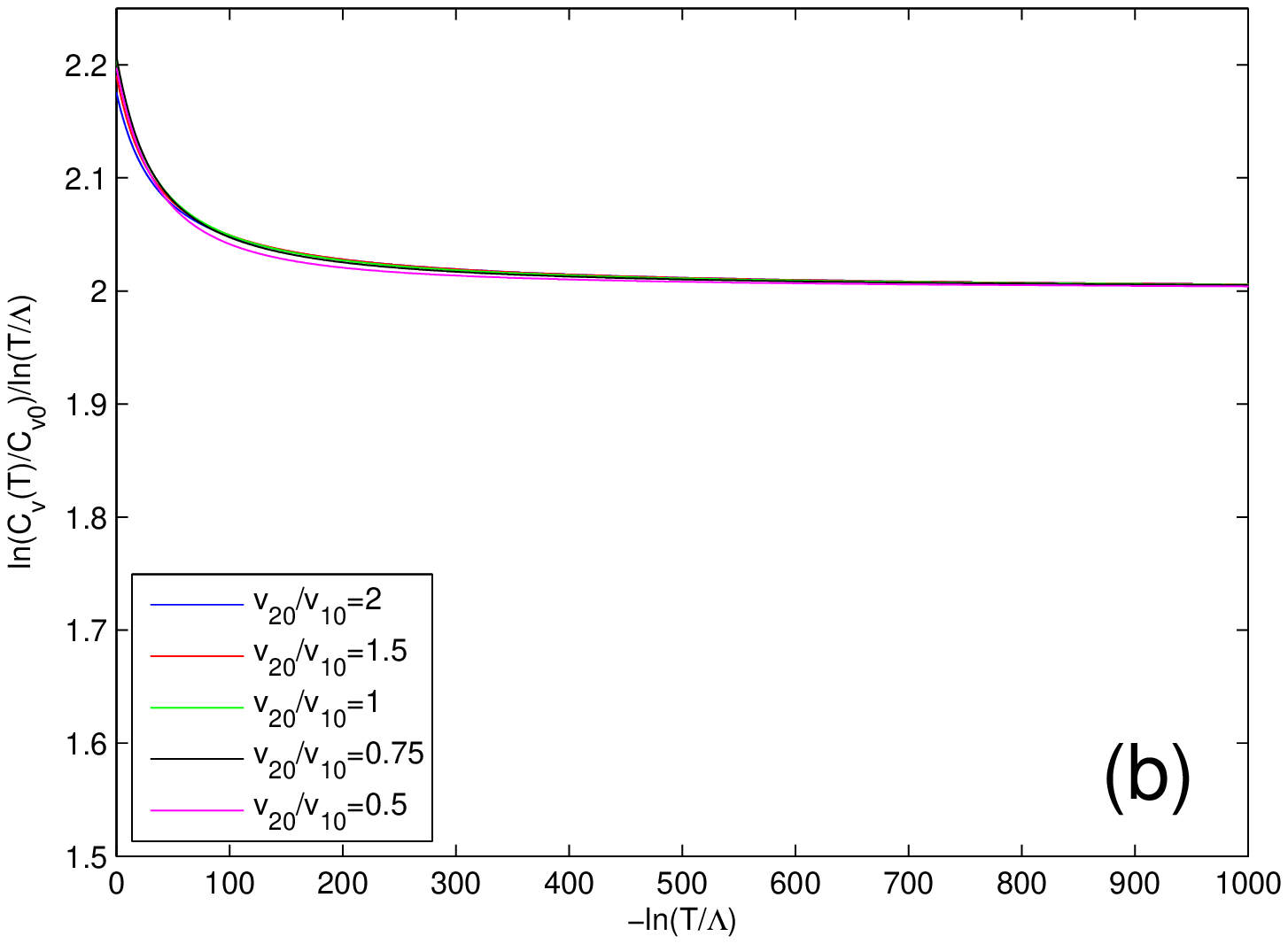}} \subfigure{
\includegraphics[width=2.8in]{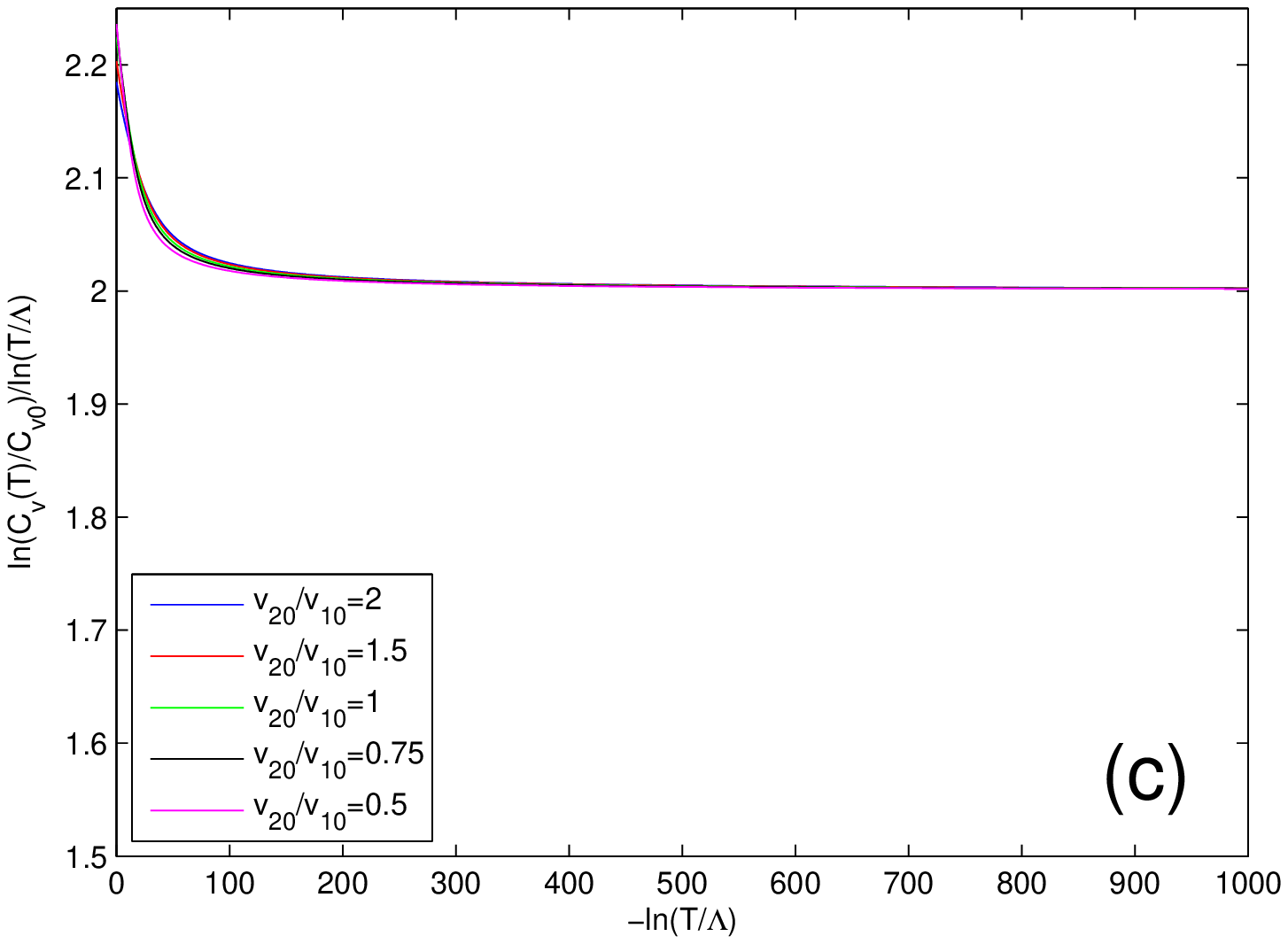}}
\caption{(a) Specific heat for different $v_{20}/v_{10}$ at fixed
coupling $\alpha_{10} = 1$ in the absence of disorders; (b) Specific
heat for different $v_{20}/v_{10}$ at fixed coupling $\alpha_{10} =
1$ in the presence of random gauge potential with
$v_{\Gamma0}^{2}\Delta/v_{10}^{2} = 0.05$; (c) Specific heat for
different $v_{20}/v_{10}$ at fixed coupling $\alpha_{10}=1$ in the
presence of random mass with $v_{\Gamma0}^{2} \Delta/v_{10}^{2} =
0.05$.}\label{Fig:Cv}
\end{figure}

\subsection{Specific heat \label{subsec:CvBehavior}}

To calculate specific heat, we also follow the method used in
Ref.~\cite{Xu}. The free energy $\mathcal{F} = T\ln\mathcal{Z}/V$
contains the following singular part
\begin{eqnarray}
\mathcal{F} = \left(\xi_{\tau}\xi_{x}\xi_{y}\right)^{-1},
\end{eqnarray}
where $\xi_{\tau}\sim1/T$, $\xi_{x}\sim v_{1}\xi_{\tau}$, and
$\xi_{y}\sim v_{2}\xi_{\tau}$. In an interacting anisotropic
graphene, the free energy is found to depend on $T$ as
\begin{eqnarray}
\mathcal{F} \sim \frac{1}{v_{1}v_{2}}T^3.
\end{eqnarray}
The corresponding specific heat is given by
\begin{eqnarray}
C_{V} = -T\frac{\partial^2 \mathcal{F}}{\partial
T^2}\sim\frac{1}{v_{1}v_{2}}T^2.
\end{eqnarray}
Here, the interaction effects are reflected in the nontrivial flows
of $v_{1}$ and $v_{2}$. After performing calculations shown in
Appendix~\ref{sec:AppendixD}, we find that $C_{V}$ varies with $T$
as
\begin{eqnarray}
\frac{d \ln C_{V}}{d\ln T} \sim 2 + \frac{2C_{0} -
C_{1}-C_{2}-2C_{g}}{1-C_{0}+C_{1}+C_{g}}
\end{eqnarray}
for $v_{1}>v_{2}$, and
\begin{eqnarray}
\frac{d \ln C_{V}}{d\ln T} \sim 2 +
\frac{2C_{0}-C_{1}-C_{2}-2C_{g}}{1-C_{0}+C_{2}+C_{g}}
\end{eqnarray}
for $v_{2}>v_{1}$.

In clean graphene, the specific heat $C_{V}(T)$ for different
parameters is shown in Fig.~\ref{Fig:Cv}(a). We can see that
$C_{V}(T)/T^2 \rightarrow0$ in the limit of $\omega \rightarrow 0$.
More concretely, we find that
\begin{eqnarray}
\frac{C_{V}(T)}{T^2}\sim\frac{1}{\ln(T)},
\end{eqnarray}
which is consistent with the results obtained in
Ref.~\cite{Vafek07}.

In the presence of random gauge potential or random mass, the
corresponding specific heat $C_{V}(T)$ for different parameters are
shown in Fig.~\ref{Fig:Cv} (b) and (c). $C_{V}(T)$ behaves as
\begin{eqnarray}
C_{V}(T) \sim T^2
\end{eqnarray}
in the limit $T \rightarrow 0$. General scaling analysis show that
the specific heat should satisfy $C_{V}(T)\propto T^{d/z}$
\cite{Lohneysen07, Continentinobook}, where $d$ is the spatial
dimension and $z$ is the dynamical exponent. In the present case,
the fermion velocities are saturated to finite values and the
dynamical exponent $z \rightarrow 1$ in the low energy regime, so
$C_{V}(T)\propto T^2$. It might seem strange that, the  DOS is not
linear in $\omega$ as shown in Eq.~(\ref{eqn:DOSDisGPMass}) but the
specific heat still exhibits quadratic $T$-dependence. This can be
understood as follows. When the Dirac fermion propagator acquires a
finite positive anomalous dimension $\eta$, the Landau damping rate
of the fermion takes non-Fermi liquid behavior and the DOS is no
longer linear in energy. However the anomalous dimension does not
change the $T$-dependence of specific heat. To better understand
this problem, we make a brief discussion in
Appendix~\ref{sec:AppendixE}.

\section{Dynamical gap generation in anisotropic graphene \label{Sec:Gap}}

Recently, a number of theoretical and numerical works
\cite{Khveshchenko01, Gorbar02, Khveshchenko04, Liu09,
Khveshchenko09, Gamayun10, Hands08, Drut09, CastroNetoPhys, Zhang11,
LiuWang, Wang12} have predicted that the long-range Coulomb
interaction between massless Dirac fermions in graphene may generate
a dynamical gap by forming excitonic pairs and consequently lead to
semimetal-insulator transition. This gap-generating mechanism is of
great interest to theorists because it can be considered as a
concrete condensed-matter realization of the non-perturbative
phenomenon of dynamical chiral symmetry breaking, which was first
proposed by Nambu and Jona-Lasinio \cite{Nambu} and has played a
significant role in the development of modern particle physics
\cite{Miransky, Roberts}. From a technological point of view, a
gapped graphene is more promising than gapless one as a candidate
for manipulating novel electronic devices \cite{CastroNetoPhys,
Geim07}. For these reasons, the mechanism of dynamical gap
generation and the resultant semimetal-insulator transition have
stimulated considerable effort in recent years.

\begin{figure}[htbp]
\center
\includegraphics[width=3.3in]{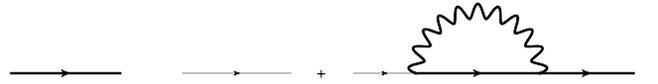}
\caption{Diagrams for fermion self-energy in DS equation approach.
Thick solid line denotes dressed fermion propagator.}
\label{Fig:FermionSFConsistent}
\end{figure}

Earlier calculations carried out using the DS equation approach have
reached an optimistic conclusion that a dynamical gap is generated
by Coulomb interaction in suspended clean graphene
\cite{Khveshchenko01, Gorbar02, Khveshchenko04, Liu09,
Khveshchenko09, Gamayun10}. However, a number of approximations are
adopted in these works, which more or less lowers the reliability of
this conclusion (see Ref.~\cite{Wang12} for a brief review of these
approximations). More recently, we have revisited this problem by
improving most of these approximations and found that the Coulomb
interaction in suspended graphene is indeed not strong enough to
open a dynamical gap \cite{Wang12}. At the same time,
experimentalists have measured the energy spectrum of suspended
graphene at ultra-low temperatures and observed no evidence of
insulating behavior \cite{Elias11, Mayorov12}. A key factor that
weakens the effective Coulomb interaction is the unusual
renormalization of fermion velocity. It is known that the Coulomb
interaction coupling $\alpha \propto e^2/v$ with $v$ being the
universal fermion velocity in isotropic graphene. As $v$ diverges at
the lowest energy, $\alpha$ tends to vanish, which means the
effective interaction strength is significantly reduced. Apparently,
fermion velocity renormalization plays a crucial role in this
problem, and needs to be carefully treated.

As shown in the last sections, the renormalization of fermion
velocities in anisotropic graphene can be different from that
in the case of isotropic graphene. Such an important difference may
lead to remarkable change of the effective strength of Coulomb
interaction. It is therefore very interesting to investigate how
dynamical gap generation is affected by the velocity anisotropy. In
this section, we are particularly interested in whether the velocity
anisotropy enhances or suppresses dynamical gap generation in clean
graphene. The non-perturbative DS equation approach
\cite{Khveshchenko01, Gorbar02, Khveshchenko04, Liu09,
Khveshchenko09, Gamayun10, Zhang11, LiuWang, Nambu, Miransky,
Roberts, WangLiu13} will be used to address this issue since the
conventional perturbative expansion is unable to tackle the
non-perturbative phenomenon of dynamical gap generation. Moreover,
in this section we will not consider the effects of disorders, which
are technically difficult to be incorporated in the self-consistent
DS equation \cite{Liu09, LiuWang}.

After including the interaction induced self-energy corrections, the
free fermion propagator, given in Eq.~(\ref{eqn:FreeFermion}), will
be renormalized to the following full propagator
\begin{eqnarray}
G(i\omega,\mathbf{k}) = \frac{1}{-i\omega A_{0}\gamma_{0} +
v_{1}k_{1}A_{1}\gamma_{1} + v_{2}k_{2}A_{2}\gamma_{2} + m},
\label{eqn:DressedFermion}
\end{eqnarray}
where $m$ represents a finite dynamical gap and $A_{0,1,2}$ are the
three components of wave function renormalization. According to the
Feynman diagram shown in Fig.~\ref{Fig:FermionSFConsistent}, the
dressed fermion propagator is related to the free one via the
following DS equation,
\begin{eqnarray}
G^{-1}(i\varepsilon,\mathbf{p}) &=&
G_{0}^{-1}(i\varepsilon,\mathbf{p}) + \int\frac{d\omega}{2\pi}
\frac{d^2\mathbf{k}}{(2\pi)^{2}} \gamma_{0}
G(i\omega,\mathbf{k})\gamma_{0} \nonumber \\
&& \times D\left(i(\varepsilon-\omega),\mathbf{p}-\mathbf{k}\right).
\label{eqn:DysonEquation}
\end{eqnarray}
To the lowest order of $1/N$ expansion, we take $A_{0} = A_{1} =
A_{2} = 1$ for simplicity and substitute
Eq.~(\ref{eqn:DressedFermion}) into Eq.~(\ref{eqn:DysonEquation}).
After straightforward calculations, we obtain an integral equation
for the dynamical gap $m$,
\begin{eqnarray}
m(i\varepsilon,p_{1},p_{2}) &=& \int\frac{d\omega}{2\pi}
\int\frac{dk_{x}}{2\pi} \int\frac{dk_{y}}{2\pi}
\nonumber \\
&& \times \frac{m(i\omega,k_{1},k_{2})}{\omega^2+v_{1}^{2}k_{1}^2 +
v_{2}^{2}k_{2}^{2} + m^{2}(i\omega,k_{1},k_{2})}
\nonumber \\
&&\times\frac{1}{\frac{|\mathbf{q}|}{\frac{2\pi e^{2}}{\epsilon}} +
\frac{N}{8v_1v_2}\frac{v_{1}^{2}q_{1}^{2}+v_{2}^{2}q_{2}^{2}}
{\sqrt{\Omega^2+v_{1}^{2}q_{1}^{2}+v_{2}^{2}q_{2}^{2}}}},
\end{eqnarray}
where $\Omega=\epsilon-\omega$, $q_{1}=p_{1}-k_{1}$ and
$q_{2}=p_{2}-k_{2}$. This nonlinear equation is very complicated and
needs to be numerically solved. A finite fermion gap is generated by
the Coulomb interaction once this equation develops a non-trivial
solution. In the anisotropic case, the equation of $m$ depends on
energy and two components of momentum separately, which makes it
difficult to solve the integral equation numerically. In order to
simplify the numerical computations, we adopt several frequently
used approximations, and then compare the results obtained under
these approximations. If some common features can be extracted from
all the results, then we can qualitatively judge whether spatial
anisotropy is in favor of dynamical gap generation or not.

We will consider six different approximations. First, we consider
the instantaneous approximation which drops the energy dependence of
polarization function \cite{Khveshchenko01, Gorbar02,
Khveshchenko04} as follows
\begin{eqnarray}
&&\frac{1}{\frac{|\mathbf{q}|}{\frac{2\pi e^{2}}{\epsilon}} +
\frac{N}{8v_1v_2}\frac{v_{1}^{2}q_{1}^{2}+v_{2}^{2}q_{2}^{2}}
{\sqrt{\Omega^2+v_{1}^{2}q_{1}^{2}+v_{2}^{2}q_{2}^{2}}}}
\nonumber \\
\rightarrow && \frac{1}{\frac{|\mathbf{q}|}{\frac{2\pi
e^{2}}{\epsilon}} + \frac{N}{8v_1v_2}\sqrt{v_{1}^{2}q_{1}^{2} +
v_{2}^{2}q_{2}^{2}}}.\label{eqn:ApproxInstan}
\end{eqnarray}
Now the gap equation is simplified to
\begin{eqnarray}
m(p_{1},p_{2}) &=& \frac{1}{2}\int\frac{dk_{1}}{2\pi}
\int\frac{dk_{2}}{2\pi} \frac{m(k_{1},k_{2})}{\sqrt{k_{1}^2 +
\delta^{2}k_{2}^{2} + m^{2}(k_{1},k_{2})}} \nonumber \\
&&\times\frac{1}{\frac{\sqrt{q_{1}^2+q_{2}^2}}{2\pi\alpha_{1}} +
\frac{N}{8\delta}\sqrt{ q_{1}^{2} +\delta^{2}q_{2}^{2}}},\label{eqn:GapInstan}
\end{eqnarray}
where $\delta=v_{2}/v_{1}$. In the derivation of this gap equation,
we have performed the re-scaling transformations
\begin{eqnarray}
v_{1}p_{1,2}\rightarrow p_{1,2}, \quad v_{1}k_{1,2} \rightarrow
k_{1,2}.
\end{eqnarray}
Such transformation will also be used in the calculations to be
performed below.

Second, we utilize the following approximation \cite{Khveshchenko09}
\begin{eqnarray}
&&\frac{1}{\frac{|\mathbf{q}|}{\frac{2\pi e^{2}}{\epsilon}} +
\frac{N}{8v_1v_2}\frac{v_{1}^{2}q_{1}^{2}+v_{2}^{2}q_{2}^{2}}
{\sqrt{\Omega^2+v_{1}^{2}q_{1}^{2}+v_{2}^{2}q_{2}^{2}}}}
\nonumber \\
\rightarrow && \frac{1}{\frac{|\mathbf{q}|}{\frac{2\pi
e^{2}}{\epsilon}}+\frac{N}{8\sqrt{2}v_1v_2}
\sqrt{v_{1}^{2}q_{1}^{2}+v_{2}^{2}q_{2}^{2}}}.\label{eqn:ApproxKhv}
\end{eqnarray}
The corresponding gap equation has the form
\begin{eqnarray}
m(p_{1},p_{2}) &=& \frac{1}{2}\int\frac{dk_{1}}{2\pi}
\int\frac{dk_{2}}{2\pi} \frac{m(k_{1},k_{2})}{\sqrt{k_{1}^2 +
\delta^{2}k_{2}^{2} + m^{2}(k_{1},k_{2})}}
\nonumber \\
&&\times\frac{1}{\frac{\sqrt{q_{1}^2+q_{2}^2}}{2\pi\alpha_{1}} +
\frac{N}{8\sqrt{2}\delta}\sqrt{ q_{1}^{2} +\delta^{2}q_{2}^{2}}}.\label{eqn:GapKhv}
\end{eqnarray}

Third, we consider the approximation used in Ref.~\cite{Gamayun10},
which assumes that $m(i\epsilon,\mathbf{p})$ is energy-independent,
i.e.,
\begin{equation}
m(i\epsilon,p_{1},p_{2}) \rightarrow m(p_{1},p_{2}).\label{eqn:ApproxGusynin}
\end{equation}
Applying this approximation leads to
\begin{eqnarray}
m(p_{1},p_{2}) &=& \alpha_{1}\int\frac{dk_{1}}{2\pi}
\int\frac{dk_{2}}{2\pi}\frac{1}{\sqrt{q_{1}^2 + q_{2}^2}}
\nonumber \\
&& \times \frac{m(k_{1},k_{2})J(d,g)}{\sqrt{k_{1}^2 +
\delta^{2}k_{2}^{2} + m^{2}(k_{1},k_{2})}},\label{eqn:GapGusynin}
\end{eqnarray}
where
\begin{eqnarray}
J(d,g) &=& \frac{ \left(d^{2}-1\right)\left[\pi - gc(d)\right] +
dg^{2}c(g)}{d^{2}+g^{2}-1},
\end{eqnarray}
with
\begin{eqnarray}
c(x)=\left\{
\begin{array}{ll}
\frac{2}{\sqrt{1-x^{2}}}\cos^{-1}\left(x\right) & x<1
\\
\\
\frac{2}{\sqrt{x^{2}-1}}\cosh^{-1}\left(x\right) & x>1
\\
\\
2 & x=1
\end{array}
\right.,
\end{eqnarray}
\begin{eqnarray}
d&=&\sqrt{\frac{k_{1}^2 + \delta^{2}k_{2}^{2} +
m^{2}(k_{1},k_{2})}{q_{1}^{2} +\delta^{2}q_{2}^{2}}},
\end{eqnarray}
\begin{eqnarray}
g&=&\frac{N\pi \alpha_{1}\sqrt{q_{1}^{2} +\delta^{2}q_{2}^{2}}}
{4\delta\sqrt{q_{1}^2+q_{2}^2}}.
\end{eqnarray}

In these approximations, the fermion velocities $v_1$ and $v_2$ are
assumed to take bare values. However, both $v_1$ and $v_2$ are
indeed strongly renormalized by the Coulomb interaction. To
incorporate the feedback effects of strong velocity renormalization
on the DS generation, we can replace the bare fermion velocities by
the renormalized, momentum-dependent velocities
\cite{Khveshchenko09}, $v_{1,2} \rightarrow v_{1,2}(k)$, which are
determined by the solutions of Eq.~(\ref{eqn:RGV1AsCl}) and
Eq.~(\ref{eqn:RGV2AsCl}), and then solve the new gap equations.

\begin{figure*}[hbtp]
\centering
   \subfigure{
   \includegraphics[width=2.1in]{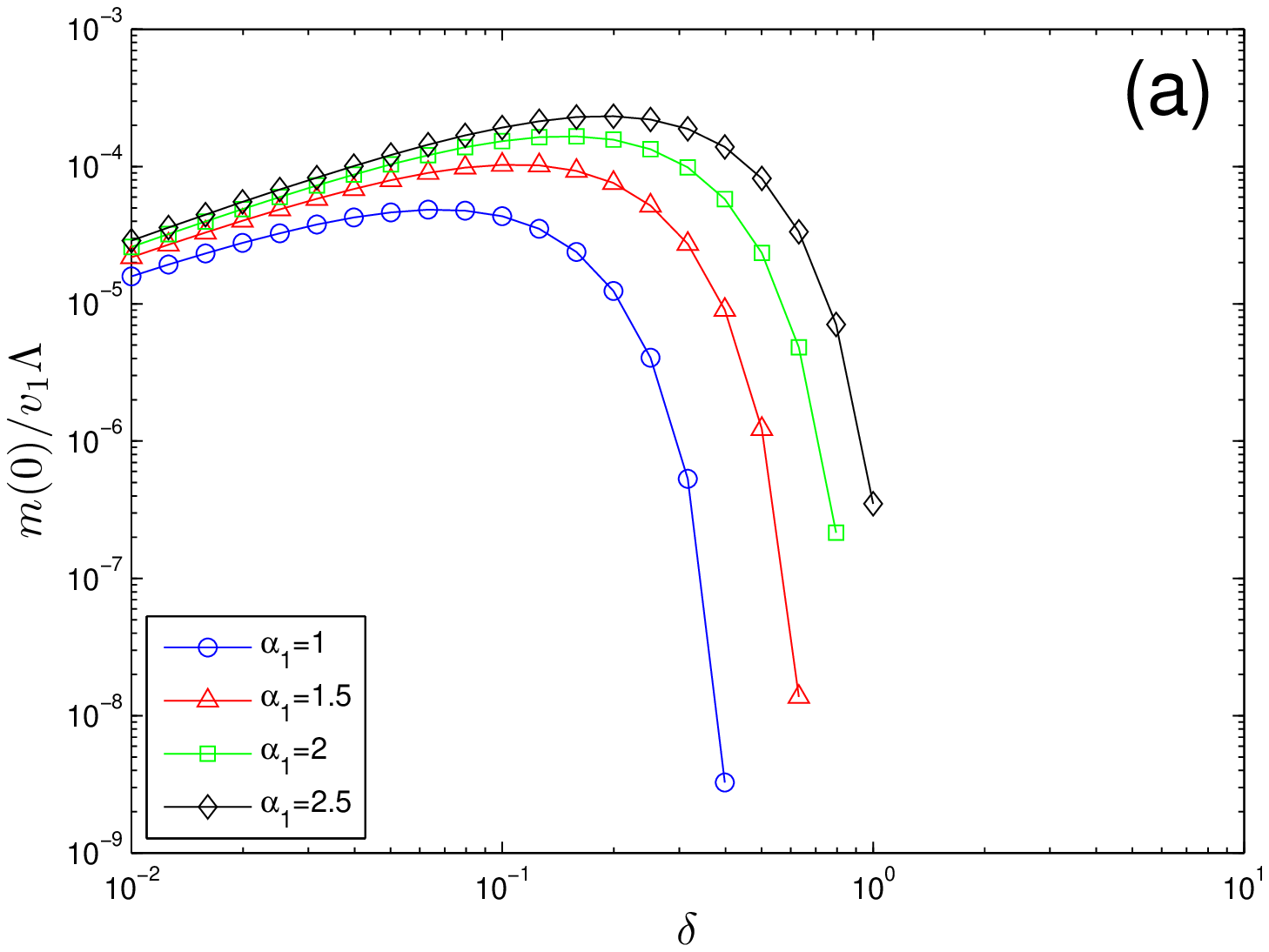}}
   \subfigure{
   \includegraphics[width=2.1in]{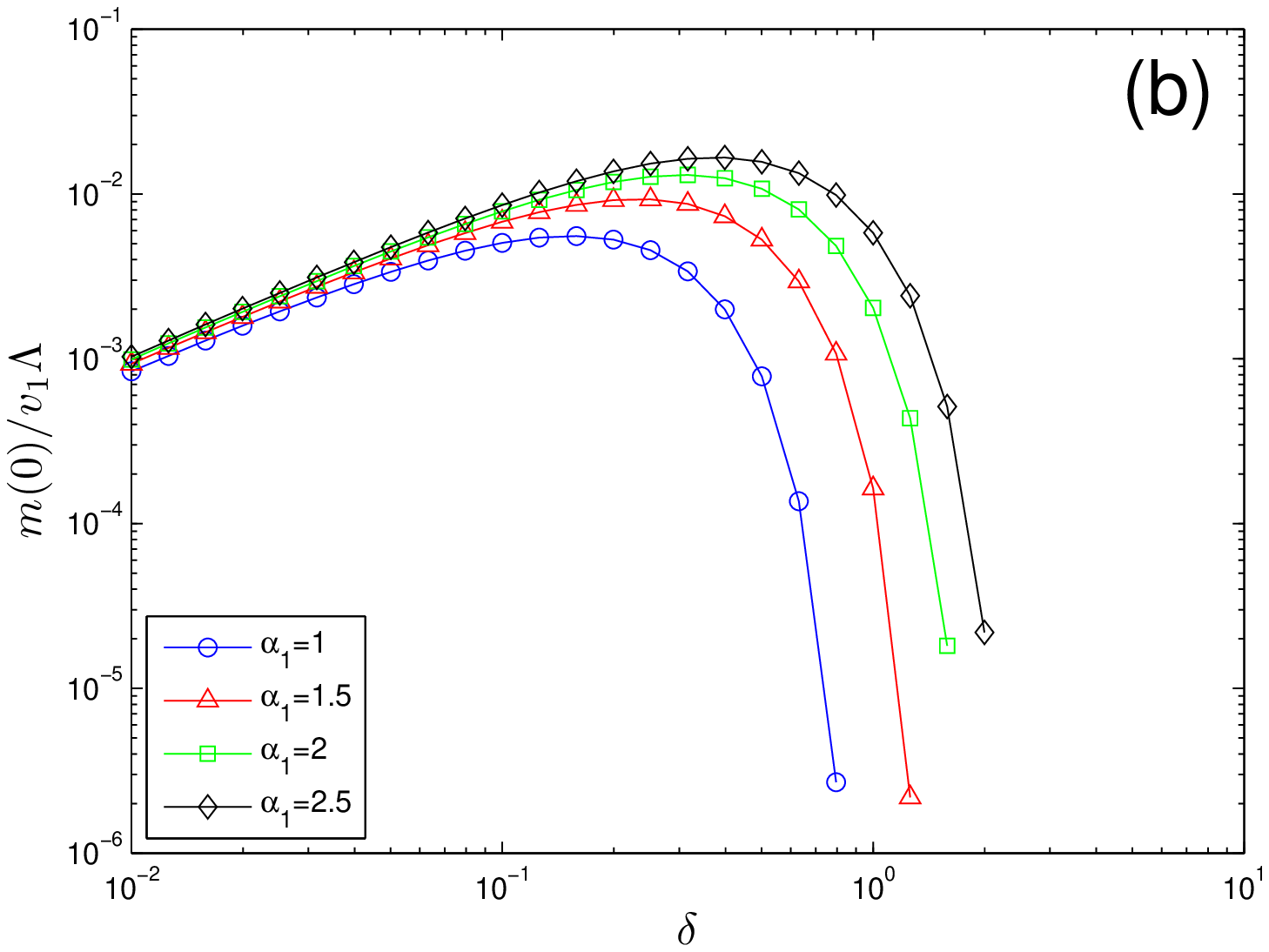}}
   \subfigure{
   \includegraphics[width=2.1in]{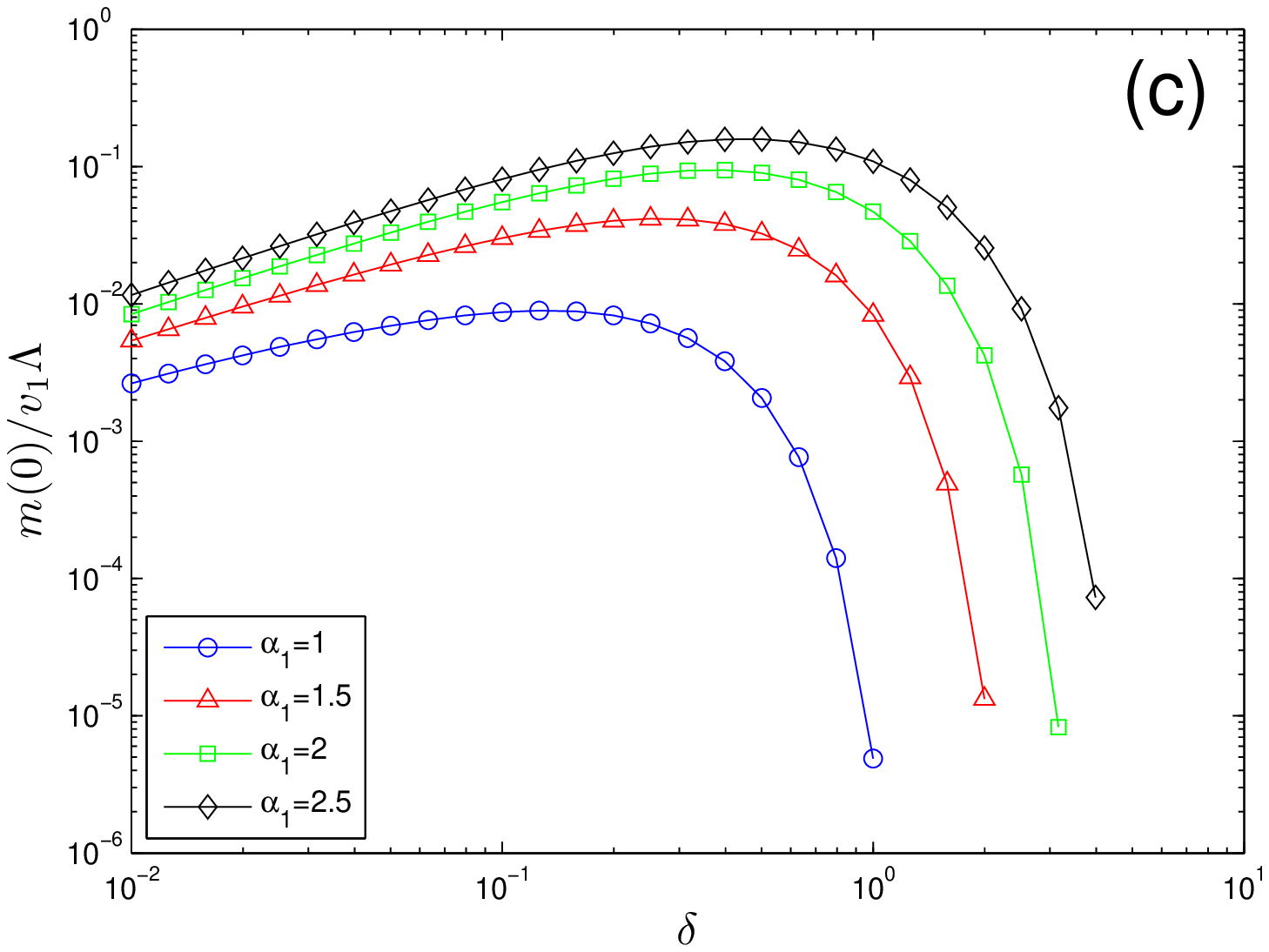}}
   \subfigure{
   \includegraphics[width=2.1in]{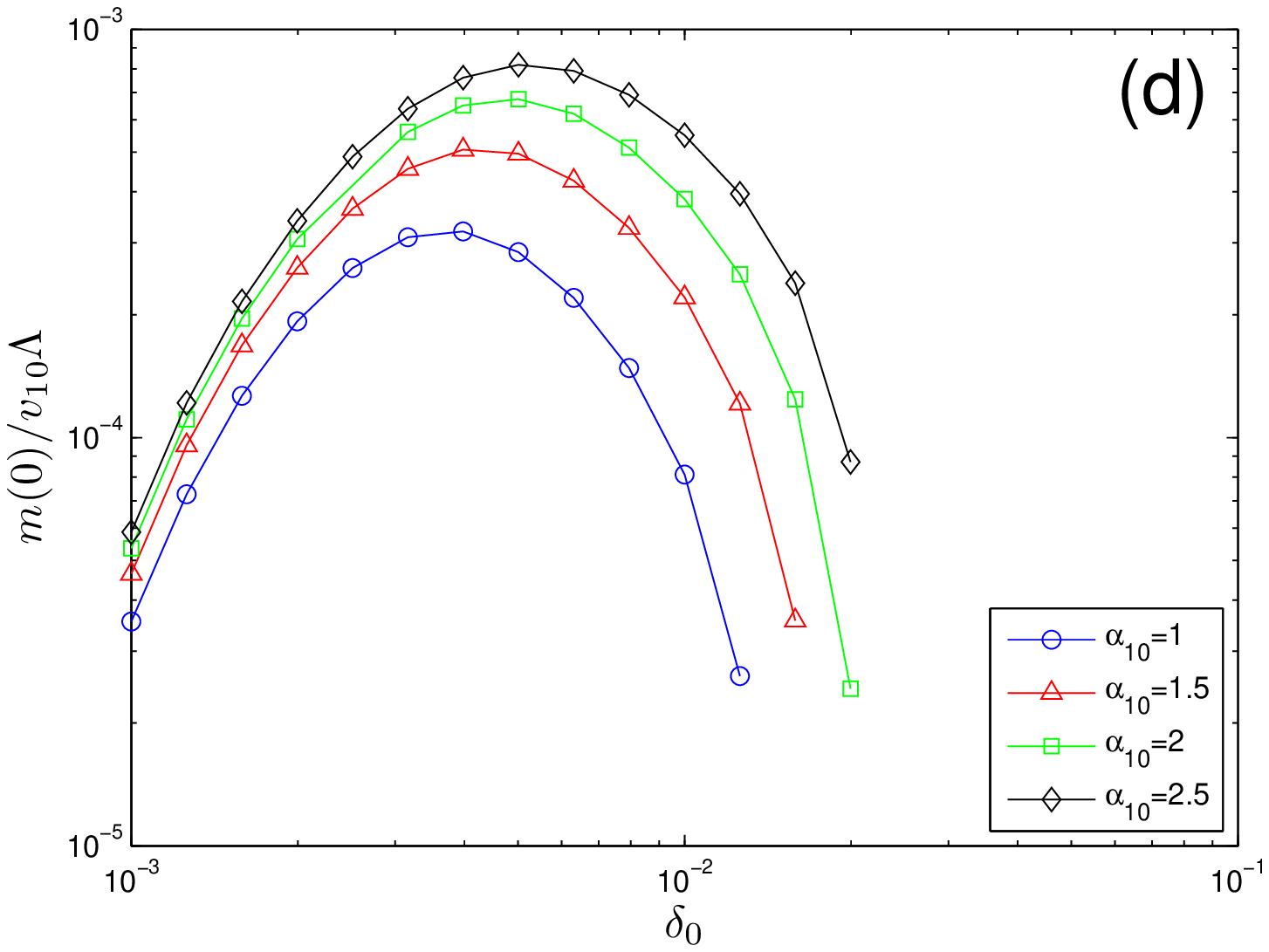}}
   \subfigure{
   \includegraphics[width=2.1in]{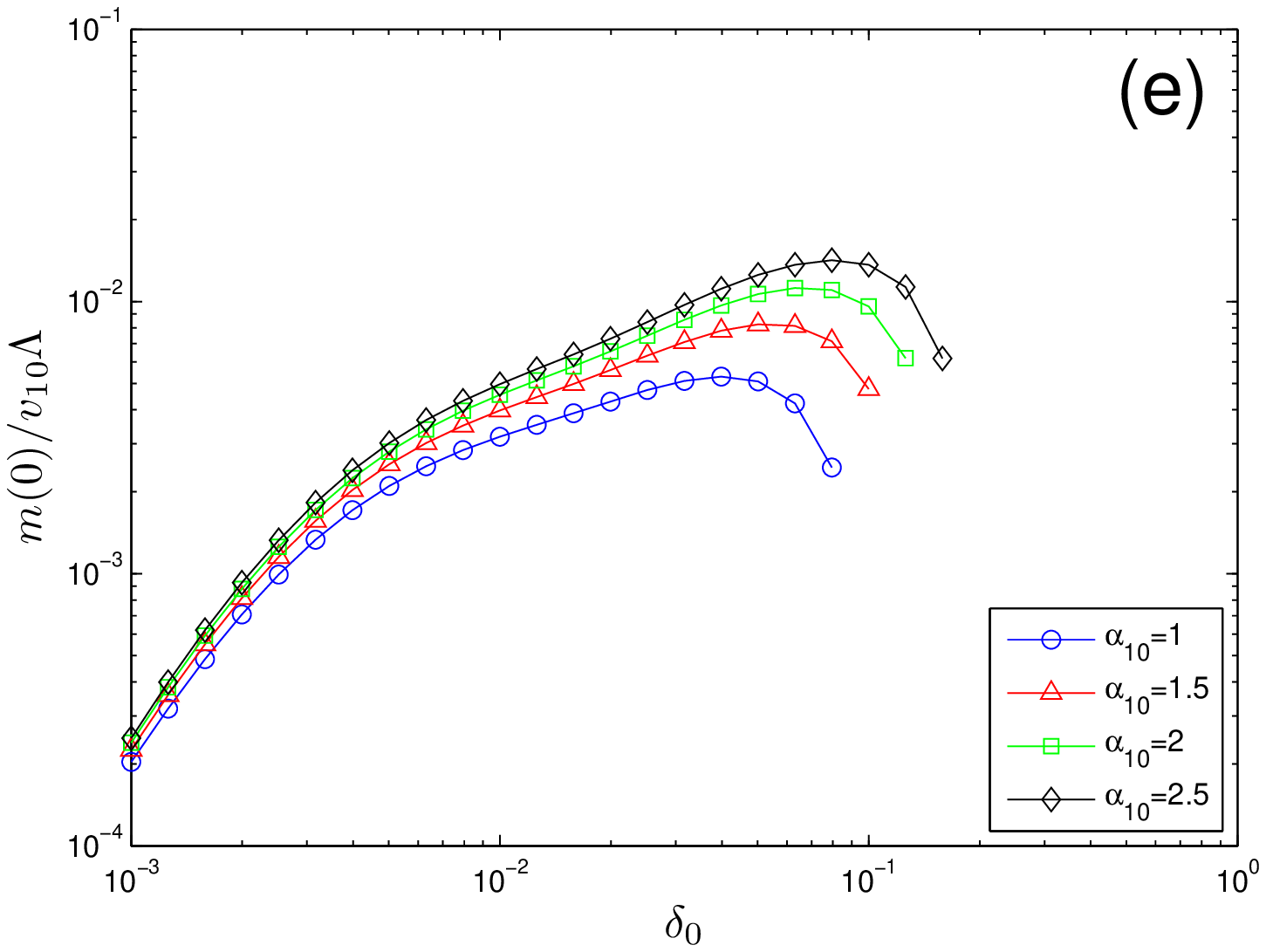}}
   \subfigure{
   \includegraphics[width=2.1in]{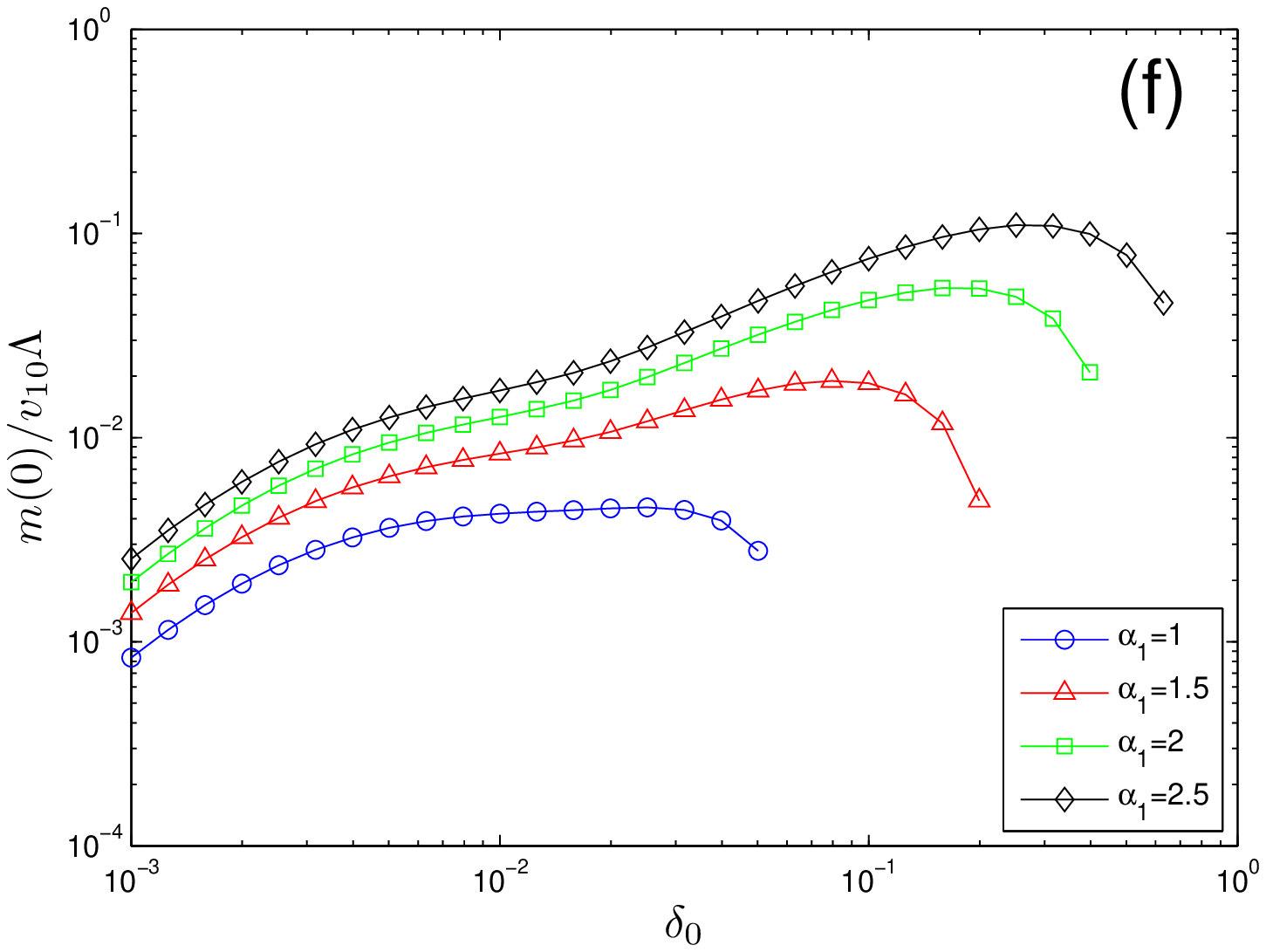}}
\caption{Dependence of dynamical gap $m(0)$ on bare velocity ratio
$\delta$ obtained under a series of approximations: (a)
Approximation (\ref{eqn:ApproxInstan}); (b) Approximation
(\ref{eqn:ApproxKhv}); (c) Approximation (\ref{eqn:ApproxGusynin}).
In (d), (e) and (f), the bare fermion velocities used to obtain the
results of (a), (b) and (c) are replaced by the renormalized
velocities obtained from the solutions of (\ref{eqn:RGV1AsCl}) and
(\ref{eqn:RGV2AsCl}).} \label{Fig:Gap0}
\end{figure*}

We present the numerical results for the dependence of dynamical gap
$m(0)$ on $\delta$ and $\alpha_{1}$ obtained by applying the first
three approximations in (a), (b) and (c) of Fig.~\ref{Fig:Gap0},
respectively. We then replace the bare velocities appearing in
Eq.~(\ref{eqn:GapInstan}), Eq.~(\ref{eqn:GapKhv}) and
Eq.~(\ref{eqn:GapGusynin}) by the corresponding renormalized
velocities, and show the results in (d), (e) and (f) of
Fig.~\ref{Fig:Gap0}, respectively. From these six figures, we see
that the quantitative results of dynamical gap are very sensitive to
the concrete approximations and significantly differ from each
other.

Strictly speaking, all the results of $m(0)$ presented in
Fig.~\ref{Fig:Gap0} may not correspond to the precise values of the
dynamical gap. Nevertheless, one can extract a common feature from
the results obtained in all these six cases: at some fixed coupling
$\alpha_{1} = e^{2}/v_{1}\epsilon$, the dependence of $m(0)$ on bare
velocity ratio $\delta$ is not monotonic. As $\delta$ is growing
from zero, the dynamical gap $m(0)$ first increases, then reaches
its maximal value at certain critical ratio $\delta_c$, and finally
decreases rapidly. This common feature is independent of the
concrete magnitudes of the coupling constant $\alpha_{1}$, provided
that $\alpha_{1}$ is sufficiently large. Certainly, the precise
positions of the peaks of $m(0)$ are strongly case dependent.

We first look at the results presented in Fig.~\ref{Fig:Gap0}(a-c).
At certain fixed ratio $\delta$, we see that the dynamical gap is
always enhanced as the coupling $\alpha_{1}$ increases, which in
turn drives both $v_{1}$ and $v_{2}$ to decrease for a given
$\varepsilon$. If we fix the value of $\alpha_{1}$ and increase the
ratio $\delta$, the dynamical gap is initially enhanced but then
gets suppressed once $\delta$ exceeds some critical value. For fixed
$\alpha_{1}$, the deceasing of $\delta$ from $1$ results in two
effects: reduction of $v_{2}$ and enhancement of velocity
anisotropy. The non-monotonic dependence of the dynamic gap on
$\delta$ implies that these two effects are competing with each
other. Since the first effect always enhances dynamical gap, the
second effect should always suppress dynamical gap. At fixed
coupling $\alpha_{1}$, increasing $\delta$ from $1$ also leads to
two effects: growth of $v_{2}$ and enhancement of velocity
anisotropy. Both of these two effects are capable of suppressing the
dynamical gap. Indeed, Fig.~\ref{Fig:Gap0}(a-c) clearly shows that
the dynamical gap is always suppressed as $\delta$ increases from
$\delta = 1$.

It is also interesting to make a comparison between
Fig.~\ref{Fig:Gap0}(a) and Fig.~\ref{Fig:Gap0}(d). For smaller values
of bare ratio $\delta_0$, the velocity renormalization promotes the
happening of dynamical gap generation. However, for relatively
larger values of $\delta_0$, the velocity renormalization suppresses
the dynamical gap. The same conclusion can be drawn if we compare
Fig.~\ref{Fig:Gap0}(b) with Fig.~\ref{Fig:Gap0}(e), and compare
Fig.~\ref{Fig:Gap0}(c) with Fig.~\ref{Fig:Gap0}(f).

In conclusion, our calculations have shown that, the dynamical gap
is enhanced (suppressed) as the fermion velocities decrease
(increase), but is always suppressed as the velocity anisotropy
increases. Apparently, the velocity anisotropy turns out to be a
negative factor for the happening of dynamical gap generation.

In this section, we have acquired only the $\delta$-dependence of
dynamical gap for several fixed values of coupling $\alpha_{1}$.
Unfortunately, it is difficult to obtain a quantitatively reliable
$\delta$-dependence of critical coupling $\alpha_{1c}$ which
separates the semimetal and insulating phases, primarily because of
the complexity of anisotropic DS equation. However, the unusual
$\delta$-dependence of dynamical gap presented in
Fig.~\ref{Fig:Gap0} suggests that it is both interesting and
necessary to solve the anisotropic DS equation more precisely. We
expect large scale Monte Carlo simulation \cite{Hands08, Drut09} can
be performed to investigate this issue and clarify some crucial
problems.

\section{Summary and discussions \label{Sec:Summ}}

In this paper, we have investigated the influence of long-range
Coulomb interaction on various properties of Dirac fermions in the
context of graphene with a spatial anisotropy by performing detailed
RG calculations based on $1/N$ expansion. We find the renormalized
fermion velocities increase monotonously as the energy scale
decreases and the system approaches a stable isotropic fixed point
in the low-energy regime.

The effects of three types of static disorders, including random
chemical potential, random gauge potential, and random mass, are
also examined using RG techniques. We have shown that the interplay
of Coulomb interaction and fermion-disorder coupling leads to a
series of unusual behaviors. In the case of random chemical
potential, the anisotropic system approaches an isotropic fixed
point for weak Coulomb interaction. However, when the Coulomb
interaction is sufficiently strong, the fermion velocities are
driven to vanish in finite energy scale and the system is very
likely an anisotropic insulator. On the other hand, both random
gauge potential and random mass turn the anisotropic system to a
stable isotropic fixed point in the low-energy regime, at an
efficiency higher than that in the case of clean anisotropic
graphene. An apparent conclusion is that random chemical potential
leads to very different behaviors compared with random gauge
potential and random mass.

In order to understand the unusual behaviors produced by Coulomb
interaction and disorders more explicitly, we have calculated
several physical quantities, including Landau damping rate, wave
renormalization factor, DOS, and specific heat, after taking into
account singular renormalization of fermion velocities. These
quantities exhibit non-Fermi liquid behaviors in many cases. Once
again, the random chemical potential is found to result in
qualitatively different behaviors of these quantities compared to
the other two disorders.

We have further studied the non-perturbative effects of Coulomb
interaction and included the velocity anisotropy into the DS gap
equation. We have acquired the dependence of dynamical gap on the
coupling $\alpha_{1}$ and the velocity ratio $\delta$, at several
different approximations of the DS equation. Our results demonstrate
that the decreasing (increasing) fermion velocities can enhance
(suppress) the dynamic gap. In addition, increasing velocity
anisotropy tends to weaken the effective strength of Coulomb
interaction and therefore suppressed the dynamical gap.

Recently, with the help of symmetry considerations, Herbut \emph{et
al.} have studied semimetal-insulator transition \cite{Herbut06,
Herbut09} and semimetal-superconductor transition \cite{Roy10,
Roy13} in honeycomb lattices, and systematically considered the
corresponding quantum critical behaviors. It would be interesting to
generalize these studies to the case of anisotropic graphene.

\section{Acknowledgement}

We thank Jing Wang for helpful discussions. J.R.W. acknowledges the
support by the MPG-CAS doctoral promotion programme. G.Z.L.
acknowledges the support by the National Natural Science Foundation
of China under grants No. 11074234 and No. 11274286.

\appendix

\section{Fermion self-energy correction due to Coulomb interaction \label{Sec:AppendixA}}

In this appendix, we provide the details for the calculations of
fermion self-energy due to Coulomb interaction. The self-energy is
given by
\begin{eqnarray}
\Sigma_{C}(i\omega,\mathbf{k}) &=&
-\int\frac{d^2{\mathbf{q}}}{(2\pi)^{2}}
\int\frac{d\Omega}{2\pi}\gamma_{0}
G_{0}(i(\Omega+\omega),\mathbf{q}+\mathbf{k})\gamma_{0} \nonumber
\\
&& \times V(i\Omega,\mathbf{q}).
\end{eqnarray}
An ultraviolet cutoff is introduced by multiplying both fermion
propagator and boson propagator by a smooth cutoff function
$\mathcal{K}(\mathbf{k}^2/\Lambda^2)$. Here $\mathcal{K}(y)$ is an
arbitrary function with $\mathcal{K}(0)=1$. It falls off rapidly
with $y$ at $y\sim1$, e. g., $\mathcal{K}(y)=e^{-y}$. However, the
results will be independent of the particular choices of
$\mathcal{K}(y)$. Now the self-energy becomes
\begin{eqnarray}
\Sigma_{C}(i\omega,\mathbf{k})
&=&-\int\frac{d^2{\mathbf{q}}}{(2\pi)^{2}}
\int\frac{d\Omega}{2\pi}\gamma_{0}
G_{0}(i(\Omega+\omega),\mathbf{q}+\mathbf{k})\gamma_{0}\nonumber
\\
&&\times V(i\Omega,\mathbf{q})
\mathcal{K}\left(\frac{(\mathbf{q}+\mathbf{k})^2}{\Lambda^2}\right)
\mathcal{K}\left(\frac{\mathbf{q}^2}{\Lambda^2}\right).
\end{eqnarray}
Namely,
\begin{eqnarray}
\Sigma_{C}(K)&=&-\int\frac{d^3Q}{(2\pi)^3}F(Q+K)V(Q)\mathcal
{K}\left(\frac{(\mathbf{q}+\mathbf{k})^2}{\Lambda^2}\right)\nonumber
\\
&&\times\mathcal {K}\left(\frac{\mathbf{q}^2}{\Lambda^2}\right),
\end{eqnarray}
where
\begin{eqnarray}
&& F(\Omega+\omega,\mathbf{q}+\mathbf{k}) \nonumber \\
&& = \frac{i\left(\Omega+\omega\right)\gamma_{0} +
v_{1}(q_{1}+k_{1})\gamma_{1} +
v_{2}(q_{2}+k_{2})\gamma_{2}}{\left(\Omega +
\omega\right)^2+v_{1}^{2} \left(q_{1} + k_{1}\right)^2 +
v_{2}^{2}\left(q_{2}+k_{2}\right)^2}
\end{eqnarray}
and $K\equiv (\omega,\mathbf{k})$ and $Q \equiv
(\Omega,\mathbf{q})$ are $3$-momenta. One can make the
following expansion to the first order of $K_{\mu}$,
\begin{eqnarray}
&& F(Q+K)\mathcal{K} \left(\frac{(\mathbf{q} +
\mathbf{k})^2}{\Lambda^2}\right) \nonumber \\
&\approx&K_\mu\left[\frac{\partial F(Q)}{\partial Q_\mu} \mathcal
{K}\left(\frac{\mathbf{q}^2}{\Lambda^2}\right) +F(Q)\frac{2q_\mu
}{\Lambda^2}\mathcal
{K'}\left(\frac{\mathbf{q}^2}{\Lambda^2}\right)\right],
\end{eqnarray}
where $K_{\mu}=(\omega,\mathbf{k})$, $k_{\mu}=(0,\mathbf{k})$.
Therefore, the self-energy is rewritten as
\begin{eqnarray}
\Sigma_{C}(K) &=& -K_{\mu}\int\frac{d^3Q}{(2\pi)^3}
\left[\frac{\partial F(Q)}{\partial Q_\mu} V(Q)\mathcal
{K}^{2}\left(\frac{\mathbf{q}^2}{\Lambda^2}\right)\right.
\nonumber \\
&& \left.+F(Q)V(Q)\frac{2q_\mu }{\Lambda^2}\mathcal
{K}\left(\frac{\mathbf{q}^2}{\Lambda^2}\right)\mathcal
{K'}\left(\frac{\mathbf{q}^2}{\Lambda^2}\right)\right],
\end{eqnarray}
which yields
\begin{eqnarray}
\frac{d \Sigma_{C}(K)}{d \ln\Lambda}
&=&K_{\mu}\int\frac{d^3Q}{(2\pi)^3}\left\{\left[\frac{\partial
F(Q)}{\partial Q_\mu} \frac{4\mathbf{q}^2}{\Lambda^2}
+F(Q)\frac{4q_\mu }{\Lambda^2}\right]\right.\nonumber
\\
&&\times V(Q)\mathcal
{K}\left(\frac{\mathbf{q}^2}{\Lambda^2}\right)\mathcal
{K'}\left(\frac{\mathbf{q}^2}{\Lambda^2}\right)\nonumber
\\
&&+F(Q)V(Q)\frac{4\mathbf{q}^2q_\mu
}{\Lambda^4}\left[\mathcal
{K}'^{2}\left(\frac{\mathbf{q}^2}{\Lambda^2}\right)\right.\nonumber
\\
&&\left.\left.+\mathcal
{K}\left(\frac{\mathbf{q}^2}{\Lambda^2}\right)\mathcal
{K''}\left(\frac{\mathbf{q}^2}{\Lambda^2}\right)\right]\right\}.
\end{eqnarray}
Converting to cylindrical co-ordinates by defining
\begin{eqnarray}
Q_\mu&=&y\Lambda(v_1x,\cos\theta,\sin\theta)\label{eqn:NewCoordinatesDefA},
\\
\hat{Q}_\mu&=&(v_1x,\cos\theta,\sin\theta)\label{eqn:NewCoordinatesDefB},
\\
q_\mu&=&y\Lambda(0,\cos\theta,\sin\theta)\label{eqn:NewCoordinatesDefC},
\\
\hat{q}_\mu&=&(0,\cos\theta,\sin\theta)\label{eqn:NewCoordinatesDefD},
\\
d^3Q&=&y^2\Lambda^3v_1 dxdyd\theta\label{eqn:NewCoordinatesDefE},
\end{eqnarray}
we have
\begin{eqnarray}
&&\frac{d \Sigma_{C}(K)}{d \ln\Lambda}
\nonumber \\
&=&K_{\mu}\frac{v_{1}}{2\pi^3}\int_{-\infty}^{+\infty}dx\int_{0}^{2\pi}
d\theta\left\{\left[\frac{\partial F(\hat{Q})}{\partial \hat{Q}_\mu}
+F(\hat{Q})\hat{q}_{\mu}\right]\right.\nonumber
\\
&&\times V(\hat{Q})\int_{0}^{+\infty}dyy\mathcal
{K}\left(y^2\right)\mathcal
{K'}\left(y^2\right)+F(\hat{Q})V(\hat{Q})\hat{q}_{\mu}\nonumber
\\
&&\left.\times\int_{0}^{+\infty}dyy^3\left[\mathcal
{K}'^{2}\left(y^2\right)+\mathcal {K}\left(y^2\right)
\mathcal{K''}\left(y^2\right)\right]\right\}.
\end{eqnarray}
Since
\begin{eqnarray}
\int_{0}^{+\infty}dyy\mathcal {K}\left(y^2\right)\mathcal
{K'}\left(y^2\right) = -\frac{1}{4}, \\
\int_{0}^{+\infty}dyy^3\left[\mathcal {K}'^{2}\left(y^2\right)
+\mathcal {K}\left(y^2\right)\mathcal {K''}\left(y^2\right)\right] =
\frac{1}{4},
\end{eqnarray}
one can further obtain
\begin{eqnarray}
\frac{d \Sigma_{C}(K)}{d\ln\Lambda} &=& -\frac{v_{1}K_{\mu}}{8\pi^3}
\int_{-\infty}^{+\infty}dx\int_{0}^{2\pi}d\theta\frac{\partial
F(\hat{Q})}{\partial \hat{Q}_\mu} V(\hat{Q}), \nonumber \\
\end{eqnarray}
where
\begin{eqnarray}
F(\hat{Q}) &=&\frac{1}{v_{1}}\frac{ix\gamma_{0}+\cos\theta\gamma_{1}
+(v_{2}/v_{1})\sin\theta\gamma_{2}}{x^2+\cos^2\theta
+(v_{2}/v_{1})^{2}\sin^2\theta},
\\
V(\hat{Q}) &=&v_{1}\mathcal{G}(x,\theta),
\end{eqnarray}
with
\begin{eqnarray}
\mathcal {G}^{-1}(x,\theta)&=&\frac{1}{\frac{2\pi e^{2}}{\epsilon
v_{1}}} +\frac{N}{8(v_2/v_{1})}\nonumber
\\
&&\times\frac{\cos^2\theta+(v_{2}/v_{1})^{2}\sin^{2}\theta}
{\sqrt{x^2+\cos^{2}\theta+(v_{2}/v_{1})^{2}\sin^{2}\theta}}. \nonumber \\
\end{eqnarray}
Using these expressions, we can finally obtain the self-energy
correction given by the Eqs.~(\ref{eqn:SelfEnergyCloumb})-(\ref{eq:C2}).\\

\section{Fermion-Disorder vertex correction due to Coulomb
interaction \label{Sec:AppendixB}}

The correction to fermion-disorder vertex due to Coulomb interaction
is given by
\begin{eqnarray}
V_{C} &=& -\int\frac{d\Omega}{2\pi}
\int\frac{d^2\mathbf{q}}{(2\pi)^2}\gamma_{0}G_{0}(i\Omega,\mathbf{q})v_\Gamma \Gamma
G_{0}(i\Omega,\mathbf{q})\gamma_{0}
\nonumber \\
&& \times V(\Omega,\mathbf{q}).
\end{eqnarray}
One can impose a momentum cutoff by multiplying both fermion and
boson propagators by a smooth function $\mathcal{K}(k^2/\Lambda^2)$,
and then obtain
\begin{eqnarray}
V_{C} = -v_\Gamma\int\frac{d^3Q}{(2\pi)^2}\gamma_{0}G_{0}(Q) \Gamma
G_{0}(Q)\gamma_{0} V(Q) \mathcal{K}^{3}
\left(\frac{\mathbf{q}^2}{\Lambda^2}\right). \nonumber \\
\end{eqnarray}
Therefore, we have
\begin{eqnarray}
\frac{dV_{C}}{d\ln\Lambda} &=& 6v_\Gamma\int\frac{d^3Q}{(2\pi)^3}
\gamma_{0}G_{0}(Q) \Gamma G_{0}(Q)\gamma_{0} V(Q)\nonumber
\\
&& \times \mathcal{K}^{2}\left(\frac{\mathbf{q}^2}{\Lambda^2}\right)
\mathcal{K}'\left(\frac{\mathbf{q}^2}{\Lambda^2}\right)
\left(\frac{\mathbf{q}^2}{\Lambda^2}\right).
\end{eqnarray}
After converting to cylindrical coordinates defined by
Eq.~(\ref{eqn:NewCoordinatesDefA})-(\ref{eqn:NewCoordinatesDefE}),
it is easy to get
\begin{eqnarray}
\frac{dV_{C}}{d\Lambda} &=&v_{\Gamma}
\frac{3v_{1}}{4\pi^3\Lambda}\int_{-\infty}^{+\infty}dx
\int_{0}^{2\pi}d\theta \gamma_{0}G(\hat{Q})\Gamma
G(\hat{Q})\gamma_{0} V(\hat{Q})
\nonumber \\
&& \times \int_{0}^{+\infty}dyy\mathcal{K}^{2}
\left(y^{2}\right)\mathcal{K}'\left(y^{2}\right),
\end{eqnarray}
where
\begin{eqnarray}
\int_{0}^{+\infty}dyy\mathcal{K}^{2}\left(y^{2}\right)
\mathcal{K}'\left(y^{2}\right) = -\frac{1}{6}.
\end{eqnarray}
Finally, we obtain
\begin{eqnarray}
\frac{dV_{C}}{d\ln\Lambda} &=&
-v_{\Gamma}\frac{v_{1}}{8\pi^3}\int_{-\infty}^{+\infty}dx
\int_{0}^{2\pi}d\theta H(\hat{Q}),
\end{eqnarray}
where
\begin{eqnarray}
H(\hat{Q})=\gamma_{0}G_{0}(\hat{Q})\Gamma
G_{0}(\hat{Q})\gamma_{0}V(\hat{Q}).
\end{eqnarray}

\section{Symmetric form of $C_{0,1,2}$ and $\mathcal{G}$\label{Sec:AppendixC}}

The parameters $C_{0,1,2}$ nd $\mathcal{G}$ appearing in RG equations can also be written in the following symmetric
form

\begin{eqnarray}
C_{0}&=&\frac{1}{8\pi^3}\int_{0}^{2\pi}d\theta\int_{-\infty}^{+\infty}dx\nonumber
\\
&&\times\frac{x^2-(v_{1}/v_{2})\cos^2\theta-(v_{2}/v_{1})\sin^2\theta}
{\left(x^2+(v_{1}/v_{2})\cos^2\theta+(v_{2}/v_{1})\sin^2\theta\right)^{2}}
\mathcal{G}(x,\theta),\nonumber
\\ \label{eq:C1Sys}
\\
C_{1}&=&\frac{1}{8\pi^3}\int_{0}^{2\pi}d\theta\int_{-\infty}^{+\infty}dx\nonumber
\\
&&\times\frac{x^2-(v_{1}/v_{2})\cos^2\theta+(v_{2}/v_{1})\sin^2\theta}
{\left(x^2+(v_{1}/v_{2})\cos^2\theta+(v_{2}/v_{1})\sin^2\theta\right)^{2}}
\mathcal{G}(x,\theta),\nonumber
\\
\\ \label{eq:C2Sys}
C_{2}&=&\frac{1}{8\pi^3}\int_{0}^{2\pi}d\theta\int_{-\infty}^{+\infty}dx\nonumber
\\
&&\times\frac{x^2+(v_{1}/v_{2})\cos^2\theta-(v_{2}/v_{1})\sin^2\theta}
{\left(x^2+(v_{1}/v_{2})\cos^2\theta+(v_{2}/v_{1})\sin^2\theta\right)^{2}}
\mathcal{G}(x,\theta),\nonumber
\\ \label{eq:C3Sys}
\end{eqnarray}
with
\begin{eqnarray}
\mathcal{G}^{-1}(x,\theta)&=&\frac{\epsilon \sqrt{v_{1}v_{2}}}{2\pi e^2}+\frac{N}{8}\nonumber
\\
&&\times\frac{(v_{1}/v_{2})\cos^2\theta+(v_{2}/v_{1})\sin^2\theta}
{\sqrt{x^2+(v_{1}/v_{2})\cos^2\theta+(v_{2}/v_{1})\sin^2\theta}}.\nonumber
\\ \label{eq:GDefSys}
\end{eqnarray}
The expressions for $C_{0,1,2}$ and $\mathcal{G}$ shown in Eqs.~(\ref{eq:C1Sys})-(\ref{eq:GDefSys})
are indeed equivalent to those in Eqs.~(\ref{eq:C0})-(\ref{eq:GDef}).

\section{Derivation for the differential equations of DOS and specific heat\label{sec:AppendixD}}

We now study the influence of Coulomb interaction on DOS using the method employed by Xu \emph{et. al.}
\cite{Xu}. The DOS $\rho(\omega)$ is defined as
\begin{eqnarray}
\rho(\omega)&=&N\int\frac{dk_{1}dk_{2}}{(2\pi)^{2}}
\mathrm{Tr}\left\{\mathrm{Im}
\left[G^{R}(\omega,v_{1}k_{1},v_{2}k_{2})\right] \right\}
\nonumber \\
&=& \frac{N}{v_{1}v_{2}}\int\frac{dk_{1}'dk_{2}'}{(2\pi)^{2}}
\mathrm{Tr}\left\{\mathrm{Im}
\left[G^{R}(\omega,k_{1}',k_{2}')\right] \right\}\label{eqn:RhoExpresssion},
\end{eqnarray}
where $G^{R}(\omega,k_1,k_2)$ is the retarded propagator of Dirac fermions.
According to the method in Xu \emph{et. al.} \cite{Xu}, DOS $\rho(\omega)$ and
specific heat $C_{V}(T)$ can be calculated through the differential equations $\frac{d\ln\rho}{d\ln\omega}$
and $\frac{d\ln C_{V}}{d\ln T}$ respectively. The qualitative behavior of
$\rho(\omega)$ is related to both the fermion anomalous dimension, $\eta_{f}=-C_{0}+C_{g}$,
and the dynamical exponents $z_{1,2}$ which are encoded in the flow of fermion velocities $v_{1,2}$.
However, the qualitative behavior of  $C_{V}(T)$ is only related to the dynamical
exponents $z_{1,2}$. Therefore, one should represent $\frac{d\ln\rho}{d\ln\omega}$
in terms of the fermion anomalous dimension $\eta_{f}$ and  the RG equations of $v_{1,2}$,
but express $\frac{d\ln C_{V}}{d\ln T}$ only through the RG equations of $v_{1,2}$. In
Sec.~\ref{Subsec:DerivationRG}, we have obtained the $l$-dependence of the RG equations of $v_{1,2}$,
as well as $C_{0,g}$. In order to get $\frac{d\ln\rho}{d\ln\omega}$
and $\frac{d\ln C_{V}}{d\ln T}$, we need to replace $\frac{d}{d\ln\omega}$ or $\frac{d}{d\ln T}$ with
$\frac{d}{d l}$.

At certain given energy $\omega$, the corresponding momentum scale should
be determined  by the larger component of the fermion velocities\cite{Xu}, namely
\begin{eqnarray}
\tilde{p}=\frac{\omega}{\max\left(v_{1},v_{2}\right)},
\end{eqnarray}
which leads to
\begin{eqnarray}
&&\frac{d\ln\omega}{d\ln\tilde{p}} = 1 +
\frac{d\ln\max\left(v_{1},v_{2}\right)}{d\ln\tilde{p}}.
\end{eqnarray}
Now the scaling equation for $\rho(\omega)$ takes the form as
\begin{eqnarray}
\frac{d\ln\rho}{d\ln\omega} &=&
\frac{d\ln\rho}{d\ln\tilde{p}\frac{d\ln\omega}{d\ln\tilde{p}}}
\nonumber \\
&=& \frac{d\ln\rho}{d\ln\tilde{p}\left(1+\frac{d\ln
\max\left(v_{1},v_{2}\right)}{d\ln\tilde{p}}\right)}.
\end{eqnarray}
Since $\tilde{p}\sim\Lambda e^{-l}$ which leads to
$d\ln\tilde{p}\sim -dl$, we can get
\begin{eqnarray}
\frac{d\ln\rho}{d\ln\omega}&=&- \frac{d\ln\rho}{dl\left(1 -
\frac{d\ln \max\left(v_{1},v_{2}\right)}{dl}\right)},
\end{eqnarray}
namely
\begin{eqnarray}
\frac{d\ln\rho}{d\ln\omega}&=&-\frac{-(1 + C_{0} - C_{g}) +
\frac{d\ln\left(\frac{1}{v_{1}v_{2}}\right)}{dl}}{\left(1 -
\frac{d\ln \max\left(v_{1},v_{2}\right)}{dl}\right)}. \label{eqn:RhoExpressionMiddleA}
\end{eqnarray}
The first term $-(1+C_{0}-C_{g})$ in the numerator comes form
the scaling exponent of the fermion propagator in Eq.~(\ref{eqn:RhoExpresssion}).
The second term $d\ln(\frac{1}{v_{1}v_{2}})/dl$ is induced by the prefactor $\frac{1}{v_{1}v_{2}}$ in
Eq.~(\ref{eqn:RhoExpresssion}). Eq.~(\ref{eqn:RhoExpressionMiddleA}) can be
further written as
\begin{eqnarray}
\frac{d\ln\rho}{d\ln\omega}&=&\frac{1 + C_{0} - C_{g} +
\frac{d\ln v_{1}}{dl}+\frac{d\ln v_{2}}{dl}}{\left(1 -
\frac{d\ln \max\left(v_{1},v_{2}\right)}{dl}\right)}. \label{eqn:RhoExpressionMiddleB}
\end{eqnarray}
Using the RG equations of $v_{1}$ and $v_{2}$, i.e. Eq.~(\ref{eqn:RGV1}) and Eq.~(\ref{eqn:RGV2}),
the above equation can be simplified to
\begin{eqnarray}
\frac{d\ln\rho}{d\ln\omega} &=&
\frac{1+3C_{0}-C_{1}-C_{2}-3C_{g}}{1-C_{0}+C_{1}+C_{g}}
\end{eqnarray}
for $v_{1}>v_{2}$, and
\begin{eqnarray}
\frac{d\ln\rho}{d\ln\omega} &=&
\frac{1+3C_{0}-C_{1}-C_{2}-3C_{g}}{1-C_{0}+C_{2}+C_{g}}
\end{eqnarray}
for $v_{2}>v_{1}$.

To calculate the specific heat, we also follow the method used in
Ref.~\cite{Xu}. The free energy $\mathcal{F} = T\ln\mathcal{Z}/V$
contains the following singular part
\begin{eqnarray}
\mathcal{F} = \left(\xi_{\tau}\xi_{x}\xi_{y}\right)^{-1},
\end{eqnarray}
where $\xi_{\tau}\sim1/T$, $\xi_{x}\sim v_{1}\xi_{\tau}$, and
$\xi_{y}\sim v_{2}\xi_{\tau}$. In a interacting anisotropic graphene, the free
energy if found to behave like
\begin{eqnarray}
\mathcal{F} \sim \frac{1}{v_{1}v_{2}}T^3.
\end{eqnarray}
The corresponding specific heat is given by
\begin{eqnarray}
C_{V} = -T\frac{\partial^2 \mathcal{F}}{\partial
T^2}\sim\frac{1}{v_{1}v_{2}}T^2.
\end{eqnarray}
After taking differentiation with respect to $T$, we get
\begin{eqnarray}
\frac{d\ln C_{V}}{d\ln T}=2+\frac{d\ln\left(\frac{1}{v_{1}v_{2}}\right)}{d\ln T}\label{eqn:SpecificHeatMiddleA}.
\end{eqnarray}
At certain given temperature $T$,
the corresponding momentum scale should be determined by the larger component
of the fermion velocities as \cite{Xu}
\begin{eqnarray}
\tilde{p}=\frac{T}{\max(v_{1},v_{2})},
\end{eqnarray}
which leads to
\begin{eqnarray}
\frac{d\ln T}{d\ln\tilde{p}} = 1 + \frac{d\ln
\max\left(v_{1},v_{2}\right)}{d\ln\tilde{p}}.
\end{eqnarray}
Now the scaling equation for $C_{V}$ becomes
\begin{eqnarray}
\frac{d \ln C_{V}}{d\ln T} &=& 2 +
\frac{d\ln\left(\frac{1}{v_{1}v_{2}}\right)}{d\ln\tilde{p}
\frac{d\ln T}{d\ln\tilde{p}}}\nonumber
\\
&=& 2 +\frac{d\ln\left(\frac{1}{v_{1}v_{2}}\right)}{d\ln\tilde{p}
\left(1+\frac{d\ln \max\left(v_{1},v_{2}\right)}{d\ln
\tilde{p}}\right)}.
\end{eqnarray}
Using the expression $d\ln\tilde{p}\sim -dl$, it is easy to get
\begin{eqnarray}
\frac{d \ln C_{V}}{d\ln T}
&=& 2-\frac{d\ln\left(\frac{1}{v_{1}v_{2}}\right)}{dl
\left(1-\frac{d\ln \max\left(v_{1},v_{2}\right)}{dl}\right)}\nonumber
\\
&=& 2+\frac{\frac{d\ln v_{1}}{dl}+\frac{d\ln v_{2}}{dl}}{
\left(1-\frac{d\ln \max\left(v_{1},v_{2}\right)}{dl}\right)}.\label{eq:CVMiddle}
\end{eqnarray}
After substituting  Eq.~(\ref{eqn:RGV1}) and Eq.~(\ref{eqn:RGV2}) into (\ref{eq:CVMiddle}) ,
we finally obtain
\begin{eqnarray}
\frac{d \ln C_{V}}{d\ln T} = 2 + \frac{2C_{0} -
C_{1}-C_{2}-2C_{g}}{1-C_{0}+C_{1}+C_{g}}
\end{eqnarray}
for $v_{1}>v_{2}$, and
\begin{eqnarray}
\frac{d \ln C_{V}}{d\ln T} =2 +
\frac{2C_{0}-C_{1}-C_{2}-2C_{g}}{1-C_{0}+C_{2}+C_{g}}
\end{eqnarray}
for $v_{2}>v_{1}$.

\section{Different dependence of DOS and specific heat on a positive
anomalous dimension\label{sec:AppendixE}}

In this appendix, we would like to demonstrate the different
influence of a finite positive anomalous dimension on DOS and
specific heat. This may help us to understand the results obtained
in Sec.~\ref{Subec:DOSBehavior} and \ref{subsec:CvBehavior}. For
this purpose, it is convenient to consider a generic model of
interacting Dirac fermions.

Let us start from a free Dirac fermion propagator with an isotropic
dispersion,
\begin{eqnarray}
G_{0}(i\omega_{n},\mathbf{k}) &=&
\frac{1}{i\omega_{n}\gamma_0-v_{F}\mathbf{\gamma\cdot k}}
\nonumber \\
&=& \frac{-i\omega_{n}\gamma_0+v_{F}\mathbf{\gamma\cdot
k}}{\omega_{n}^2+v_{F}^{2}k^2},
\end{eqnarray}
where $\omega_{n}=(2n+1)\pi T$ is the Matsubara frequency. Carrying
out analytic continuation $i\omega_{n} \rightarrow \omega+i\delta$,
we can get the retarded propagator
\begin{eqnarray}
G_{0}^{R}(\omega,\mathbf{k})&=&
\left[\mathcal{P}\frac{1}{\omega^2-v_{F}^{2}k^2}
-i\pi\mathrm{sgn}(\omega)\delta(\omega^2-v_{F}^{2}k^2)\right]
\nonumber \\
&&\times \left(\omega\gamma_{0} -
v_{F}\mathbf{\gamma}\cdot\mathbf{k}\right).
\end{eqnarray}
From this propagator, it is easy to get a spectral function
\begin{eqnarray}
A_{0}(\omega,\mathbf{k})&=&-\frac{1}{\pi}
\mathrm{Tr}\left[\gamma_0\mathrm{Im}G_{0}^{R}(\omega,\mathbf{k})\right]
\nonumber \\
&=&2\left[\delta\left(v_{F}k-|\omega|\right) +
\delta\left(v_{F}k+|\omega|\right)\right].
\end{eqnarray}
The fermion DOS can be computed directly, i.e.,
\begin{eqnarray}
\rho_{0}(\omega)&=&
N\int\frac{d^2\mathbf{k}}{(2\pi)^2}A_{0}(\omega,\mathbf{k})
=\frac{N}{\pi v_{F}^{2}}|\omega|.\label{eqn:DOSFree}
\end{eqnarray}
The free energy for free fermions is therefore
\begin{eqnarray}
F_{0}(T) &=& 4NT\sum_{\omega_{n}} \int\frac{d^2\mathbf{k}}{(2\pi)^2}
\ln\left[\left(\omega_{n}^2 +
v_{F}^{2}k^2\right)^{\frac{1}{2}}\right]\nonumber
\\
&=&2N\int\frac{d^2k}{(2\pi)^2} T\sum_{\omega_{n}}
\ln\left[\omega_{n}^2+v_{F}^{2}k^2\right].
\end{eqnarray}
The summation over frequency $\omega_{n}$ can be easily performed, leading to
\begin{eqnarray}
F_{0}(T) =2N\int\frac{d^2\mathbf{k}}{(2\pi)^2}
\left[v_{F}k-2T\ln\left(1+e^{-\frac{v_{F}k}{T}}\right)\right],
\end{eqnarray}
which is clearly divergent. In order to get a finite free energy, we
redefine $F_{0}(T)-F_{0}(0)$ as $F_{0}(T)$ and get
\begin{eqnarray}
F_{0}(T)
&=&-4NT\int\frac{d^2\mathbf{k}}{(2\pi)^2}
\ln\left[1+e^{-\frac{v_{F}k}{T}}\right]
\nonumber \\
&=& -\frac{3N\zeta(3)}{2\pi v_{F}^{2}}T^3.
\label{eq:SpecificHeatFreeDef}
\end{eqnarray}
The corresponding specific heat is
\begin{eqnarray}
C_{V0} = -T\frac{\partial^2 F_{0}(T)}{\partial T^2}
=\frac{9N\zeta(3)}{\pi v_{F}^{2} }T^2. \label{eqn:SpecificHeatFree}
\end{eqnarray}

Now suppose the fermion propagator acquires a finite positive
anomalous dimension $\eta$ due to some interaction \cite{Herbut09,
KhveshchenkoPasske, Franz02, Sachdev10}, yielding
\begin{eqnarray}
G(i\omega_{n},\mathbf{k}) &=& \frac{1}{(i\omega_{n}\gamma_0 -
v_{F}\mathbf{\gamma\cdot k})\left(\frac{\sqrt{\omega_{n}^2 +
v_{F}^{2}k^2}}{v_{F}\Lambda}\right)^{-\eta}}
\nonumber \\
&=&\frac{-(i\omega_{n}\gamma_0-v_{F}\mathbf{\gamma\cdot
k})}{(v_{F}\Lambda)^{\eta}(\omega_{n}^2 +
v_{F}^{2}k^2)^{1-\frac{\eta}{2}}}.
\end{eqnarray}
The retarded propagator is therefore given by
\begin{eqnarray}
&&G^{R}(\omega,\mathbf{k})\nonumber
\\
&=&\theta(v_{F}k-|\omega|)\nonumber
\\
&&\times\left[\mathcal{P}\frac{1}{\omega^2-v_{F}^{2}k^2}
-i\pi\mathrm{sgn}(\omega)\delta(\omega^2-v_{F}^{2}k^2)\right]
\nonumber \\
&&\times\frac{(\omega\gamma_0 - v_{F}\mathbf{\gamma\cdot
k})}{(v_{F}\Lambda)^{\eta}
\left(\sqrt{v_{F}^{2}k^2-\omega^2}\right)^{-\eta}}
\nonumber \\
&& +\theta(|\omega|-v_{F}k)\nonumber \\
&&\times\left[\mathcal{P}\frac{1}{\omega^2-v_{F}^{2}k^2} -
i\pi\mathrm{sgn}(\omega)\delta(\omega^2-v_{F}^{2}k^2)\right]
\nonumber \\
&&\times\frac{(\omega\gamma_0 - v_{F}\mathbf{\gamma\cdot
k})}{(v_{F}\Lambda)^{\eta}
\left(\sqrt{\omega^2-v_{F}^{2}k^2}\right)^{-\eta}}
\nonumber \\
&&\times\left(\cos\left(\frac{\pi\eta}{2}\right)
-\mathrm{sgn}(\omega)i\sin\left(\frac{\pi\eta}{2}\right)\right),
\end{eqnarray}
which results in the following spectral function
\begin{eqnarray}
A(\omega,\mathbf{k})&=&-\frac{1}{\pi}
\mathrm{Tr}\left[\gamma_0\mathrm{Im}G^{R}(\omega,\mathbf{k})\right]
\nonumber \\
&=& \frac{4}{\pi} \frac{\theta(|\omega|-v_{F}k)|\omega|
\sin\left(\frac{\pi\eta}{2}\right)}{(v_{F}\Lambda)^{\eta}
\left(\sqrt{\omega^2-v_{F}^{2}k^2}\right)^{2-\eta}} .
\end{eqnarray}
Now the fermion DOS depends on $\eta$ as
\begin{eqnarray}
\rho(\omega) &=& N\int\frac{d^2\mathbf{k}}{(2\pi)^2}
A(\omega,\mathbf{k})
\nonumber \\
&=& \frac{2N}{\pi^2}\frac{1}{\eta}
\sin\left(\frac{\pi\eta}{2}\right)\frac{|\omega|^{1+\eta}}
{v_{F}^{2}(v_{F}\Lambda)^{\eta}}, \label{eqn:DOSAnormalousDimension}
\end{eqnarray}
where $\eta$ modifies the $\omega$-dependence of $\rho(\omega)$.
However, the free energy of interacting fermions is
\begin{eqnarray}
F(T) &=& 4NT\sum_{\omega_{n}}\int\frac{d^2\mathbf{k}}{(2\pi)^2}
\ln\left[\left(\omega_{n}^2+v_{F}k^2\right)^{\frac{1}{2} -
\frac{\eta}{2}}\right]
\nonumber \\
&=&\left(1-\eta\right)2N\int\frac{d^2k}{(2\pi)^2}
\nonumber \\
&&\times T \sum_{\omega_{n}}
\ln\left[\omega_{n}^2+v_{F}^{2}k^2\right].
\\
&=&\left(1-\eta\right)2N\int\frac{d^2\mathbf{k}}{(2\pi)^2}
\nonumber \\
&& \times \left[v_{F}k-2T\ln\left(1 +
e^{-\frac{v_{F}k}{T}}\right)\right],
\end{eqnarray}
which is also divergent. Similar to Eq.~(\ref{eq:SpecificHeatFreeDef}),
the finite redefined $F(T)$ has the form
\begin{eqnarray}
F(T)
&=&-\left(1-\eta\right)4NT\int\frac{d^2\mathbf{k}}{(2\pi)^2}
\ln\left[1+e^{-\frac{v_{F}k}{T}}\right]\nonumber
\\
&=& -(1-\eta)\frac{3N\zeta(3)}{2\pi v_{F}^{2}}T^3,
\end{eqnarray}
which leads to the specific heat
\begin{eqnarray}
C_{V}=-T\frac{\partial^2 F(T)}{\partial T^2} =
(1-\eta)\frac{9N\zeta(3)}{\pi v_{F}^{2}}T^2.
\label{eqn:SpecificHeatAnormalousDimension}
\end{eqnarray}

Comparing  Eq.~(\ref{eqn:DOSFree}) with
Eq.~(\ref{eqn:DOSAnormalousDimension}), we see that the fermion DOS
is linear in energy for the free system, but the linear dependence
on energy of DOS is changed once a finite positive anomalous
dimension $\eta$ is generated in the fermion propagator
\cite{Gusynin03, Zhong13}. The quadratic $T$-dependence of specific
heat does not change even if $\eta\neq0$ \cite{Herbut09, Zhong13,
Kaul08}. From Eq.~(\ref{eqn:SpecificHeatAnormalousDimension}), we
see that $\eta$ enters into the specific heat only in the prefactor
of $T^2$.

Now let us go back to the interacting model considered in this
paper. Due to the interplay of Coulomb interaction and random gauge
potential (random mass), the fermion DOS is no longer linear in
$\omega$ in the limit $\omega \rightarrow 0$, but the specific heat
still exhibits quadratic $T$-dependence in the limit $T \rightarrow
0$. The reason for this behavior is that, the fermion velocities
$v_{1}$ and $v_{2}$ approach a constant at the lowest energy, which
means the dynamical exponent $z \rightarrow 1$, whereas the fermion
propagator acquires a finite positive anomalous dimension $\eta =
\lim_{l \rightarrow \infty}(-C_{0}(l) + C_{g}(l))$.

\end{document}